\documentclass{iau}

\usepackage{hyperref}
\usepackage{amsmath}
\usepackage{graphicx}
\usepackage{multirow}
\usepackage{wrapfig}

\def\arcsec{\hbox{$^{\prime\prime}$}}

\def\farcs{\hbox{$.\!\!^{\prime\prime}$}}
\def\gsim{\mathrel{\hbox{\rlap{\lower.55ex \hbox {$\sim$}}
                   \kern-.3em \raise.4ex \hbox{$>$}}}}
\def\lsim{\mathrel{\hbox{\rlap{\lower.55ex \hbox {$\sim$}}
                   \kern-.3em \raise.4ex \hbox{$<$}}}}

\def\he2{\hbox{He\,{\sc ii} $\lambda$4686}}

\def\RL1{\hbox{{$R_{L_{1}}$}}}

\begin{document}
\lefttitle{P.J. Groot et al.}
\righttitle{The Time Domain Telescope }

\jnlPage{}{}
\jnlDoiYr{}
\doival{}

\aopheadtitle{Proceedings IAU Symposium}
\editors{}

\title{The Time Domain Telescope: \\ A Global Spectroscopic Array}

\author{P.J. Groot$^{1,2,3}$, Th. Kupfer$^{4,5}$, K. O'Brien$^6$, S. Scaringi$^{7,8}$, A. Brown$^4$, T. Morris$^{6}$, S. Bloemen$^{1}$, B.T. G\"ansicke$^{9,10,11}$, S. Nissanke$^{12,13}$, and the TDT Collaboration$^{*}$}
\affiliation{
$^1$ Department of Astrophysics/IMAPP, Radboud University, P.O. Box 9010, 6500 GL, Nijmegen, The Netherlands \\
$^2$ Department of Astronomy, University of Cape Town, Private Bag X3, 7701, Rondebosch, South Africa \\
$^3$ South African Astronomical Observatory, P.O. Box 9, 7935 Observatory, South Africa \\
$^4$ Hamburger Sternwarte, University of Hamburg, Gojensbergweg X, Bergesdorf, Germany \\
$^5$ Department of Physics and Astronomy, Texas Tech University,  2500 Broadway,  Lubbock, TX 79409, USA \\
$^6$ Centre for Advanced Instrumentation, Department of Physics, Durham University, South Road, Durham, DH1 3LE, United Kingdom \\
$^7$ Centre for Extragalactic Astronomy, Department of Physics, Durham University, South Road, Durham, DH1 3LE, United Kingdom \\
$^8$ INAF-Osservatorio Astronomico di Capodimonte, Salita Moiariello 16, Naples, 80131, Italy \\
$^9$ Department of Physics, University of Warwick, Coventry, CV4 7AL, UK \\
$^{10}$ Instituto de Astrofisica de Canarias, E-38205 La Laguna, Tenerife, Spain \\
$^{11}$ Centre for Exoplanets and Habitability, University of Warwick,
  Gibbet Hi ll Road, Coventry, CV4 7AL, UK \\
$^{12}$ Deutsches Elektronen-Synchrotron DESY, Platanenallee 6, 15738 Zeuthen, Germany \\
$^{13}$ Deutsche Zentrum für Astrophysik DZA, Postplatz 1, 02826 Görlitz, Germany\\
$^{*}$ Full author list given at the end of the article.\\
\url{www.tdtelescope.org}
}

\begin{abstract}
The Time Domain Telescope (TDT) is an initiative in the context of the ESO Expanding Horizons program. The TDT project aims for a global telescope array for spectroscopic time-resolved and time-critical observations in the optical-near-infrared, with a core at the European Southern Observatory and partners around the world. The TDT core consists of 100 unit telescopes of 2-meter diameter each that are fully Adaptive Optics-assisted in the blue, to deliver an enclosed energy of $\geq$80\% within $\sim$0.12\arcsec at 400nm. The unit telescopes are coupled to a set of spectrographs covering spectral resolutions $\Re$=10$^3$ - 10$^5$ over the wavelength range 0.35 - 1.8 $\mu$m. The combination of extreme adaptive optics with near-read-noise-free detectors will allow the TDT to reach unprecedented depths and/or high cadence. When all 100 units target a single object a depth of 25 mag can be reached in 30 minutes at a signal-to-noise ratio of 10 per spectral resolution element at a spectral resolution of $\Re$=5000. With this the TDT will spectroscopically match the imaging depth of the Vera Rubin Observatory and the {\sl Euclid} and {\sl Roman} spacecrafts, the expected brightness of electromagnetic counterparts discovered by the {\sl LISA} space mission, as well as by the next-generation gravitational wave detectors. The science of the TDT covers a large part of astronomy, and ranges from the outer Solar System, to stellar and binary systems to extragalactic transients and the growth of supermassive black holes all the way back to the expansion history of the Universe. The TDT is transformational, unique and highly complementary to the current and near-future ESO facilities, in particular by providing the possibility for time-resolved spectroscopic observations well below the brightness of the optical night sky.  
\end{abstract}


\maketitle

\begin{figure}[h]
    \includegraphics[width=\textwidth]{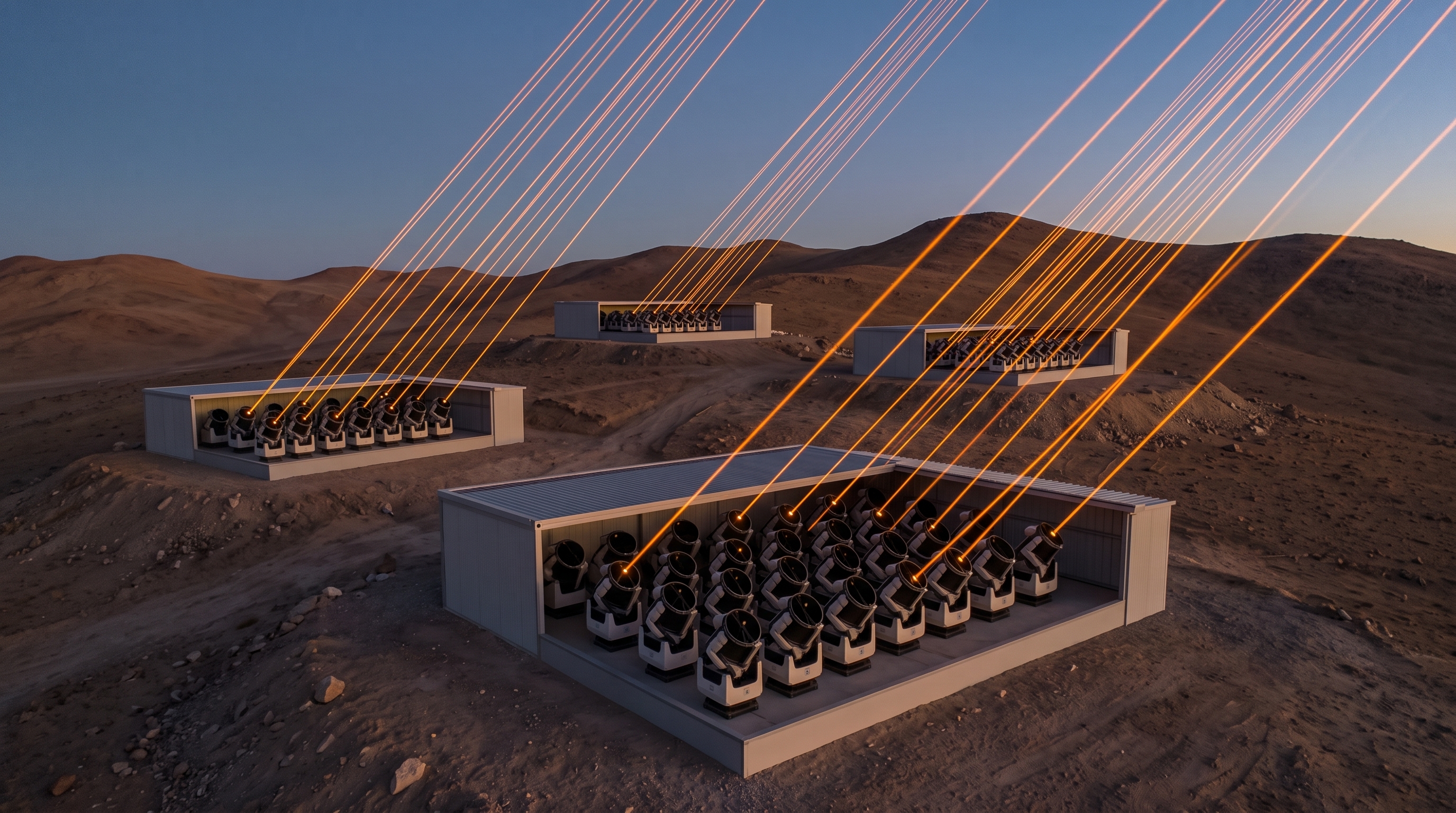}
    \caption{Conceptual rendering of the core of Time Domain Telescope array at a central ESO site
    \label{fig:TDTConcept}}
\end{figure}

\section{Time Domain Astronomy}

The use of time variability to understand the Universe and the fundaments of physics is a tool that has been used since mankind first looked up at the stars. Using modern detectors, as well as advances in compute, storage and information technologies, `time-domain astronomy' has taken off in the last decades, often using wide-field (semi-)autonomous telescopes, see e.g. \cite{PTF09}. The study of cosmic explosions, exoplanets, the solar system, stellar structure and evolution, and the evolution of the Universe all use time-domain as a tool. The high energy X-ray and $\gamma$-ray communities, and more recently the optical communities, are using wide-field monitoring to discover and study astrophysical phenomena such as X-ray binaries (e.g. \citealt{BlackCat16}), supernovae \citep{2017hsn..book.....A}, gamma-ray bursts \citep{GRBLevan}, all varieties of `novae' (e.g. \citealt{Scaringi22}), tidal disruption events (\citealt{TDEGezari}) and, in the last decade, multi-messenger phenomena such as gravitational wave mergers (\citealt{GW150915, GW170817}) and neutrino events (\citealt{stein21}). In the field of exoplanets, the use of time-domain studies has been inherent from the start with the two main discovery tools being radial-velocity variations of the host star (e.g. \citealt{51Peg95}), and photometric transits induced by the exoplanet passing in front of the host star (\citealt{Kepler10}). In stellar structure and evolution, both asteroseismology \citep{2010aste.book.....A} as well as single and binary star research extensively use spectrophotometric time series to understand stellar interior structures as well as orbital characteristics. For instance in the most extreme, ultracompact binary systems with orbital periods of no more than a few minutes (\citealt{roelofs2010, Chakraborty24}), that are also the strongest, known sources for the upcoming LISA mission for gravitational waves in the mHz regime (\citealt{LISA23}). In extragalactic systems both individual transient/explosive events as well as the time-variable nature of accretion onto the central supermassive black holes at the centres of galaxies are used to constrain fundamental physics, the supermassive black hole history, the circumnuclear-environment properties and the evolution of galaxies themselves. 

Large-scale facilities are studying the time-resolved Universe, but primarily photometrically. In the optical these include the Vera Rubin Observatory (\citealt{LSST19}), the Zwicky Transient Facility (\citealt{ZTF19}), the PanStarrs (\citealt{PS1DR2}), ATLAS (\citealt{ATLAS18}), GOTO (\citealt{GOTO22}), BlackGEM/MeerLICHT (\citealt{BG24}), SkyMapper (\citealt{SMDR4}), VST (\citealt{1998Msngr..93...30A}) and WFST (\citealt{WFST23}) telescopes from the ground, and the Gaia (\citealt{Gaia16}), TESS (\citealt{TESS15}), and soon PLATO (\citealt{PLATO25}) satellites in space. Across the electromagnetic spectrum these include the Fermi (\citealt{2009ApJ...697.1071A}), Einstein Probe (\citealt{EP22}), SVOM (\citealt{SVOM26}), UltraSat (\citealt{Ultrasat24}) and UVEX (\citealt{UVEX21}) satellites, as well as the CHIME {\citealt{2021ApJS..257...59C}, MeerKAT (\citealt{2016mks..confE...1J}), ASKAP (\citealt{2009IEEEP..97.1507D}), LOFAR (\citealt{2013A&A...556A...2V}), MWA (\citealt{2013PASA...30....7T}) and soon the DSA (\citealt{2023ApJ...949L...3R} and SKA (\citealt{2019arXiv191212699B}) radio telescopes and the CTA $\gamma$-ray observatory (\citealt{2019scta.book.....C}). In the multi-messenger regime these are joined by the LIGO/Virgo/KAGRA (\citealt{LIGO15, Virgo15, KAGRA19}) laser interferometers for gravitational waves and the IceCube (\citealt{IceCube21})  and Antares/KM3NET (\citealt{Antares11,KM3NET19}) detectors for neutrinos. Combined these facilities produce hundreds of alerts per night, and build up a long-term record of the brightness of billions of objects over time, with increasing photometric cadences. 

A well-known and well-recognised bottle neck in the study of the time-variable Universe is the (un)availability of (optical-near-infrared) spectroscopy to further characterise and study objects found photometrically \citep{SRK20}. Of all transient alerts, at the present day, no more than 10\% gets spectroscopically identified and no more than 1\% is the target of a detailed campaign \citep{Mogawana26}. This bottleneck is caused by the limited number of `target-hours' available on the global set of optical-near-infrared spectroscopic facilities. A target-hour is the combination of required flux (i.e. the size/spatial resolution of the telescope offset against the spectral resolution), as well as the number of hours available to study individual objects. 
Inherent to the study of the time-domain Universe is the need for {\sl repeated} observations, making this costly in target-hours. This is compounded by the fact that the sky surface density of targets-of-interest is generally low, i.e. $<$1 per square degree per night. The specification per unit time is of the essence here as there is often only a limited amount of time, or a particular (orbital/transient) phase, when spectroscopic observations are essential or most insightful. For instance, studying the flash ionisation resulting from the shock break-out phenomena in supernovae to study the density, composition and mass-loss history of the progenitor requires spectroscopy within minutes to hours of the explosion \citep{WaxmanKatz17}. Continued spectroscopic monitoring is required to probe physical processes that are evident at different time scales, from the early flash-ionisation signatures, to the evolution of circumstellar interactions to late-time nebular emission. 
In tidal-disruption events (TDEs), as in supernovae, a regular spectroscopic monitoring is needed to derive the event's properties. 
Similarly, transmission spectroscopy during an exoplanet transit is required during a very specific epoch and not at a random phase of the orbit, which also holds true for stellar occultations. Hence, time-domain spectroscopy is most often {\sl time-critical} spectroscopy. This time criticality also implies that a facility addressing time-domain spectroscopy is a {\sl global} facility which achieves a 24/7 duty cycle by spreading observing stations around the world. 

The Time Domain Telescope (TDT) addresses the critical bottle neck in the need for repeated and time-critical spectroscopy to study the time-variable Universe. It is a novel and ground-breaking concept where developments in adaptive optics, low read-noise detectors, and replication techniques are combined with increased computational capabilities towards flexible scheduling and the running of a robotic observatory. The TDT will allow to go far below the natural-seeing optical sky-background brightness which is often limiting current facilities for point sources, while at the same time providing the multiplicity needed for repeated monitoring of individual events scattered around the sky. The concept is being developed in the context of the Expanding Horizons initiative of the European Southern Observatory (ESO) and foresees a core component located at ESO with partner stations around the globe to achieve the combination of depth and round-the-clock coverage. It will complement current and near-future capabilities available to the ESO community which include the Very Large Telescope (Interferometer), the Atacama Large Millimeter Array, the La Silla facilities including the NTT+SOXS combination coming into operations now (\citealt{2019frap.confE..78C}), and in the near future the Cherenkov Telescope Array and the Extremely Large Telescope, as well as facilities and space missions around the globe.   

\section{The TDT Science}

The science case for the TDT spans virtually all fields in astronomy and astrophysics and is captured in four science themes: 

\begin{enumerate}
\item {\bf Stellar Deaths and Cosmic Elements}\\
From a Universe consisting of only hydrogen, helium and a trace of lithium just after the Big Bang, we now live in a Universe enriched by all the other elements in the periodic table. Where did these come from? When were which (groups of) elements formed? How do they lead to the complexity we see around us, ranging from crystallised white dwarfs to rocky planets, to life requiring many higher-mass elements, such as carbon, oxygen and phosphorus, to even exist. Without silicon, iron and uranium we could not have built our civilisation. How much of the trans-iron elements are made in cosmic explosions such as supernovae? How much is made in cosmic mergers such as colliding neutron stars witnessed in gravitational wave events? Which elements, and how much of each, are made in stellar furnaces and how are they spread over a full galaxy? When are the various elements made, and when was the enrichment sufficient for the formation of planets and, possibly, life?  This first theme contains many sub-questions such as `What is the cosmic rate of gravitational wave mergers and gamma-ray bursts, and how do they lead to the formation and spread of heavy elements?', `Which star leads to which type of supernova, and what remnant does it leave?', `What role do the various types of supernovae play in the chemical enrichment of the Universe?', `What is the role and physics of stellar furnaces in the chemical enrichment, and what role do binaries and higher-order systems play in this?' \\

The TDT will stand out by the combination to react on time scales of minutes to external alerts, and the ability to go down to $g\sim$25 within 30 minutes at a signal-to-noise ratio of 10 per resolution element, to cover the full optical-near-infrared spectrum at resolutions of $\Re\sim$5\,000, and to monitor transients and progenitor systems on the required time scales. A recent, very nearby, example of these kind of studies is exemplified by SN2023ixf in M101 \citep{2024Natur.627..759Z}. The TDT will be able to extend these studies to systems far outside of the very-nearby Universe.
The expected rate of binary neutron star mergers with a good sky localisation ($<$10 deg$^2$)
detectable to the Einstein Telescope out to a distance of 1.5 Gpc ($z\sim$0.3) is estimated to be 10s-100s per year; $\approx$1 per week \citep{2026MNRAS.tmp.1798K}. Scaling from GW170817, the peak brightness of these events will be $g\sim$24: detectable with the current and upcoming generation of imaging telescopes, but out of reach for routine time-resolved spectroscopic observations. Fig.\ \ref{fig:uvz} shows which rest-frame UV-lines enter the TDT wavelength range of 0.35 - 1.8 $\mu$m over which redshift range for extragalactic transients of a given absolute magnitude. 
\vspace{2mm}

\begin{figure}[h]
    \includegraphics[width=\textwidth]{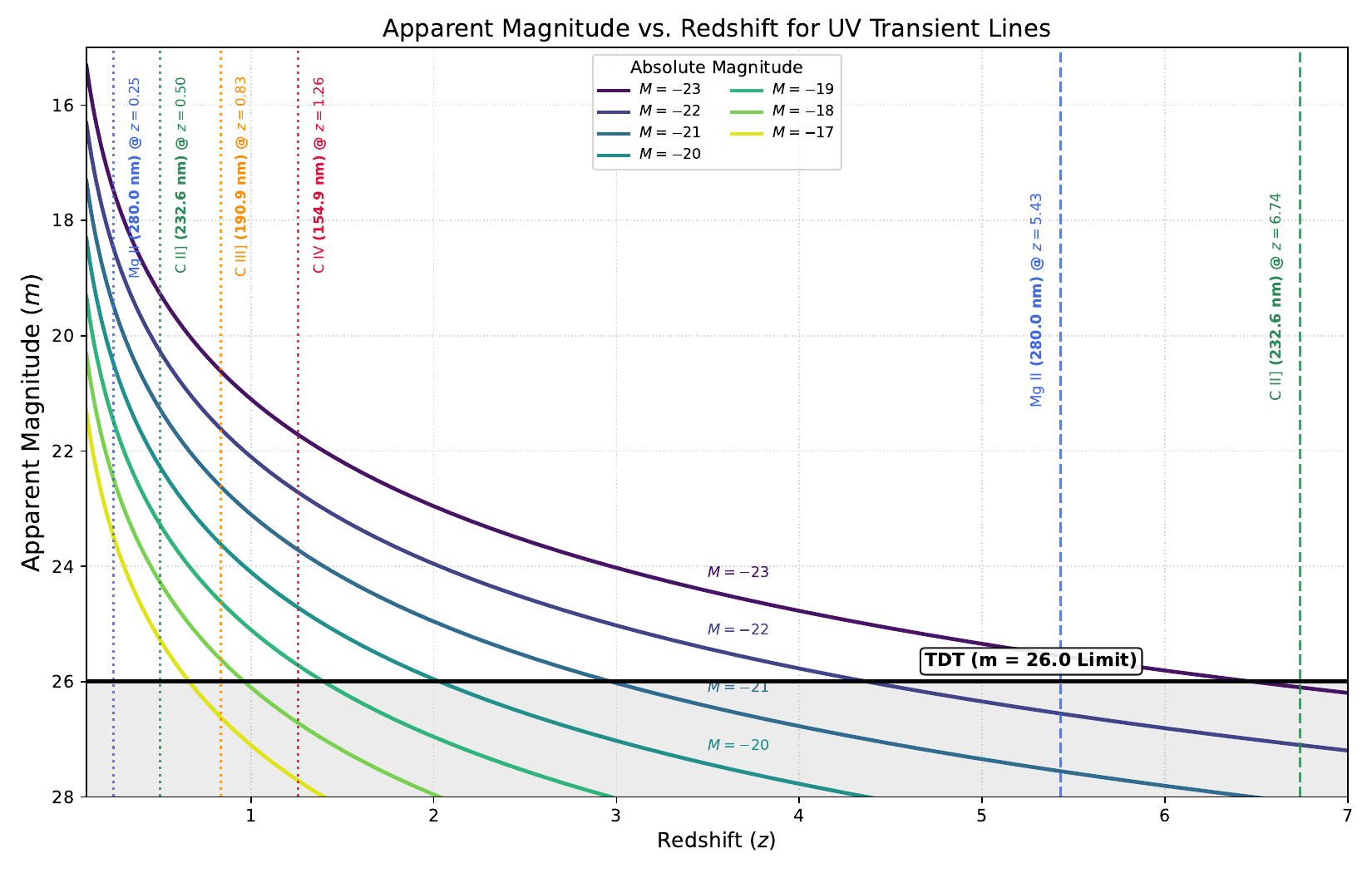}
    \caption{For bright, extragalactic transients, an overview of which rest-frame UV lines, commonly detected in bright transients, are observable within the TDT wavelength range of 350 - 1800 nm, depending on redshift, down to a practical limit of $g$=26 mag for a given set of absolute magnitudes for the (bright) event, assuming a flat SED. Marked dashed lines show at which redshift a particular line enters and exits the TDT wavelength coverage, e.g. the C {\sc II]} 232.6nm line can be observed between in the redshift range 0.50 $\leq z \leq 6.74$. A cosmology with $H_0$=70 km s$^{-1}$ Mpc$^{-1}$, $\Omega_m$= 0.3, $\Omega_\Lambda$ = 0.7 was used.}
    \label{fig:uvz}
\end{figure}

\item{\bf Powering the Universe} \\
Over the lifetime of the Universe accretion is arguably the most powerful source of electromagnetic radiation. Most often it proceeds via an accretion disk in settings that vary strongly in size and time scales, ranging from (planet forming) young stellar objects, to compact binaries and extragalactic (nuclear) transients, and to Active Galactic Nuclei. What governs the physics of accretion in disks? Is the physics of accretion (disks) universal? How does accretion proceed in stellar mergers, and how does this determine observed phenomena such as systems seen in gravitational waves? How does the growth of supermassive black holes in the center of galaxies proceed, and can e.g. changing-look quasars be directly compared to tidal-disruption events or accreting compact binaries? Can we translate the understanding gained from spectroscopic monitoring of disk outbursts and state changes in compact (X-ray) binaries to variations seen in AGNs and/or Tidal Disruption Events and vice versa?\\

The TDT will be essential to probe all the time scales involved, from microseconds in systems or stellar origin, to decades in extragalactic systems, both for monitoring as well as event-driven observations of e.g. tidal-disruption events and stellar mergers, that are intrinsically rare ($\ll1$ per square degree per night), and require fast response times. Stellar mergers with an absolute peak magnitude of $M_g$=--15 will be spectroscopically observable to a Gpc ($z\approx0.2$). TDEs with a peak magnitude of $M_g$=--20 will be spectroscopically traceable out to at least 10 Gpc ($z\geq1.2$), of fundamental importance to expand out of the local Universe and study them in the context of cosmic history. For both classes the spatial resolution of the TDT is required to isolate the events against the background glare of their host galaxies. 
\vspace{2mm}

\item{\bf The History of the Universe} \\
Arguably the biggest question of our time, besides the existence of life elsewhere, is the origin and history of our Universe. How did expansion start and what is dark energy that drives the current epoch of acceleration? The lower and most uncertain rungs on the cosmic distance ladder invariably depend on time-variable objects: pulsating stars and supernovae Type Ia (SN Ia) explosions. What is the physics governing both classes and how does e.g. metallicity and dust change the population characteristics? Is the expansion of the Universe homogeneous in time and isotropic in space? The only way to answer this is to use tracers such as SN Ia as a function of both time and space. How can gravitational wave sirens \citep{Schutz86} be placed on the cosmic distance ladder to allow a tracing even further back into cosmic history when the next generation of gravitational wave detectors allow detections throughout cosmic time? Besides dark energy the evolution of the Universe is dominated by dark matter: what is the true nature of dark matter? Can we use wide binary systems to constrain the nature of dark matter and distinguish between a particle nature and a modification of Newtonian gravity? \\

The TDT will be unique in addressing a number of these questions: to investigate if the Universe expands isotropically as well as homogeneously requires the monitoring of thousands of SN Ia along various cosmic lines of sight, and not just all-sky, over a significant fraction of cosmic time. Precision studies of SN Iae over a look-back time of $\geq$10 Gyr allow a view into the Universe when dark energy did {\sl not} yet dominate cosmic expansion. A SN Ia at peak absolute magnitude of $M_g = -19.3$ will be detectable to the TDT out to redshifts of $z\sim$1.25. In other words: the TDT will allow to spectroscopically trace the onset and evolution of dark energy over time, beyond the cosmic inflection point using thousands of SN Ia (Fig.\ \ref{fig:SNIa}). 
Using wide binaries \citep{2024MNRAS.528.4720H, 2025ApJ...992..102Y} the TDT will be able to test the reality of modified theories of gravity, such as MOND \citep{1997ApJ...480..492S}, as an alternative to dark matter particles.
\vspace{2mm}

\begin{figure}[h]
    \includegraphics[width=\textwidth]{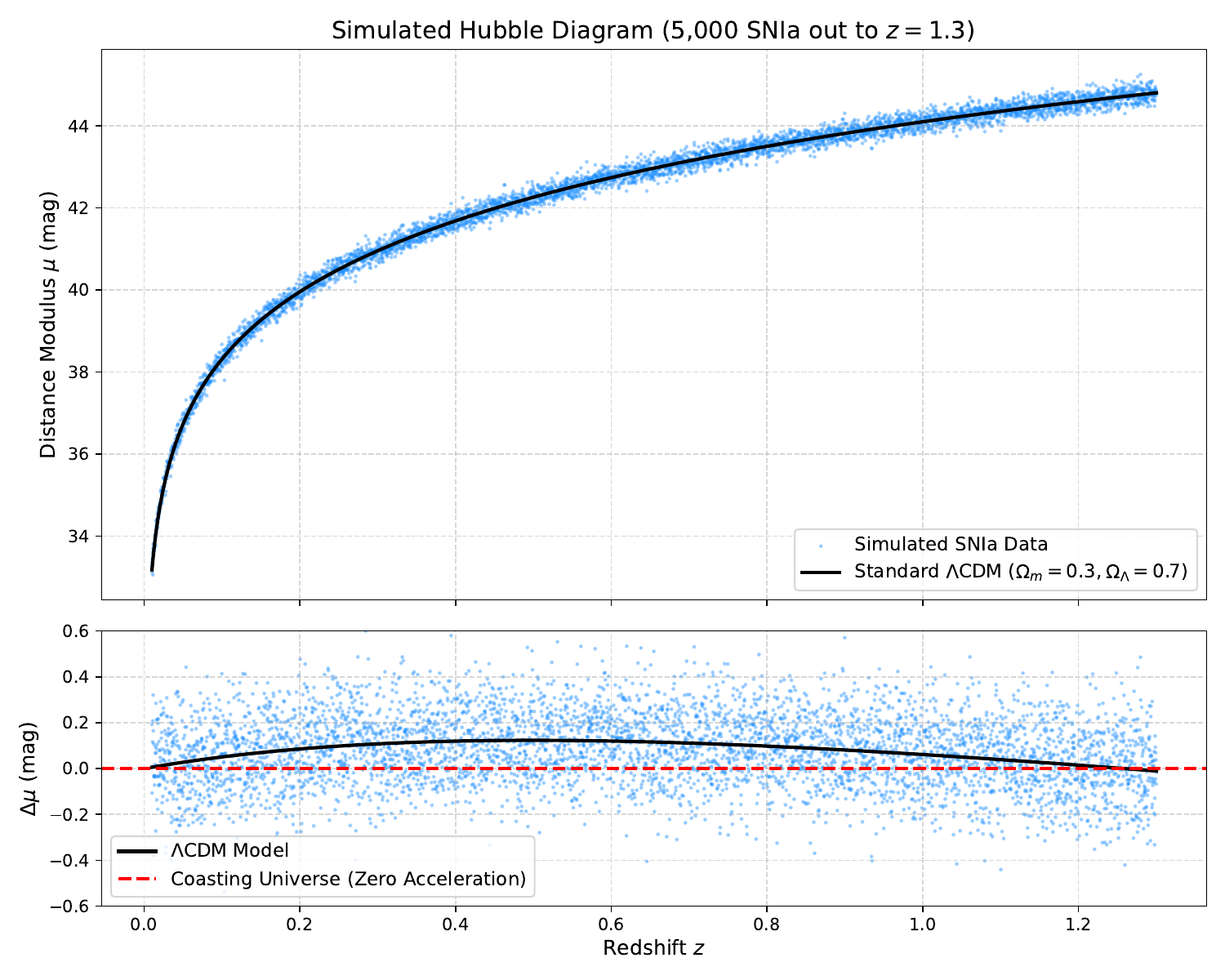}
    \caption{Simulated, illustrative, Hubble diagram using 5000 SN Ia, equally spread in redshift space, out to a redshift of $z\approx1.3$, showing in the bottom plot, the inflection point in cosmic expansion around $z\approx0.6$ due to the onset of dark energy. A cosmology with $H_0$=70 km s$^{-1}$ Mpc$^{-1}$, $\Omega_m$= 0.3, $\Omega_\Lambda$ = 0.7 was used.}
    \label{fig:SNIa}
\end{figure}

\item{\bf Exoplanets and the Galaxy} \\
How have the properties of planetary systems and their habitability been impacted by the (star) formation history of the Galaxy? 
The last three decades have seen a dramatic increase in our knowledge, and exploration, of exoplanetary systems. We are now entering a new phase, which is the characterisation of exoplanetary systems in their Galactic environment. Radial-velocity studies will remain a cornerstone of exoplanet studies, also with targets from Gaia and PLATO becoming available in the next few years. Transmission spectroscopy from the ground and in space is revealing the composition of exoplanetary atmospheres \citep{2010Natur.465.1049S}. Multi-planetary systems are becoming more and more common and a quantitative comparison with the Solar System is raising questions on our uniqueness in the Galaxy. Habitability of exoplanetary systems and the dependence on their host star's activity cycles will be explored in depth. Planets around stellar remnants are allowing a glimpse into the future of the Solar System, and reveal stunningly variable environments with (planetary) remnants being torn apart and accreted onto white dwarfs (e.g \citealt{2026NatAs.tmp..221W}). By combining stellar evolution, the star formation history of the Milky Way and the diversity of exoplanetary systems, the question on exoplanets occurrence and their habitability as a function of place and age in the Milky Way Galaxy can now be addressed. 

The revolution in our mapping of the Solar System through time-domain surveys is in itself raising many questions on the structure and history of the Outer Solar System, where the Vera Rubin Observatory's (VRO) observations are already uncovering a completely new population of minor bodies. Recently, the VRO announced the discovery of its first, newly-detected dwarf planet, 2026 ED 138 \citep{2026arXiv261038693H}, which, at $g\approx20.3$, may be the brightest of a completely new set of systems.\\

The TDT will bring the capability to probe planetary atmospheres through transmission spectroscopy across a range of settings, and with a range of host stars, and in particular in Solar-system analogues. This will also allow the study of exoplanet weather and climate cycles. The strict time scheduling allowed by the TDT provides an effective path forward to probe a large number of exoplanets repeatedly in radial-velocity studies. Strict, simultaneous photometric monitoring is essential, which is enabled by equipping (at least) a number of the TDT units in a cluster with sensitive photometers. For precision radial velocity studies, at $\Re$=10$^5$ and using the full 100-unit array, the TDT will be able to reach SNR=200 per resolution element in 1 hour of integration on a $g$=16 host star. This matches the photometric depth of the TESS and PLATO missions. For transmission spectroscopy and habitability studies, micro-turbulence on the host star, as well as the presence of star spots, and the surface inhomogeneity of the planet, will require multiple transits to be covered, making this a multi-year endeavour even for a single target. The transit of an Earth-like planet around a Sun-like star lasts $\approx$13 hours, and therefore the global nature of the TDT, and its guaranteed long-time existence as an ESO facility, is essential to cover transits multiple times. 

\end{enumerate}

What all four science themes have in common is that to answer them requires optical -- near-infrared spectroscopy to use chemical compositions, spectral line strengths and variations in both shape and position to trace changes over (cosmic) time, unveiling the underlying physics and chemistry. All four fields will see an explosion of targets-of-interest in the next decade, coming from the exoplanetary missions and time domain surveys named above, as well as, in the near future, the Argus array \citep{2022PASP..134c5003L}, the GOTTA network \citep{2026SPIE14147E..0NL} and the Digital Telescope; in addition to data from systematic mappings of the Universe using current and future space telescopes such as with Gaia, Euclid (\citealt{Euclid25}) and Roman (\citealt{Roman19}) and new facilities and missions uncovering the multi-messenger Universe such as Ligo/Virgo/KAGRA, IceCube/KM3Net, CTAO (\citealt{cta19}), Einstein Telescope (\citealt{et20}) and/or Cosmic Explorer (\citealt{CE21}), LISA (\citealt{LISA23}) and possibly TianQin/Taiji (\citealt{TianQin16, Taiji20}). 

What all four science themes also have in common is that to answer them one needs to push deeper and/or at higher cadence and with stricter time constraints than is currently possible. For instance, for the binary neutron star merger case this would realistically require a spectrum of a GW event every day with a flexible scheduling. No current or upcoming facility will be able to manage this at $g_{\rm peak}\geq24$, except the TDT. 

The science case of the TDT has been detailed in a series of white papers outlining the need for a facility dedicated to time-domain astronomy: see \cite{2025arXiv251214546N, 2025arXiv251214832C, 2025arXiv251215010S, 2025arXiv251217404G, 2025arXiv251213802S, 2025arXiv251213960M, 2025arXiv251215821E, 2025arXiv251215832V, 2025arXiv251220265A, 2026arXiv260102028C, 2025arXiv251214800P, 2025arXiv251214802B, 2025arXiv251215814U, 2025arXiv251216988K, 2025arXiv251216991K, 2026arXiv260117179M}

\section{The TDT Concept}
The TDT uses the concept of a distributed array of unit telescopes to achieve the combination of depth and sky coverage. A number of unit telescopes (a `cluster') are fiber-coupled into a set of low/medium/high-resolution spectrographs, where the fibers are arranged into a pseudo-slit. This way the same spectrograph and detector hardware is shared between the unit telescopes, which can still be operated individually. Multiple clusters make up the full array of $\mathcal{O}(100)$ units. Depending on future technological developments a set of spectrographs per telescope may be feasible/affordable, eliminating the need for (long) fibers. A conceptual rendering is shown in Fig.\ \ref{fig:TDTConcept}. The history of using spectroscopic arrays to increase collecting area goes back to early concepts such as FLOAT \citep{FLOAT77}, and more modern systems such as MAST (\citealt{MAST26}), MARVELS (\citealt{MARVEL20}), MINERVA \citep{MINERVA19} and LFAST \citep{LFAST22}. What the TDT brings is the combination of a large number of medium-sized apertures as pioneered in the \cite{FLOAT77} study, in combination with extreme adaptive optics, agentic scheduling, low-read-noise detectors and a global array. This allows the TDT to significantly surpass any other concept in depth, precision and flexibility. 

\subsection{Unit telescopes}
The units of the TDT will be 2-m class telescopes. On first analysis, this appears to be close to optimal in a trade-off between cost, manufacturability, diameter required for the AO performance, and the spatial resolution required to resolve out the sky background flux. Smaller unit telescopes will be easier and cheaper to manufacture but may add a large maintenance overhead and have diffraction limits that are not sufficient to resolve out the sky background to levels allowing very deep observations. Larger unit telescopes will be substantially more costly, although easing the burden on maintenance and scheduling complexity, and it will complicate the optical-blue AO system with a higher number of required actuators. Due to the lower number of units it will also limit the simultaneous multiplexing factor for single targets.

\subsection{Detectors} 
Near-read-noise-free detectors (RON$<$0.1 e$^{-}$) are needed to avoid the read-out penalty that would normally come with using many pixels on the detectors to record the combined light of the unit telescopes. Current day qCMOS detectors are already close to this limit at $\sim$0.3 e$^{-}$ (e.g. \citealt{ORCA26}), and different detector array technologies such as MKIDs \citep{KIDspec} and SPADs \citep{10.3389/fphy.2022.906671} are on the horizon. With specialised controllers the detectors will be split up in dedicated rows/pixels for each unit telescope, taking away the need for synchronised start/stop times between the unit telescopes, creating immense flexibility in rearranging the full array. 

\subsection{Extreme AO in the optical blue}
The application of Extreme AO in the optical blue is at the core of the Time Domain Telescope. On a dark site, the sky background in the optical amounts to $g\lsim$22 per square arcsecond (e.g. \citealt{2007PASP..119..687K}). Most spectrographs on the current generation of 8-10m class telescopes, such as X-Shooter on the VLT, are designed for observations in 1\arcsec\ conditions \citep{XSH11}. This limits the performance of these facilities in two ways: the sky background dominates the SNR when observing targets fainter than the background limit, and, the ability to resolve out targets at angular distances smaller than 1\arcsec. The same is true for current and upcoming multi-object spectrographs (MOS) which often have fiber entrances even exceeding 1\arcsec (see Table\ \ref{tab:mof}).
Both the sky background as well as the limited spatial resolution are crucial limitations that need to be overcome for future studies. Already current-day facilities such as the VRO and {\sl Euclid} chart the optical sky down to magnitudes $g>25$, and with (centroid) resolutions $\ll1\arcsec$. In practice, no current facility is able to perform spectroscopy at levels deeper than $g\sim$23 on a regular basis and these do not allow to resolve out sources in crowded regions, or only at the cost of spectral coverage. Two of the current, state-of-the-art instruments, X-Shooter and MUSE \citep{MUSE10}, both on the VLT, are limited by this in their use for (faint) time-domain astronomy.  In addition to this science limitation, instruments designed for seeing-limited conditions are, by necessity, bulky, and therefore expensive, in contrast to instruments designed for diffraction-limited conditions, where the sizes scale linearly with the spatial resolution ratio, and hence the volume as the cube of the ratio. 

Extreme-AO provides a way to break this deadlock: by applying this in the optical blue (400nm) for point sources, the diffraction limit and not the seeing becomes the limiting factor. For a 2m-class telescope in the optical blue, this shifts the limit from $\sim$1\arcsec\ to $\sim$0.1\arcsec: a dilution of a factor of 100 in the sky background flux and a decrease in instrumentation size of a linear factor of (D$_1$/D$_2$)($\sigma_1$/$\sigma_2$) = 4$\times$10 = 40,  when going from an 8-m class telescope ($D_1$) at $\sigma_1$ = 1\arcsec\, spatial resolution to a 2-m class telescope ($D_2$) working at $\sigma_2$ = 0.1\arcsec\, resolution, for the same spectrograph set up. Current technology can achieve Strehl-ratios of $\sim15\%$ at 400nm. Therefore a development path towards an enclosed energy of $\sim$80\% within the diffraction limited core at 400nm will be needed. AO systems with more actuators running at higher frequencies and with more powerful laser guide star systems (due to the more limited return flux) will be needed to make this feasible on a 2m-class telescope. 

In addition, with 100 telescopes running at the same time on the same site, the AO system needs to be very robust and robotic. Although robotic optical-red AO systems exist, primarily the RoboAO system \citep{RoboAO13} developed at the University of Hawai'i, a step up will be needed for the TDT. 

\subsection{Spectrographs}
\begin{wrapfigure}{r}{0.5\textwidth}
    \includegraphics[width=0.48\textwidth]{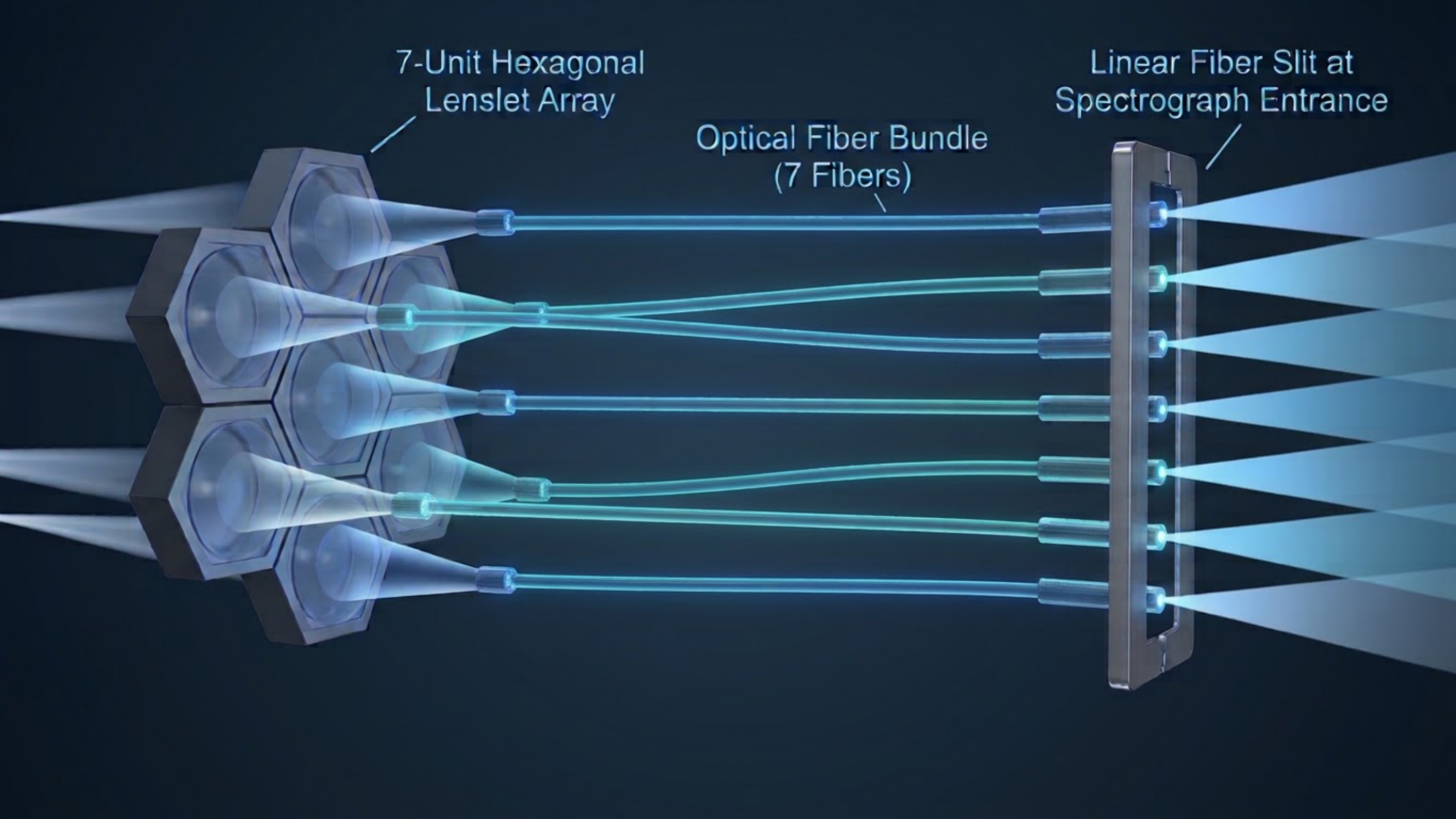}
    \caption{Illustration of the focal plane coverage with a 7-unit lenslet array coupled to a pseudo-slit at the spectrograph entrance}
    \label{fig:lenslet}
\end{wrapfigure}
The AO-corrected focal plane of each telescope will be sampled with a seven-fiber bundle, with fibers $\sim0.12$\arcsec\, in the blue, and larger in the (infra)red, due to the larger diffraction limit (Fig.\ \ref{fig:lenslet}). A set of spectrographs will cover spectral resolutions of $\Re$ = 10$^3$ - 10$^5$. The aim of the TDT is to cover the wavelength range of 350 - 1800 nm ($u-H$-band) simultaneously up to resolutions of $\sim$5\,000 (similar to X-Shooter), and in selected wavelength ranges at higher resolutions. In the basic TDT concept a number of unit telescopes (a `cluster') will be fiber-connected to a central set of spectrographs, necessitating a number of spectroscopic units. However, for the lower spectral resolutions, with the much decreased size of a 2-m AO-corrected beam, a solution of a spectrograph per telescope unit may be feasible using photonic lanterns and/or spectrographs-on-a-chip, which will decrease the restrictions on fiber throughput and length, in particular in the blue.  
For high-precision radial-velocity studies the fibers can be coupled into a thermally-stabilised spectrograph, similar to a smaller version of HARPS \citep{HARPS03}. Also here the strong decrease in required linear size may facilitate a `per unit' spectrograph. A completely new type of set-up could come from a spectrograph based on KIDS sensors to provide a medium-resolution echelle spectrograph, as described by \cite{KIDspec}. 

\subsection{Scheduling and Operations}
The TDT as envisioned above is an extremely versatile facility. The science user will have a choice between the number of unit telescopes and the number of targets, the spectral resolution and, of course, the cadence or time-criticality of the observations. As such, the scheduling of the TDT will require a development path itself, as no facility in the world comes close to this level of complexity. All observations are treated as time critical, and therefore scheduling conflicts are to be expected. This is partly alleviated by the large number of unit telescopes, but we foresee a high demand for full-array observations to maximise on either depth or cadence. 

Given the time-variable nature of the field itself, the full array needs to be reconfigurable on time scales of a minute, to be able to maximise the science return of near-real-time transient detections that are expected to be the standard in the 2040s. All of the above will require an agentic approach, where various AI agents are in charge of aspects of the facility. One can therefore have an agent per telescope (`telescope manager agents'), an agent for the scheduling (`scheduling agent'), an agent for the data reduction process (`data reduction agent') and for (preventive) maintenance (`maintenance agent'). Taking the humans out of the loop not only is a strong diminishing of the operational costs, but also allows a much more agile system, where agent-based decisions can be taken in a matter of seconds (or faster), taking away a significant human overhead, and allowing simultaneous multi-tasking. In this approach it is possible that the scheduling agent communicates with the telescope manager agent of unit 52 that it needs to switch to a new target in 10 seconds, as it has just become available from an monitored external trigger system. The scheduling agent will then also have to reschedule the remainder of the array, while taking the time-criticality and depth constraints of upcoming proposals into account. As such the agents are in charge of resource management across the array within strictly-defined optimisation parameters. Running such an agentic telescope is not without its own running costs, and e.g. energy consumption. The TDT will be designed to be carbon-neutral in operations. 

The TDT is foreseen as a classic `PI-driven' facility, much like e.g the VLT is now. Individual users can propose for observations through a central time allocation committee that ranks the proposals. Users will ask for `unit-hours', which can be in any configuration. Flexibility and agility are key within the TDT concept. Although the design of the TDT is focused on its suitability for time-domain science, it can of course be used for any spectroscopic observations of individual objects, variable or not, and as such will extend the capabilities currently covered by VLT instruments such as X-Shooter and FORS. 

\section{Global array}
The TDT collaboration operates under the slogan ``{\sl Spectra when you need them. Anywhere, anytime}". To make true on this slogan requires a global, 24/7, coverage. As well known from similar global arrays, such as the Las Cumbres Observatory or the combined network of ZTF, PanStarrs, ATLAS, GOTO and BlackGEM/MeerLICHT, one needs at least six stations to cover all possible locations in the sky at any given time. From the science case such global coverage is essential to capture those aspects that happen within the first few hours of e.g. a transient event: supernova shock break-outs or the onset of electromagnetic emission from gravitational wave events which is a sensitive probe of the optical depth through recently formed $r$-processed material, or the amount of asymmetry in the outflow from polarisation measurements. 
Placed at currently operating observatories one can achieve a 100\% sky coverage with a redundancy factor of 2.4 to buffer against weather and technical downtime. With some compromise on sky coverage, a three station, Southern Hemisphere array would provide 24/7 coverage of 70\% of the sky with a redundancy factor of 1.2. This is considered to be a bare minimum. 

Where the core at ESO would be a full configuration of $\mathcal{O}{(100)}$ unit telescopes, a trade-off study will have to show if the other stations could be smaller clusters, to provide time coverage, but at the cost of depth per station. These can be under the Expanding Horizons program or in a direct partnership with other communities. 

\section{Expected performance \label{sec:etc}} 
Using version 1.0 of the \href{https://www.tdtelescope.org/etc}{TDT Exposure Time Calculator}, Fig.\ \ref{fig:etc} shows the expected signal-to-noise ratio per spectral resolution element of the TDT at $\Re$=5\,000 with representative curves for 1, 10 and 100 units, each 2m in diameter for a 1800s exposure. It is assumed that the AO performance is at the diffraction limit at 400nm, that the detectors have a RON=0.1 e$^{-}$ per pixel, that the overall efficiency of the facility from top of the atmosphere to the detector is 15\% and that the sky background brightness is $g_{\rm sky}$=22 mag per square arcsecond, typical of New Moon conditions on a dark site. Figure 1 shows that, in an 1800s exposure, a SNR=10 is obtained for $g=20$, $g=22.4$ and $g=24.8$ for a 1, 10 or 100-unit TDT array for a spectral resolution of $\Re=5\,000$. 

\begin{figure}[h]
    \includegraphics[width=\textwidth]{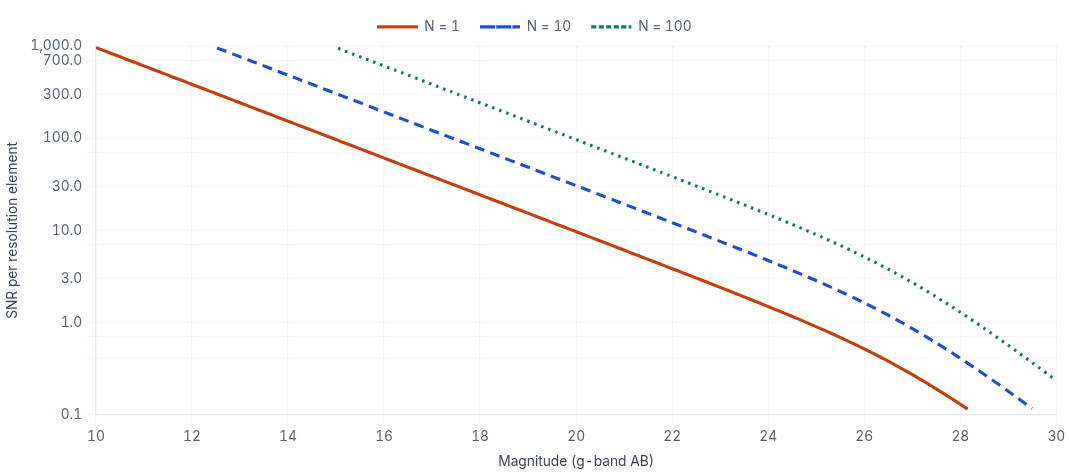}
    \caption{Expected performance of the TDT in an 1800s exposure for a 1, 10 and 100-unit configuration at a spectral resolution $\Re$=5000.}
    \label{fig:etc}
\end{figure}

In Fig\ \ref{fig:etc2} we compare the TDT performance to the current `gold standard', X-Shooter on the VLT, and the predicted Extremely Large Telescope (ELT) performance, making use of the \href{https://www.eso.org/etc}{ESO Exposure Time Calculators}. We note that the ELT is not optimised for optical-blue performance and that high Strehl-ratio AO corrections in the optical blue are also not foreseen at this time. 


Fig\ \ref{fig:etc2} shows that a single TDT unit is, in the sky-background-limited regime, comparable in performance to the VLT+X-Shooter, and therefore the gains over the VLT start as soon as the second TDT unit becomes operational. With respect to the ELT in a with Laser Tomography/ Multi-conjugate (LTAO/MCAO) Adaptive Optics mode, the ELT performance is similar to a TDT 30-unit mode, depending on the wavelength. We note that an ELT-SCAO/XAO comparison would be more appropriate but this is currently not available on the ESO ETC. Therefore, in this configuration, the point-source efficienty of the full TDT is three times that of the ELT at these wavelengths.

\begin{figure}[h]
    \includegraphics[width=\textwidth]{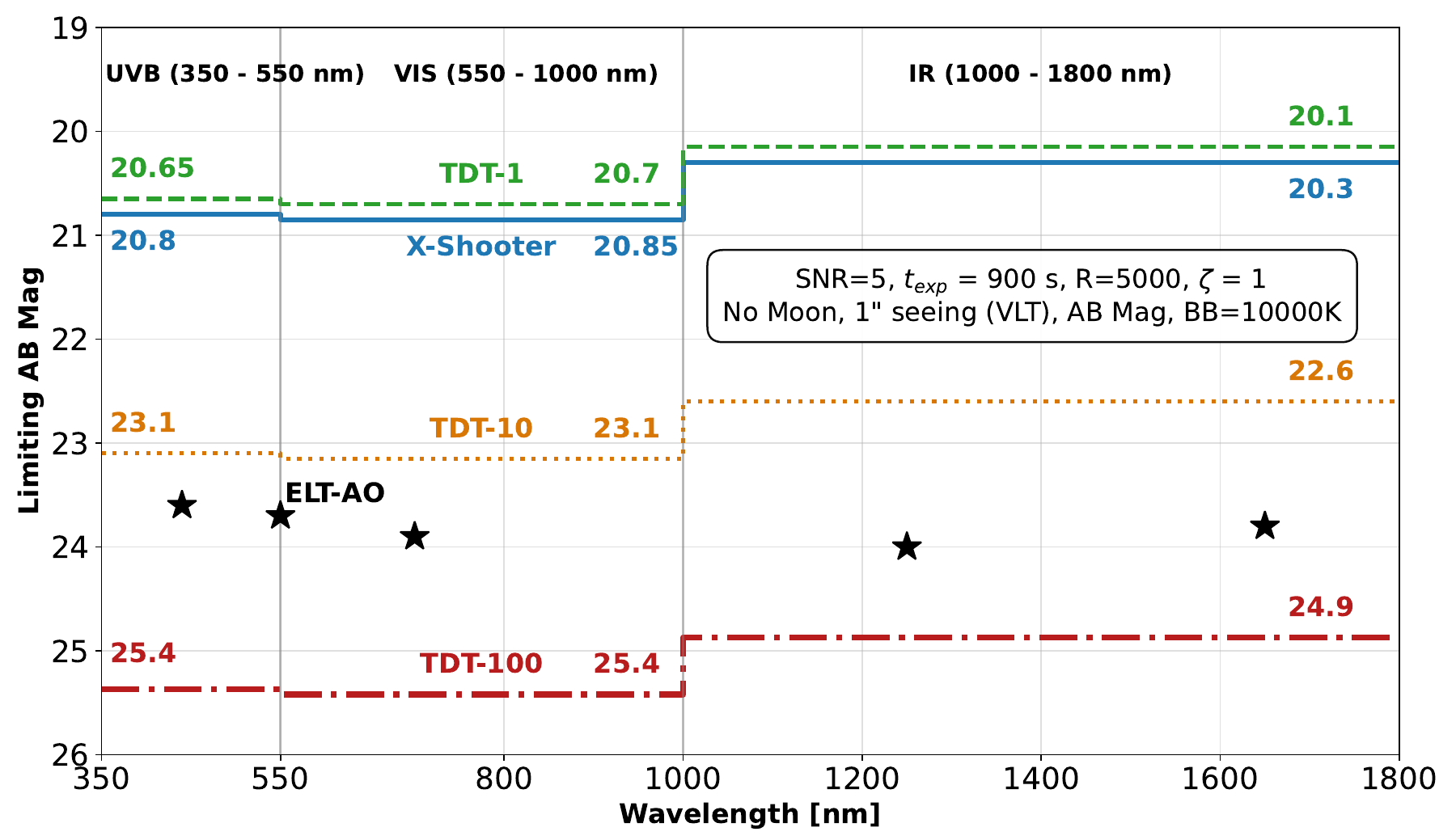}
    \caption{Expected limiting magnitudes in the AB system between VLT-X-Shooter, ELT and a TDT-1, TDT-10, and TDT-100 array of 2-m telescopes each, assuming an 10\,000 K Blackbody spectrum for $\Re$=5\,000, a SNR=5, 1\arcsec\, seeing for VLT-X-Shooter, and 0\farcs1 fibers for the TDT.  X-Shooter magnitudes are taken at peak blaze wavelength performance. The UVB and IR arms for the TDT have been scaled from X-Shooter. The ELT-AO performance uses the ESO-ELT ETC v6.4.0 in the Laser-AO/MultiConjugate AO mode at the B-H reference wavelengths with a 100 mas reference area for the SNR calculation. }
    \label{fig:etc2}
\end{figure}

\subsection{Multiplexing capabilities of the TDT}

The flexibility of the TDT allows the user to make a choice to either combine all telescopes on a single target to achieve ultimate depth, or to put each telescope on a different target, that can be scattered around the observable sky, as well as a subset of telescopes working in unison to observe a target, possibly at different spectral resolutions.  To compare the TDT to existing and possible future facilities we introduce a figure-of-merit, the Multi-Object-Flexibility (MOF), defined as:
\begin{equation}
    MOF = N_{\rm object}\, (\Omega_{\rm sky}/{4\pi})\, (D/\sigma)^2, 
    \label{eq:mof}
\end{equation}
where $N_{\rm object}$ is the total number of independent objects that
can be simultaneously observed,  $\Omega_{\rm sky}$ is the instantaneous area of the 
sky over which the $N_{\rm
  object}$ targets can be positioned, $D$ is the diameter of the
telescope, and $\sigma$ is the effective spatial resolution of the
facility. In the case of the instruments listed in Tab.\ \ref{tab:mof}
this is the fiber size in arcseconds. The quantity ($D/\sigma)^{2}$ is
used because in the sky-background-limited regime the signal-to-noise
ratio scales as the square of the diameter of the telescope over the
spatial resolution. MOF values of various multi-object spectrograph
facilities that operate (approximately) at a spectral resolution of
$\Re$=5000 are given in Tab\ \ref{tab:mof} and Fig\ \ref{fig:mof}. Due
to the unique distributed nature of the TDT, Eq.\ \ref{eq:mof}
produces a unique number for the TDT, irrespective of whether it is
configured with 100 telescopes on a single target, or all 100 units on
an individual target as $D \propto \sqrt{N_{\rm tel}}$, and therefore
$MOF_{\rm TDT}= constant$. In essence MOF is the survey speed as a
fraction of the total sky available to distribute the targets
over. Fig.\ \ref{fig:mof} shows that in the combination of number of
targets, available sky angle and instantaneous sensitivity the TDT is
exceptional. The modest amount of fibers/targets, with respect to e.g. the Wide-field Spectroscopic Telescope (WST, \citealt{2026arXiv260808297S}) or the Extremely-large Spectroscopic Survey Telescope (ESST, \citealt{10.1117/12.3104704}) is more than
compensated for by the available sky angle and the instantaneous
depth, which is due to its high spatial resolution. 
Table\ \ref{tab:mof} and Fig.\ \ref{fig:mof} also show the
complementarity of the TDT with the ELT: whereas the ELT provides
high-precision multi-object capabilities on a very narrow sky angle,
the TDT complements this by distributed all-sky access. 

\begin{table}[htb]
\centering
\caption{MOF values of the TDT versus multi-object spectrographs currently existing, under construction or {\sl proposed}, operating at spectral resolutions of $\Re\sim$5\,000. Columns list the facility, the (proposed) year of start of operations, the number of objects that can be targeted simultaneously, the area of the sky over which these can be distributed, the diameter of the primary mirror of the facility, the effective spatial resolution of the facility and the resulting Multi Object Flexibility (MOF) figure-of-merit following Eq.\ \ref{eq:mof}. \label{tab:mof}}
 {\tablefont\begin{tabular}{@{\extracolsep{\fill}}lrrrrrr}
    \midrule
    Facility  & Year & $N_{\rm object}$ & $\Omega_{\rm sky}$ (sqd) & D (m) & $\sigma$ (\arcsec) & MOF (sqd m$^2$ \arcsec$^{-2}$) \\
         \midrule
    ELT - MOSAIC & 2035 & 200 & 0.01 & 39.2 & 0.9 & 0.10\\
    VLT - MOONS & 2027 & 1000 & 0.14 & 8.2 & 1 &0.22 \\
    WHT - WEAVE & 2026 & 1000 & 3.14 & 4.2 & 1.3 & 0.78 \\
    VISTA - 4MOST & 2026 & 1624 & 4.1 & 4.1 & 1.45 & 1.26  \\
    LAMOST & 2014 & 4000 & 19.6 & 4 & 3.3 & 2.75\\
    VLT - X-Shooter & 2010 & 1 & 10000 & 8.2 & 1 &16.0\\
    LFAST & 2030 & 100 & 10000 & 3.5 & 1.3 & 172.6 \\
    {\sl WST - MOS LR} & 2042 & 30000 & 3 & 12 & 1 & 308.6\\
    {\sl ESST } &2030+ & 30000 & 3.14 & 12& 1 & 323.0\\
    {\sl TDT-1 } & 2042 & 100 & 10000 & 2 & 0.12 & 9523.8 \\
    {\sl TDT-10 } & 2042 & 10 & 10000 & 6.3 & 0.12 & 9523.8 \\
    {\sl TDT-100 } & 2042 & 1 & 10000 & 20 & 0.12 & 9523.8 \\   
    \midrule
    \end{tabular}}
\end{table}

\begin{figure}[h]
  \includegraphics[width=0.5\textwidth]{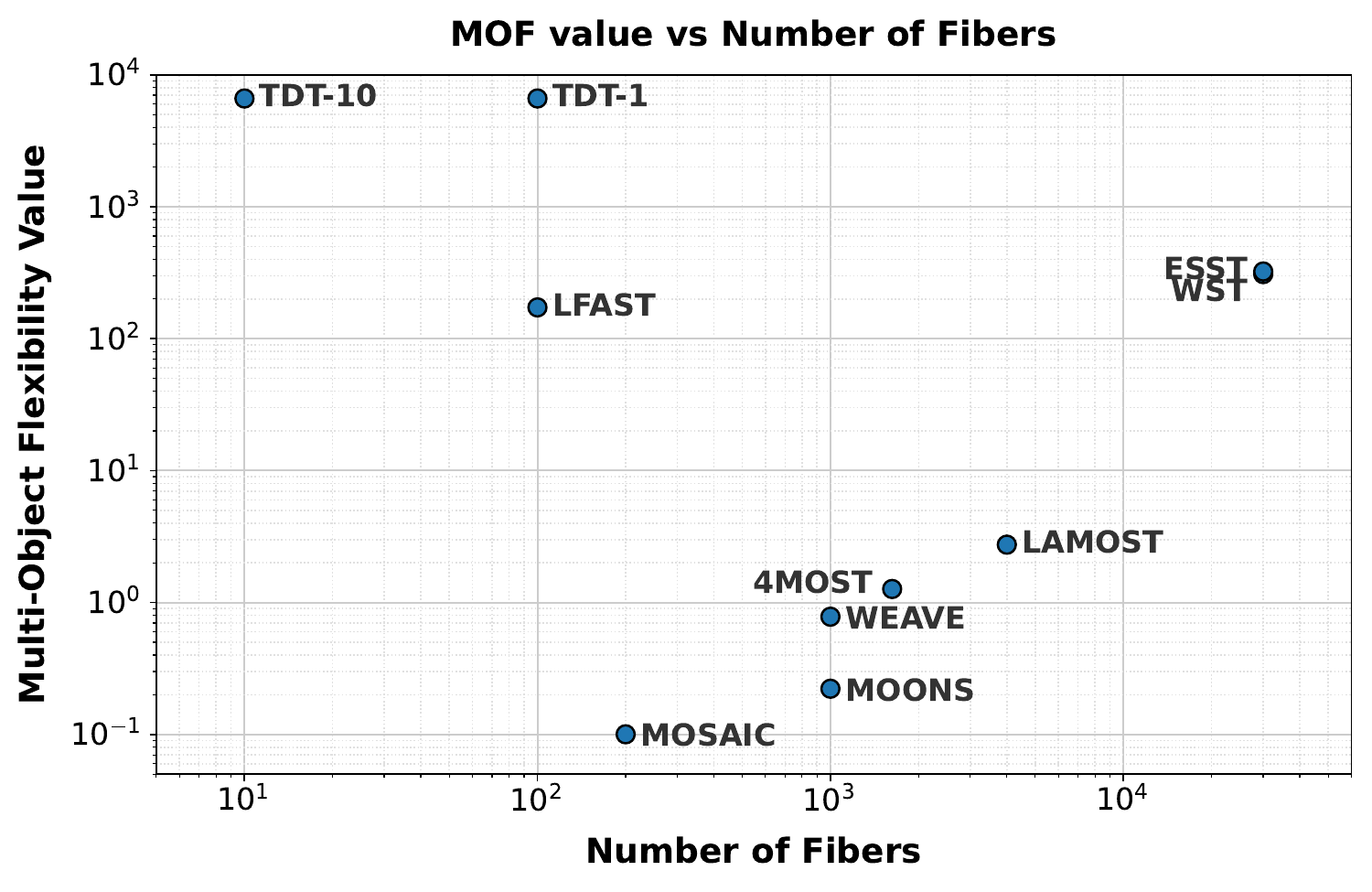}
  \includegraphics[width=0.5\textwidth]{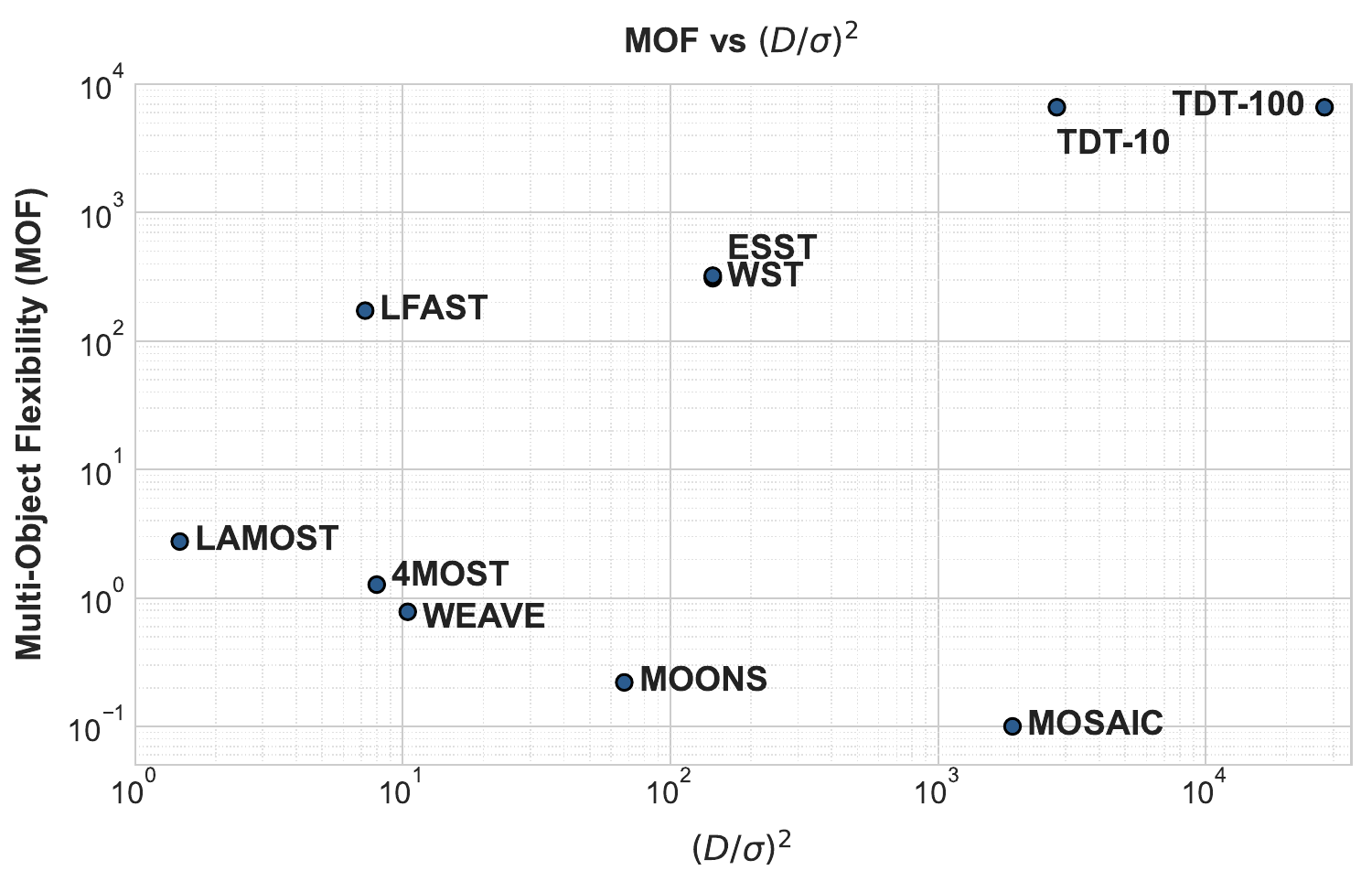}
    \caption{Multi-Object Flexibility of the TDT versus a set of other
      multi-object spectrographs. {\sl (left)} MOF as a function of the
      number of fibers (targets) that can be obtained simultaneously,
      {\sl (right)} MOF as a function of $(D/\sigma)^2$, 
      which is proportional to instantaneous sensitivity. In both
      panels points higher up and towards the right are preferred.}
    \label{fig:mof}
\end{figure}

\section{Additional capabilities: high-timing, polarimetry, trigger array, and Intensity Interferometry}
The core of the TDT is the spectroscopic array as outlined above. However, a number of component could be added to even further enhance the scientific value of the facility. First, these could be extremely-high-time-resolution photometers similar to the SiFAP2 instrument recently deployed on the TNG telescope \citep{2014SPIE.9147E..8RA}, in particular in tandem with time-resolved spectroscopy, or simultaneously with X-ray observations from e.g. the Athena mission (see \citealt{2026arXiv261018322B} for a recent example). Second could be the inclusion of polarimetry units in the spectrographs, which would allow the study of e.g. asymmetries in cosmic explosions, and would be highly complementary to X-ray polarimetry provided by IXPE, XPoSat and eXTP \citep{2022JATIS...8b6002W, 2025JATIS..11c5001V, 2019SCPMA..6229502Z}. Thirdly, the TDT relies on external triggers to provide targets. Partly these are guaranteed from upcoming facilities, but the existence of a global optical synoptic survey array such as currently existing with ATLAS/GOTO/MeerLICHT/BlackGEM/LAST facilities and the future Argus and GOTTA array is not a given. A modest expansion of the TDT project would be to add, e.g. a Digital Telescope or MezzoCielo \citep{2026SPIE14147E..0DB} station to each site to provide an independent transient-detection capability. Such an all-sky monitoring array would also function as an effective satellite and airplane detection system, feeding into the laser avoidance system. 
The set up of the TDT is also highly suitable for Intensity Interferometry \citep{2026arXiv260212717K} which requires an array distributed over the interferometric ($u,v$) plane where individual telescopes are coupled to medium-resolution spectrograph: a set-up similar to the TDT.   

\section{Technology development path}

Some key technology aspects for the TDT are not available today: blue robotic Extreme AO, sufficiently powerful laser guide star systems, near-noise-free detectors and agentic scheduling. However, all four are areas of strong development. After technical tests in Europe it is currently planned to bring a first version of the TDT (proto-TDT/FAROS) to the ESO Paranal observatory by adapting the four Auxiliary Telescopes (ATs) to also function as TDT units, besides their role in the Very Large Telescope Interferometer. This set up allows a full test of all aspects: the robotic AO, the near-read-noise-free detectors and the scheduling aspects. On top of that, with the laser guide star systems already deployed at Paranal for the VLT UTs, and the near-future operation of the Extremely Large Telescope at nearby Cerro Armazones, it also serves as a real-world test on laser avoidance and laser collision. Because of this problem, and to safeguard against weather influences, it may be beneficial to spread the TDT core array out over a number of (non-interfering) locations, e.g. a cluster at Cerro Paranal, one at Cerro Armazones and two at Cerro La Silla and its extension Cerro Vizcachas. As shown in Sect.\ \ref{sec:etc}, on its own proto-TDT/FAROS would be equivalent to a 3.8m telescope operating at near-diffraction limited spatial resolution, essentially the equivalent of NTT+SOXS, but then with the grasp of a telescope 10 times the collecting diameter of the NTT.

\section{Conclusion}
The TDT concept is a revolutionary new step in time-domain astronomy that uses the latest, and near-future, technology to make a giant leap forward in both depth as well as availability and flexibility of time-resolved spectroscopy in the optical-near-infrared. It will answer key questions in cosmology, in transient physics, in nucleocosmogenesis, in exoplanetary physics, and multi-messenger astrophysics.  With a grasp that reaches beyond $g>25$ it is the required spectroscopic counterpart to current and future wide-field imaging surveys and space-based sky mapping. With its configuration versatility and global coverage it is the required facility to characterise time-critical phenomena at all magnitudes, reaching from the brightest to the faintest. With its modular and scalable approach it allows a gradual roll-out where operations can be done with the prototype as well as the first completed TDT telescopes, delivering science back to the community from day one. The TDT has a global appeal and is highly complementary to current and near-future facilities at ESO, while being a completely novel and unique concept, requiring technologies that are within grasp on a 15-year time scale. 

\begin{figure}[h]
    \includegraphics[width=\textwidth]{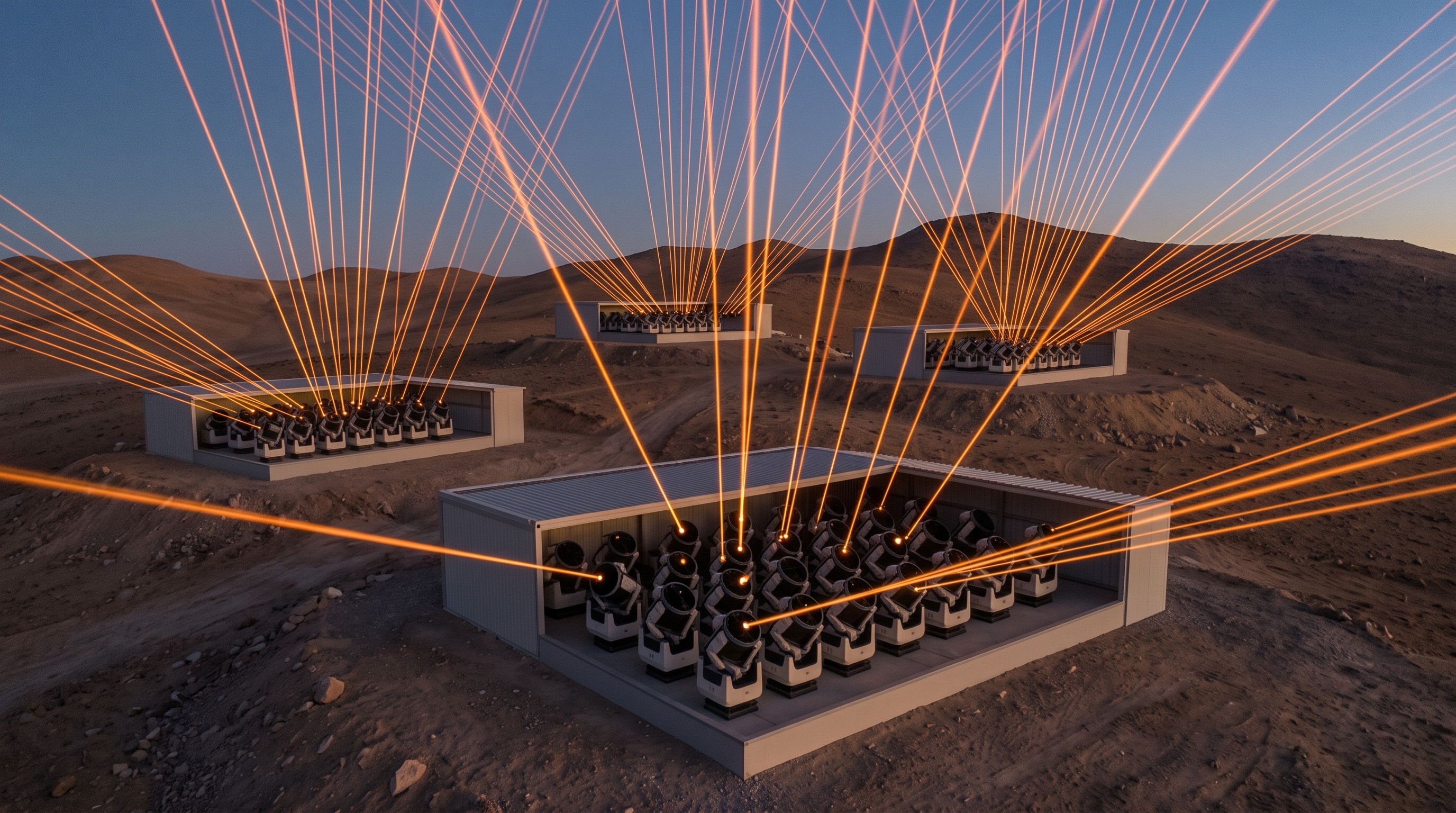}
    \caption{Conceptual rendering of the core of the TDT array in a distributed pointing mode}
    \label{fig:TDTConcept_wide}
\end{figure}

\vspace{0.6cm}
\noindent {\small \textbf{Author list continued}}
\vspace{0.2cm}\\

\noindent {\small
M.C.~Baglio$^{14}$,
N.~Blagorodnova$^{15,16,17}$,
D.~Buckley$^{3,18,19}$,
N.~Castro Segura$^{9}$,
L.~Christensen$^{20,21}$,
D.~de Martino$^{8}$,
D.~Defr\`ere$^{22}$,
M.~Del Santo$^{23}$,
A.~Ederoclite$^{24}$,
C.~Gall$^{25}$,
T.~Granzer$^{26}$,
R.~Kotak$^{27}$,
H.~Kuncarayakti$^{28,29}$,
G.~Leloudas$^{30}$,
A.~Levan$^{1,9}$,
C.~Malacaria$^{31}$,
A.~Marino$^{32,33}$,
P.F.L.~Maxted$^{34}$,
A.~Mortier$^{35}$,
F.~Onori$^{31}$,
A.F.~Pala$^{36}$,
S.~Piranomonte$^{31}$,
F.~Ragosta$^{37,8}$,
A.~Sanna$^{38}$,
N.~Sarin$^{39,40}$,
C.~Shneider$^{41}$,
F.~Tombesi$^{42,31,43}$,
S.~Toonen$^{44}$,
V.~Van Grootel$^{45}$,
A.~Veledina$^{27,46}$,
F.~Vincentelli$^{47}$,
and T.G.~Wilson$^{9}$
}

\vspace{0.3cm}
\noindent {\footnotesize
$^{14}$ INAF -- Osservatorio Astronomico di Brera, Via E.\ Bianchi 46, I-23807 Merate (LC), Italy \\
$^{15}$ Institut de Ci\`encies del Cosmos (ICCUB), Universitat de Barcelona (UB), c.\ Mart\'{\i} i Franqu\`es 1, 08028 Barcelona, Spain \\
$^{16}$ Departament de F\'{\i}sica Qu\`antica i Astrof\'{\i}sica (FQA), Universitat de Barcelona (UB), c.\ Mart\'{\i} i Franqu\`es 1, 08028 Barcelona, Spain \\
$^{17}$ Institut d'Estudis Espacials de Catalunya (IEEC), Edifici RDIT, Campus UPC, 08860 Castelldefels (Barcelona), Spain \\
$^{18}$ University of the Free State, Bloemfontein, South Africa \\
$^{19}$ National Astronomical Observatories, Chinese Academy of Sciences (NAOC), Beijing, China \\
$^{20}$ Niels Bohr Institute, University of Copenhagen, Jagtvej 128, DK-2200 Copenhagen N, Denmark \\
$^{21}$ Cosmic Dawn Center (DAWN), Denmark \\
$^{22}$ Institute of Astronomy, KU Leuven, Celestijnenlaan 200D, 3001 Leuven, Belgium \\
$^{23}$ INAF/IASF Palermo, Via Ugo La Malfa 153, I-90146 Palermo, Italy \\
$^{24}$ Centro de Estudios de F\'{\i}sica del Cosmos de Arag\'on (CEFCA), Teruel, Spain \\
$^{25}$ DARK, Niels Bohr Institute, University of Copenhagen, Jagtvej 155A, 2200 Copenhagen, Denmark \\
$^{26}$ Leibniz-Institut f\"ur Astrophysik Potsdam (AIP), 14482 Potsdam, Germany \\
$^{27}$ Department of Physics and Astronomy, FI-20014 University of Turku, Finland \\
$^{28}$ Tuorla Observatory, Department of Physics and Astronomy, FI-20014 University of Turku, Finland \\
$^{29}$ Finnish Centre for Astronomy with ESO (FINCA), FI-20014 University of Turku, Finland \\
$^{30}$ DTU Space, Department of Space Research and Space Technology, Technical University of Denmark, Elektrovej 327, 2800 Kgs.\ Lyngby, Denmark \\
$^{31}$ INAF -- Osservatorio Astronomico di Roma, Via Frascati 33, 00078 Monte Porzio Catone (RM), Italy \\
$^{32}$ Departamento de F\'{\i}sica, Universidad de Santiago de Chile (USACH), Av.\ V\'{\i}ctor Jara 3493, Estaci\'on Central, Chile \\
$^{33}$ Institute of Space Sciences (ICE, CSIC), Campus UAB, Carrer de Can Magrans s/n, E-08193 Barcelona, Spain \\
$^{34}$ Astrophysics Group, Keele University, Staffordshire, ST5 5BG, UK \\
$^{35}$ School of Physics and Astronomy, University of Birmingham, Edgbaston, Birmingham, B15 2TT, UK \\
$^{36}$ European Southern Observatory, Karl-Schwarzschild-Stra{\ss}e 2, D-85748 Garching, Germany \\
$^{37}$ Dipartimento di Fisica ``E.\ Pancini'', Universit\`a degli Studi di Napoli Federico II, Napoli, Italy \\
$^{38}$ Dipartimento di Fisica, Universit\`a degli Studi di Cagliari, SP Monserrato-Sestu km 0.7, 09042 Monserrato, Italy \\
$^{39}$ Kavli Institute for Cosmology Cambridge, Madingley Road, Cambridge, CB3 0HA, UK \\
$^{40}$ Institute of Astronomy, University of Cambridge, Madingley Road, Cambridge, CB3 0HA, UK \\
$^{41}$ Space Systems Research Group, Interdisciplinary Centre for Security, Reliability and Trust (SnT), University of Luxembourg, 29 Av.\ J.F.\ Kennedy, 1855 Luxembourg, Luxembourg \\
$^{42}$ Physics Department, Tor Vergata University of Rome, Via della Ricerca Scientifica 1, 00133 Rome, Italy \\
$^{43}$ INFN -- Rome Tor Vergata, Via della Ricerca Scientifica 1, 00133 Rome, Italy \\
$^{44}$ Anton Pannekoek Institute for Astronomy, University of Amsterdam, The Netherlands \\
$^{45}$ Space sciences, Technologies and Astrophysics Research (STAR) Institute, Universit\'e de Li\`ege, 19C All\'ee du 6 Ao\^ut, 4000 Li\`ege, Belgium \\
$^{46}$ Nordita, KTH Royal Institute of Technology and Stockholm University, Hannes Alfv\'ens v\"ag 12, SE-10691 Stockholm, Sweden \\
$^{47}$ Coventry University, Coventry, UK
}


\clearpage


\begin{thebibliography}{}
\expandafter\ifx\csname natexlab\endcsname\relax\def\natexlab#1{#1}\fi
\providecommand{\url}[1]{\href{#1}{#1}}
\providecommand{\dodoi}[1]{doi:~\href{http://doi.org/#1}{\nolinkurl{#1}}}
\providecommand{\doeprint}[1]{\href{http://ascl.net/#1}{\nolinkurl{http://ascl.net/#1}}}
\providecommand{\doarXiv}[1]{\href{https://arxiv.org/abs/#1}{\nolinkurl{https://arxiv.org/abs/#1}}}

\bibitem[{{Aartsen} {et~al.}(2021){Aartsen}, {Abbasi}, {Ackermann}, {Adams}, {Aguilar}, {Ahlers}, {Ahrens}, {Alispach}, {Allison}, {Amin}, {Andeen}, {Anderson}, {Ansseau}, {Anton}, {Arg{\"u}elles}, {Arlen}, {Auffenberg}, {Axani}, {Bagherpour}, {Bai}, {Balagopal V}, {Barbano}, {Bartos}, {Bastian}, {Basu}, {Baum}, {Baur}, {Bay}, {Beatty}, {Becker}, {Tjus}, {BenZvi}, {Berley}, {Bernardini}, {Besson}, {Binder}, {Bindig}, {Blaufuss}, {Blot}, {Bohm}, {Bohmer}, {B{\"o}ser}, {Botner}, {B{\"o}ttcher}, {Bourbeau}, {Bourbeau}, {Bradascio}, {Braun}, {Bron}, {Brostean-Kaiser}, {Burgman}, {Burley}, {Buscher}, {Busse}, {Bustamante}, {Campana}, {Carnie-Bronca}, {Carver}, {Chen}, {Chen}, {Cheung}, {Chirkin}, {Choi}, {Clark}, {Clark}, {Classen}, {Coleman}, {Collin}, {Connolly}, {Conrad}, {Coppin}, {Correa}, {Cowen}, {Cross}, {Dave}, {Deaconu}, {De Clercq}, {DeLaunay}, {De Kockere}, {Dembinski}, {Deoskar}, {De Ridder}, {Desai}, {Desiati}, {de Vries}, {de Wasseige}, {de With}, {DeYoung}, {Dharani}, {Diaz}, {D{\'\i}az-V{\'e}lez},
  {Dujmovic}, {Dunkman}, {DuVernois}, {Dvorak}, {Ehrhardt}, {Eller}, {Engel}, {Evans}, {Evenson}, {Fahey}, {Farrag}, {Fazely}, {Felde}, {Fienberg}, {Filimonov}, {Finley}, {Fischer}, {Fox}, {Franckowiak}, {Friedman}, {Fritz}, {Gaisser}, {Gallagher}, {Ganster}, {Garcia-Fernandez}, {Garrappa}, {Gartner}, {Gerhard}, {Gernhaeuser}, {Ghadimi}, {Glaser}, {Glauch}, {Gl{\"u}senkamp}, {Goldschmidt}, {Gonzalez}, {Goswami}, {Grant}, {Gr{\'e}goire}, {Griffith}, {Griswold}, {G{\"u}nd{\"u}z}, {Haack}, {Hallgren}, {Halliday}, {Halve}, {Halzen}, {Hanson}, {Hanson}, {Hardin}, {Haugen}, {Haungs}, {Hauser}, {Hebecker}, {Heinen}, {Heix}, {Helbing}, {Hellauer}, {Henningsen}, {Hickford}, {Hignight}, {Hill}, {Hill}, {Hoffman}, {Hoffmann}, {Hoffmann}, {Hoinka}, {Hokanson-Fasig}, {Holzapfel}, {Hoshina}, {Huang}, {Huber}, {Huber}, {Huege}, {Hughes}, {Hultqvist}, {H{\"u}nnefeld}, {Hussain}, {In}, {Iovine}, {Ishihara}, {Jansson}, {Japaridze}, {Jeong}, {Jones}, {Jonske}, {Joppe}, {Kalekin}, {Kang}, {Kang}, {Kang}, {Kappes}, {Kappesser},
  {Karg}, {Karl}, {Karle}, {Katori}, {Katz}, {Kauer}, {Keivani}, {Kellermann}, {Kelley}, {Kheirandish}, {Kim}, {Kin}, {Kintscher}, {Kiryluk}, {Kittler}, {Kleifges}, \& {Klein}}]{IceCube21}
{Aartsen}, M.~G., {Abbasi}, R., {Ackermann}, M., {et~al.} 2021, Journal of Physics G Nuclear Physics, 48, 060501, \dodoi{10.1088/1361-6471/abbd48}

\bibitem[{{Abbott} {et~al.}(2016){Abbott}, {Abbott}, {Abbott}, {Abernathy}, {Acernese}, {Ackley}, {Adams}, {Adams}, {Addesso}, {Adhikari}, {Adya}, {Affeldt}, {Agathos}, {Agatsuma}, {Aggarwal}, {Aguiar}, {Aiello}, {Ain}, {Ajith}, {Allen}, {Allocca}, {Altin}, {Anderson}, {Anderson}, {Arai}, {Arain}, {Araya}, {Arceneaux}, {Areeda}, {Arnaud}, {Arun}, {Ascenzi}, {Ashton}, {Ast}, {Aston}, {Astone}, {Aufmuth}, {Aulbert}, {Babak}, {Bacon}, {Bader}, {Baker}, {Baldaccini}, {Ballardin}, {Ballmer}, {Barayoga}, {Barclay}, {Barish}, {Barker}, {Barone}, {Barr}, {Barsotti}, {Barsuglia}, {Barta}, {Bartlett}, {Barton}, {Bartos}, {Bassiri}, {Basti}, {Batch}, {Baune}, {Bavigadda}, {Bazzan}, {Behnke}, {Bejger}, {Belczynski}, {Bell}, {Bell}, {Berger}, {Bergman}, {Bergmann}, {Berry}, {Bersanetti}, {Bertolini}, {Betzwieser}, {Bhagwat}, {Bhandare}, {Bilenko}, {Billingsley}, {Birch}, {Birney}, {Birnholtz}, {Biscans}, {Bisht}, {Bitossi}, {Biwer}, {Bizouard}, {Blackburn}, {Blair}, {Blair}, {Blair}, {Bloemen}, {Bock}, {Bodiya}, {Boer},
  {Bogaert}, {Bogan}, {Bohe}, {Bojtos}, {Bond}, {Bondu}, {Bonnand}, {Boom}, {Bork}, {Boschi}, {Bose}, {Bouffanais}, {Bozzi}, {Bradaschia}, {Brady}, {Braginsky}, {Branchesi}, {Brau}, {Briant}, {Brillet}, {Brinkmann}, {Brisson}, {Brockill}, {Brooks}, {Brown}, {Brown}, {Brown}, {Buchanan}, {Buikema}, {Bulik}, {Bulten}, {Buonanno}, {Buskulic}, {Buy}, {Byer}, {Cabero}, {Cadonati}, {Cagnoli}, {Cahillane}, {Bustillo}, {Callister}, {Calloni}, {Camp}, {Cannon}, {Cao}, {Capano}, {Capocasa}, {Carbognani}, {Caride}, {Diaz}, {Casentini}, {Caudill}, {Cavagli{\`a}}, {Cavalier}, {Cavalieri}, {Cella}, {Cepeda}, {Baiardi}, {Cerretani}, {Cesarini}, {Chakraborty}, {Chalermsongsak}, {Chamberlin}, {Chan}, {Chao}, {Charlton}, {Chassande-Mottin}, {Chen}, {Chen}, {Cheng}, {Chincarini}, {Chiummo}, {Cho}, {Cho}, {Chow}, {Christensen}, {Chu}, {Chua}, {Chung}, {Ciani}, {Clara}, {Clark}, {Cleva}, {Coccia}, {Cohadon}, {Colla}, {Collette}, {Cominsky}, {Constancio}, {Conte}, {Conti}, {Cook}, {Corbitt}, {Cornish}, {Corsi}, {Cortese}, {Costa},
  {Coughlin}, {Coughlin}, {Coulon}, {Countryman}, {Couvares}, {Cowan}, {Coward}, \& {Cowart}}]{GW150915}
{Abbott}, B.~P., {Abbott}, R., {Abbott}, T.~D., {et~al.} 2016, PhRL, 116, 061102, \dodoi{10.1103/PhysRevLett.116.061102}

\bibitem[{{Abbott} {et~al.}(2017){Abbott}, {Abbott}, {Abbott}, {Acernese}, {Ackley}, {Adams}, {Adams}, {Addesso}, {Adhikari}, {Adya}, {Affeldt}, {Afrough}, {Agarwal}, {Agathos}, {Agatsuma}, {Aggarwal}, {Aguiar}, {Aiello}, {Ain}, {Ajith}, {Allen}, {Allen}, {Allocca}, {Altin}, {Amato}, {Ananyeva}, {Anderson}, {Anderson}, {Angelova}, {Antier}, {Appert}, {Arai}, {Araya}, {Areeda}, {Arnaud}, {Arun}, {Ascenzi}, {Ashton}, {Ast}, {Aston}, {Astone}, {Atallah}, {Aufmuth}, {Aulbert}, {AultONeal}, {Austin}, {Avila-Alvarez}, {Babak}, {Bacon}, {Bader}, {Bae}, {Bailes}, {Baker}, {Baldaccini}, {Ballardin}, {Ballmer}, {Banagiri}, {Barayoga}, {Barclay}, {Barish}, {Barker}, {Barkett}, {Barone}, {Barr}, {Barsotti}, {Barsuglia}, {Barta}, {Barthelmy}, {Bartlett}, {Bartos}, {Bassiri}, {Basti}, {Batch}, {Bawaj}, {Bayley}, {Bazzan}, {B{\'e}csy}, {Beer}, {Bejger}, {Belahcene}, {Bell}, {Berger}, {Bergmann}, {Bernuzzi}, {Bero}, {Berry}, {Bersanetti}, {Bertolini}, {Betzwieser}, {Bhagwat}, {Bhandare}, {Bilenko}, {Billingsley}, {Billman},
  {Birch}, {Birney}, {Birnholtz}, {Biscans}, {Biscoveanu}, {Bisht}, {Bitossi}, {Biwer}, {Bizouard}, {Blackburn}, {Blackman}, {Blair}, {Blair}, {Blair}, {Bloemen}, {Bock}, {Bode}, {Boer}, {Bogaert}, {Bohe}, {Bondu}, {Bonilla}, {Bonnand}, {Boom}, {Bork}, {Boschi}, {Bose}, {Bossie}, {Bouffanais}, {Bozzi}, {Bradaschia}, {Brady}, {Branchesi}, {Brau}, {Briant}, {Brillet}, {Brinkmann}, {Brisson}, {Brockill}, {Broida}, {Brooks}, {Brown}, {Brown}, {Brunett}, {Buchanan}, {Buikema}, {Bulik}, {Bulten}, {Buonanno}, {Buskulic}, {Buy}, {Byer}, {Cabero}, {Cadonati}, {Cagnoli}, {Cahillane}, {Calder{\'o}n Bustillo}, {Callister}, {Calloni}, {Camp}, {Canepa}, {Canizares}, {Cannon}, {Cao}, {Cao}, {Capano}, {Capocasa}, {Carbognani}, {Caride}, {Carney}, {Carullo}, {Casanueva Diaz}, {Casentini}, {Caudill}, {Cavagli{\`a}}, {Cavalier}, {Cavalieri}, {Cella}, {Cepeda}, {Cerd{\'a}-Dur{\'a}n}, {Cerretani}, {Cesarini}, {Chamberlin}, {Chan}, {Chao}, {Charlton}, {Chase}, {Chassande-Mottin}, {Chatterjee}, {Chatziioannou}, {Cheeseboro},
  {Chen}, {Chen}, {Chen}, {Cheng}, {Chia}, {Chincarini}, {Chiummo}, {Chmiel}, {Cho}, {Cho}, {Chow}, {Christensen}, {Chu}, {Chua}, \& {Chua}}]{GW170817}
---. 2017, PhRL, 119, 161101, \dodoi{10.1103/PhysRevLett.119.161101}

\bibitem[{{Acernese} {et~al.}(2015){Acernese}, {Agathos}, {Agatsuma}, {Aisa}, {Allemandou}, {Allocca}, {Amarni}, {Astone}, {Balestri}, {Ballardin}, {Barone}, {Baronick}, {Barsuglia}, {Basti}, {Basti}, {Bauer}, {Bavigadda}, {Bejger}, {Beker}, {Belczynski}, {Bersanetti}, {Bertolini}, {Bitossi}, {Bizouard}, {Bloemen}, {Blom}, {Boer}, {Bogaert}, {Bondi}, {Bondu}, {Bonelli}, {Bonnand}, {Boschi}, {Bosi}, {Bouedo}, {Bradaschia}, {Branchesi}, {Briant}, {Brillet}, {Brisson}, {Bulik}, {Bulten}, {Buskulic}, {Buy}, {Cagnoli}, {Calloni}, {Campeggi}, {Canuel}, {Carbognani}, {Cavalier}, {Cavalieri}, {Cella}, {Cesarini}, {Mottin}, {Chincarini}, {Chiummo}, {Chua}, {Cleva}, {Coccia}, {Cohadon}, {Colla}, {Colombini}, {Conte}, {Coulon}, {Cuoco}, {Dalmaz}, {D'Antonio}, {Dattilo}, {Davier}, {Day}, {Debreczeni}, {Degallaix}, {Del{\'e}glise}, {Pozzo}, {Dereli}, {Rosa}, {Fiore}, {Lieto}, {Virgilio}, {Doets}, {Dolique}, {Drago}, {Ducrot}, {Endr{\H{o}}czi}, {Fafone}, {Farinon}, {Ferrante}, {Ferrini}, {Fidecaro}, {Fiori}, {Flaminio},
  {Fournier}, {Franco}, {Frasca}, {Frasconi}, {Gammaitoni}, {Garufi}, {Gaspard}, {Gatto}, {Gemme}, {Gendre}, {Genin}, {Gennai}, {Ghosh}, {Giacobone}, {Giazotto}, {Gouaty}, {Granata}, {Greco}, {Groot}, {Guidi}, {Harms}, {Heidmann}, {Heitmann}, {Hello}, {Hemming}, {Hennes}, {Hofman}, {Jaranowski}, {Jonker}, {Kasprzack}, {K{\'e}f{\'e}lian}, {Kowalska}, {Kraan}, {Kr{\'o}lak}, {Kutynia}, {Lazzaro}, {Leonardi}, {Leroy}, {Letendre}, {Li}, {Lieunard}, {Lorenzini}, {Loriette}, {Losurdo}, {Magazz{\`u}}, {Majorana}, {Maksimovic}, {Malvezzi}, {Man}, {Mangano}, {Mantovani}, {Marchesoni}, {Marion}, {Marque}, {Martelli}, {Martellini}, {Masserot}, {Meacher}, {Meidam}, {Mezzani}, {Michel}, {Milano}, {Minenkov}, {Moggi}, {Mohan}, {Montani}, {Morgado}, {Mours}, {Mul}, {Nagy}, {Nardecchia}, {Naticchioni}, {Nelemans}, {Neri}, {Neri}, {Nocera}, {Pacaud}, {Palomba}, {Paoletti}, {Paoli}, {Pasqualetti}, {Passaquieti}, {Passuello}, {Perciballi}, {Petit}, {Pichot}, {Piergiovanni}, {Pillant}, {Piluso}, {Pinard}, {Poggiani}, {Prijatelj},
  {Prodi}, {Punturo}, {Puppo}, {Rabeling}, {R{\'a}cz}, {Rapagnani}, {Razzano}, {Re}, {Regimbau}, {Ricci}, {Robinet}, {Rocchi}, {Rolland}, {Romano}, {Rosi{\'n}ska}, {Ruggi}, \& {Saracco}}]{Virgo15}
{Acernese}, F., {Agathos}, M., {Agatsuma}, K., {et~al.} 2015, Classical and Quantum Gravity, 32, 024001, \dodoi{10.1088/0264-9381/32/2/024001}

\bibitem[{{Addison} {et~al.}(2019){Addison}, {Wright}, {Wittenmyer}, {Horner}, {Mengel}, {Johns}, {Marti}, {Nicholson}, {Soutter}, {Bowler}, {Crossfield}, {Kane}, {Kielkopf}, {Plavchan}, {Tinney}, {Zhang}, {Clark}, {Clerte}, {Eastman}, {Swift}, {Bottom}, {Muirhead}, {McCrady}, {Herzig}, {Hogstrom}, {Wilson}, {Sliski}, {Johnson}, {Wright}, {Johnson}, {Blake}, {Riddle}, {Lin}, {Cornachione}, {Bedding}, {Stello}, {Huber}, {Marsden}, \& {Carter}}]{MINERVA19}
{Addison}, B., {Wright}, D.~J., {Wittenmyer}, R.~A., {et~al.} 2019, PASP, 131, 115003, \dodoi{10.1088/1538-3873/ab03aa}

\bibitem[{{Aerts} {et~al.}(2010){Aerts}, {Christensen-Dalsgaard}, \& {Kurtz}}]{2010aste.book.....A}
{Aerts}, C., {Christensen-Dalsgaard}, J., \& {Kurtz}, D.~W. 2010, {Asteroseismology}, \dodoi{10.1007/978-1-4020-5803-5}

\bibitem[{{Ageron} {et~al.}(2011){Ageron}, {Aguilar}, {Al Samarai}, {Albert}, {Ameli}, {Andr{\'e}}, {Anghinolfi}, {Anton}, {Anvar}, {Ardid}, {Arnaud}, {Aslanides}, {Assis Jesus}, {Astraatmadja}, {Aubert}, {Auer}, {Barbarito}, {Baret}, {Basa}, {Bazzotti}, {Becherini}, {Beltramelli}, {Bersani}, {Bertin}, {Beurthey}, {Biagi}, {Bigongiari}, {Billault}, {Blaes}, {Bogazzi}, {de Botton}, {Bou-Cabo}, {Boudahef}, {Bouwhuis}, {Brown}, {Brunner}, {Busto}, {Caillat}, {Calzas}, {Camarena}, {Capone}, {Caponetto}, {C{\^a}rloganu}, {Carminati}, {Carmona}, {Carr}, {Carton}, {Cassano}, {Castorina}, {Cecchini}, {Ceres}, {Chaleil}, {Charvis}, {Chauchot}, {Chiarusi}, {Circella}, {Comp{\`e}re}, {Coniglione}, {Coppolani}, {Cosquer}, {Costantini}, {Cottini}, {Coyle}, {Cuneo}, {Curtil}, {D'Amato}, {Damy}, {van Dantzig}, {de Bonis}, {Decock}, {Decowski}, {Dekeyser}, {Delagnes}, {Desages-Ardellier}, {Deschamps}, {Destelle}, {di Maria}, {Dinkespiler}, {Distefano}, {Dominique}, {Donzaud}, {Dornic}, {Dorosti}, {Drogou}, {Drouhin},
  {Druillole}, {Durand}, {Durand}, {Eberl}, {Emanuele}, {Engelen}, {Ernenwein}, {Escoffier}, {Falchini}, {Favard}, {Fehr}, {Feinstein}, {Ferri}, {Ferry}, {Fiorello}, {Flaminio}, {Folger}, {Fritsch}, {Fuda}, {Galat{\'a}}, {Galeotti}, {Gay}, {Gensolen}, {Giacomelli}, {Gojak}, {G{\'o}mez-Gonz{\'a}lez}, {Goret}, {Graf}, {Guillard}, {Halladjian}, {Hallewell}, {van Haren}, {Hartmann}, {Heijboer}, {Heine}, {Hello}, {Henry}, {Hern{\'a}ndez-Rey}, {Herold}, {H{\"o}{\ss}l}, {Hogenbirk}, {Hsu}, {Hubbard}, {Jaquet}, {Jaspers}, {de Jong}, {Jourde}, {Kadler}, {Kalantar-Nayestanaki}, {Kalekin}, {Kappes}, {Karg}, {Karkar}, {Karolak}, {Katz}, {Keller}, {Kestener}, {Kok}, {Kok}, {Kooijman}, {Kopper}, {Kouchner}, {Kretschmer}, {Kruijer}, {Kuch}, {Kulikovskiy}, {Lachartre}, {Lafoux}, {Lagier}, {Lahmann}, {Lahonde-Hamdoun}, {Lamare}, {Lambard}, {Languillat}, {Larosa}, {Lavalle}, {Le Guen}, {Le Provost}, {Levansuu}, {Lef{\`e}vre}, {Legou}, {Lelaizant}, {L{\'e}v{\'e}que}, {Lim}, {Lo Presti}, {Loehner}, {Loucatos}, {Louis},
  {Lucarelli}, {Lyashuk}, {Magnier}, {Mangano}, {Marcel}, {Marcelin}, {Margiotta}, {Martinez-Mora}, {Masullo}, {Maz{\'e}as}, {Mazure}, {Meli}, {Melissas}, {Migneco}, {Mongelli}, {Montaruli}, {Morganti}, {Moscoso}, {Motz}, {Musumeci}, {Naumann}, {Naumann-Godo}, {Neff}, {Niess}, {Nooren}, {Oberski}, \& {Olivetto}}]{Antares11}
{Ageron}, M., {Aguilar}, J.~A., {Al Samarai}, I., {et~al.} 2011, Nuclear Instruments and Methods in Physics Research A, 656, 11, \dodoi{10.1016/j.nima.2011.06.103}

\bibitem[{{Aiello} {et~al.}(2019){Aiello}, {Akrame}, {Ameli}, {Anassontzis}, {Andre}, {Androulakis}, {Anghinolfi}, {Anton}, {Ardid}, {Aublin}, {Avgitas}, {Bagatelas}, {Barbarino}, {Baret}, {Barrios-Mart{\'\i}}, {Belias}, {Berbee}, {van den Berg}, {Bertin}, {Biagi}, {Biagioni}, {Biernoth}, {Boumaaza}, {Bourret}, {Bouta}, {Bouwhuis}, {Bozza}, {Br{\^a}nza{\textcommabelow s}}, {Bruchner}, {Bruijn}, {Brunner}, {Buis}, {Buompane}, {Busto}, {Calvo}, {Capone}, {Celli}, {Chabab}, {Chau}, {Cherubini}, {Chiarella}, {Chiarusi}, {Circella}, {Cocimano}, {Coelho}, {Coleiro}, {Molla}, {Coniglione}, {Coyle}, {Creusot}, {Cuttone}, {D'Onofrio}, {Dallier}, {De Sio}, {Di Palma}, {D{\'\i}az}, {Diego-Tortosa}, {Distefano}, {Domi}, {Don{\`a}}, {Donzaud}, {Dornic}, {D{\"o}rr}, {Durocher}, {Eberl}, {van Eijk}, {El Bojaddaini}, {Eljarrari}, {Elsaesser}, {Enzenh{\"o}fer}, {Fermani}, {Ferrara}, {Filipovi{\'c}}, {Fusco}, {Gal}, {Garcia}, {Garufi}, {Gialanella}, {Giorgio}, {Giuliante}, {Gozzini}, {Gracia}, {Graf}, {Grasso}, {Gr{\'e}goire},
  {Grella}, {Hallmann}, {Hamdaoui}, {van Haren}, {Heid}, {Heijboer}, {Hekalo}, {Hern{\'a}ndez-Rey}, {Hofest{\"a}dt}, {Illuminati}, {James}, {Jongen}, {de Jong}, {de Jong}, {Kadler}, {Kalaczy{\'n}ski}, {Kalekin}, {Katz}, {Khan Chowdhury}, {Kie{\ss}ling}, {Koffeman}, {Kooijman}, {Kouchner}, {Kreter}, {Kulikovskiy}, {Kunhikannan Kannichankandy}, {Lahmann}, {Larosa}, {Le Breton}, {Leone}, {Leonora}, {Levi}, {Lincetto}, {Lonardo}, {Longhitano}, {Lopez Coto}, {Lotze}, {Maderer}, {Maggi}, {Ma{\'n}czak}, {Mannheim}, {Margiotta}, {Marinelli}, {Markou}, {Martin}, {Mart{\'\i}nez-Mora}, {Martini}, {Marzaioli}, {Mele}, {Melis}, {Migliozzi}, {Migneco}, {Mijakowski}, {Miranda}, {Mollo}, {Morganti}, {Moser}, {Moussa}, {Muller}, {Musumeci}, {Nauta}, {Navas}, {Nicolau}, {Nielsen}, {{\'O} Fearraigh}, {Organokov}, {Orlando}, {Ottonello}, {Panagopoulos}, {Papalashvili}, {Papaleo}, {P{\u{a}}v{\u{a}}la{\textcommabelow s}}, {Pellegrino}, {Perrin-Terrin}, {Piattelli}, {Pikounis}, {Pisanti}, {Poir{\'e}}, {Polydefki}, {Popa}, {Post},
  {Pradier}, {P{\"u}hlhofer}, {Pulvirenti}, {Quinn}, {Raffaelli}, {Randazzo}, {Razzaque}, {Real}, {Resvanis}, {Reubelt}, {Riccobene}, {Richer}, {Rigalleau}, {Rovelli}, {Saffer}, {Salvadori}, {Samtleben}, {S{\'a}nchez Losa}, {Sanguineti}, {Santangelo}, {Santonocito}, {Sapienza}, {Schumann}, {Sciacca}, {Seneca}, {Sgura}, {Shanidze}, {Sharma}, {Simeone}, {Sinopoulou}, {Spisso}, {Spurio}, {Stavropoulos}, \& {Steijger}}]{KM3NET19}
{Aiello}, S., {Akrame}, S.~E., {Ameli}, F., {et~al.} 2019, Astroparticle Physics, 111, 100, \dodoi{10.1016/j.astropartphys.2019.04.002}

\bibitem[{{Akeson} {et~al.}(2019){Akeson}, {Armus}, {Bachelet}, {Bailey}, {Bartusek}, {Bellini}, {Benford}, {Bennett}, {Bhattacharya}, {Bohlin}, {Boyer}, {Bozza}, {Bryden}, {Calchi Novati}, {Carpenter}, {Casertano}, {Choi}, {Content}, {Dayal}, {Dressler}, {Dor{\'e}}, {Fall}, {Fan}, {Fang}, {Filippenko}, {Finkelstein}, {Foley}, {Furlanetto}, {Kalirai}, {Gaudi}, {Gilbert}, {Girard}, {Grady}, {Greene}, {Guhathakurta}, {Heinrich}, {Hemmati}, {Hendel}, {Henderson}, {Henning}, {Hirata}, {Ho}, {Huff}, {Hutter}, {Jansen}, {Jha}, {Johnson}, {Jones}, {Kasdin}, {Kelly}, {Kirshner}, {Koekemoer}, {Kruk}, {Lewis}, {Macintosh}, {Madau}, {Malhotra}, {Mandel}, {Massara}, {Masters}, {McEnery}, {McQuinn}, {Melchior}, {Melton}, {Mennesson}, {Peeples}, {Penny}, {Perlmutter}, {Pisani}, {Plazas}, {Poleski}, {Postman}, {Ranc}, {Rauscher}, {Rest}, {Roberge}, {Robertson}, {Rodney}, {Rhoads}, {Rhodes}, {Ryan}, {Sahu}, {Sand}, {Scolnic}, {Seth}, {Shvartzvald}, {Siellez}, {Smith}, {Spergel}, {Stassun}, {Street}, {Strolger}, {Szalay},
  {Trauger}, {Troxel}, {Turnbull}, {van der Marel}, {von der Linden}, {Wang}, {Weinberg}, {Williams}, {Windhorst}, {Wollack}, {Wu}, {Yee}, \& {Zimmerman}}]{Roman19}
{Akeson}, R., {Armus}, L., {Bachelet}, E., {et~al.} 2019, arXiv e-prints, arXiv:1902.05569, \dodoi{10.48550/arXiv.1902.05569}

\bibitem[{{Alsabti} \& {Murdin}(2017)}]{2017hsn..book.....A}
{Alsabti}, A.~W., \& {Murdin}, P. 2017, {Handbook of Supernovae}, \dodoi{10.1007/978-3-319-21846-5}

\bibitem[{{Amaro-Seoane} {et~al.}(2023){Amaro-Seoane}, {Andrews}, {Arca Sedda}, {Askar}, {Baghi}, {Balasov}, {Bartos}, {Bavera}, {Bellovary}, {Berry}, {Berti}, {Bianchi}, {Blecha}, {Blondin}, {Bogdanovi{\'c}}, {Boissier}, {Bonetti}, {Bonoli}, {Bortolas}, {Breivik}, {Capelo}, {Caramete}, {Cattorini}, {Charisi}, {Chaty}, {Chen}, {Chru{\'s}li{\'n}ska}, {Chua}, {Church}, {Colpi}, {D'Orazio}, {Danielski}, {Davies}, {Dayal}, {De Rosa}, {Derdzinski}, {Destounis}, {Dotti}, {Du{\c{t}}an}, {Dvorkin}, {Fabj}, {Foglizzo}, {Ford}, {Fouvry}, {Franchini}, {Fragos}, {Fryer}, {Gaspari}, {Gerosa}, {Graziani}, {Groot}, {Habouzit}, {Haggard}, {Haiman}, {Han}, {Istrate}, {Johansson}, {Khan}, {Kimpson}, {Kokkotas}, {Kong}, {Korol}, {Kremer}, {Kupfer}, {Lamberts}, {Larson}, {Lau}, {Liu}, {Lloyd-Ronning}, {Lodato}, {Lupi}, {Ma}, {Maccarone}, {Mandel}, {Mangiagli}, {Mapelli}, {Mathis}, {Mayer}, {McGee}, {McKernan}, {Miller}, {Mota}, {Mumpower}, {Nasim}, {Nelemans}, {Noble}, {Pacucci}, {Panessa}, {Paschalidis}, {Pfister}, {Porquet},
  {Quenby}, {Ricarte}, {R{\"o}pke}, {Regan}, {Rosswog}, {Ruiter}, {Ruiz}, {Runnoe}, {Schneider}, {Schnittman}, {Secunda}, {Sesana}, {Seto}, {Shao}, {Shapiro}, {Sopuerta}, {Stone}, {Suvorov}, {Tamanini}, {Tamfal}, {Tauris}, {Temmink}, {Tomsick}, {Toonen}, {Torres-Orjuela}, {Toscani}, {Tsokaros}, {Unal}, {V{\'a}zquez-Aceves}, {Valiante}, {van Putten}, {van Roestel}, {Vignali}, {Volonteri}, {Wu}, {Younsi}, {Yu}, {Zane}, {Zwick}, {Antonini}, {Baibhav}, {Barausse}, {Bonilla Rivera}, {Branchesi}, {Branduardi-Raymont}, {Burdge}, {Chakraborty}, {Cuadra}, {Dage}, {Davis}, {de Mink}, {Decarli}, {Doneva}, {Escoffier}, {Gandhi}, {Haardt}, {Lousto}, {Nissanke}, {Nordhaus}, {O'Shaughnessy}, {Portegies Zwart}, {Pound}, {Schussler}, {Sergijenko}, {Spallicci}, {Vernieri}, \& {Vigna-G{\'o}mez}}]{LISA23}
{Amaro-Seoane}, P., {Andrews}, J., {Arca Sedda}, M., {et~al.} 2023, Living Reviews in Relativity, 26, 2, \dodoi{10.1007/s41114-022-00041-y}

\bibitem[{{Amato} {et~al.}(2025){Amato}, {Bachetti}, {Soria}, {G{\'u}rpide}, {Imbrogno}, {Salvaggio}, {Salvaterra}, {Del Santo}, {Scaringi}, {Casella}, \& {Wolter}}]{2025arXiv251220265A}
{Amato}, R., {Bachetti}, M., {Soria}, R., {et~al.} 2025, arXiv e-prints, arXiv:2512.20265, \dodoi{10.48550/arXiv.2512.20265}

\bibitem[{{Ambrosino} {et~al.}(2014){Ambrosino}, {Meddi}, {Rossi}, {Sclavi}, {Nesci}, {Bruni}, {Ghedina}, {Riverol}, \& {Di Fabrizio}}]{2014SPIE.9147E..8RA}
{Ambrosino}, F., {Meddi}, F., {Rossi}, C., {et~al.} 2014, in Society of Photo-Optical Instrumentation Engineers (SPIE) Conference Series, Vol. 9147, Ground-based and Airborne Instrumentation for Astronomy V, ed. S.~K. {Ramsay}, I.~S. {McLean}, \& H.~{Takami}, 91478R, \dodoi{10.1117/12.2064649}

\bibitem[{{Angel} {et~al.}(1977){Angel}, {Adams}, {Boroson}, \& {Moore}}]{FLOAT77}
{Angel}, J.~R.~P., {Adams}, M.~T., {Boroson}, T.~A., \& {Moore}, R.~L. 1977, ApJ, 218, 776, \dodoi{10.1086/155734}

\bibitem[{{Angel} {et~al.}(2022){Angel}, {Bender}, {Berkson}, {Didato}, {Ford}, {Gray}, {Jannuzi}, {Ketelsen}, {Kim}, {Chavez Lopez}, {Monson}, {Oh}, {Patrou}, {Rademacher}, {Schwab}, {Sisco}, {Wortley}, \& {Young}}]{LFAST22}
{Angel}, R., {Bender}, C., {Berkson}, J., {et~al.} 2022, in Society of Photo-Optical Instrumentation Engineers (SPIE) Conference Series, Vol. 12182, Ground-based and Airborne Telescopes IX, ed. H.~K. {Marshall}, J.~{Spyromilio}, \& T.~{Usuda}, 121821U, \dodoi{10.1117/12.2629655}

\bibitem[{{Arnaboldi} {et~al.}(1998){Arnaboldi}, {Capaccioli}, {Mancini}, {Rafanelli}, {Scaramella}, {Sedmak}, \& {Vettolani}}]{1998Msngr..93...30A}
{Arnaboldi}, M., {Capaccioli}, M., {Mancini}, D., {et~al.} 1998, The Messenger, 93, 30

\bibitem[{{Atwood} {et~al.}(2009){Atwood}, {Abdo}, {Ackermann}, {Althouse}, {Anderson}, {Axelsson}, {Baldini}, {Ballet}, {Band}, {Barbiellini}, \& et~al.}]{2009ApJ...697.1071A}
{Atwood}, W.~B., {Abdo}, A.~A., {Ackermann}, M., {et~al.} 2009, ApJ, 697, 1071, \dodoi{10.1088/0004-637X/697/2/1071}

\bibitem[{{Bacon} {et~al.}(2010){Bacon}, {Accardo}, {Adjali}, {Anwand}, {Bauer}, {Biswas}, {Blaizot}, {Boudon}, {Brau-Nogue}, {Brinchmann}, {Caillier}, {Capoani}, {Carollo}, {Contini}, {Couderc}, {Daguis{\'e}}, {Deiries}, {Delabre}, {Dreizler}, {Dubois}, {Dupieux}, {Dupuy}, {Emsellem}, {Fechner}, {Fleischmann}, {Fran{\c{c}}ois}, {Gallou}, {Gharsa}, {Glindemann}, {Gojak}, {Guiderdoni}, {Hansali}, {Hahn}, {Jarno}, {Kelz}, {Koehler}, {Kosmalski}, {Laurent}, {Le Floch}, {Lilly}, {Lizon}, {Loupias}, {Manescau}, {Monstein}, {Nicklas}, {Olaya}, {Pares}, {Pasquini}, {P{\'e}contal-Rousset}, {Pell{\'o}}, {Petit}, {Popow}, {Reiss}, {Remillieux}, {Renault}, {Roth}, {Rupprecht}, {Serre}, {Schaye}, {Soucail}, {Steinmetz}, {Streicher}, {Stuik}, {Valentin}, {Vernet}, {Weilbacher}, {Wisotzki}, \& {Yerle}}]{MUSE10}
{Bacon}, R., {Accardo}, M., {Adjali}, L., {et~al.} 2010, in Society of Photo-Optical Instrumentation Engineers (SPIE) Conference Series, Vol. 7735, Ground-based and Airborne Instrumentation for Astronomy III, ed. I.~S. {McLean}, S.~K. {Ramsay}, \& H.~{Takami}, 773508, \dodoi{10.1117/12.856027}

\bibitem[{{Ballocco} {et~al.}(2026){Ballocco}, {Miraval Zanon}, {Papitto}, {Burgay}, {Baglio}, {Malacaria}, {Ambrosino}, {Possenti}, {Illiano}, {Barr}, \& et~al.}]{2026arXiv261018322B}
{Ballocco}, C., {Miraval Zanon}, A., {Papitto}, A., {et~al.} 2026, arXiv e-prints, arXiv:2610.18322, \dodoi{10.48550/arXiv.2610.18322}

\bibitem[{{Baranec} {et~al.}(2013){Baranec}, {Riddle}, {Law}, {Ramaprakash}, {Tendulkar}, {Bui}, {Burse}, {Chordia}, {Das}, {Davis}, {Dekany}, {Kasliwal}, {Kulkarni}, {Morton}, {Ofek}, \& {Punnadi}}]{RoboAO13}
{Baranec}, C., {Riddle}, R., {Law}, N.~M., {et~al.} 2013, Journal of Vibration Engineering, 72, 50021, \dodoi{10.3791/50021}

\bibitem[{{Bellm} {et~al.}(2019){Bellm}, {Kulkarni}, {Graham}, {Dekany}, {Smith}, {Riddle}, {Masci}, {Helou}, {Prince}, {Adams}, {Barbarino}, {Barlow}, {Bauer}, {Beck}, {Belicki}, {Biswas}, {Blagorodnova}, {Bodewits}, {Bolin}, {Brinnel}, {Brooke}, {Bue}, {Bulla}, {Burruss}, {Cenko}, {Chang}, {Connolly}, {Coughlin}, {Cromer}, {Cunningham}, {De}, {Delacroix}, {Desai}, {Duev}, {Eadie}, {Farnham}, {Feeney}, {Feindt}, {Flynn}, {Franckowiak}, {Frederick}, {Fremling}, {Gal-Yam}, {Gezari}, {Giomi}, {Goldstein}, {Golkhou}, {Goobar}, {Groom}, {Hacopians}, {Hale}, {Henning}, {Ho}, {Hover}, {Howell}, {Hung}, {Huppenkothen}, {Imel}, {Ip}, {Ivezi{\'c}}, {Jackson}, {Jones}, {Juric}, {Kasliwal}, {Kaspi}, {Kaye}, {Kelley}, {Kowalski}, {Kramer}, {Kupfer}, {Landry}, {Laher}, {Lee}, {Lin}, {Lin}, {Lunnan}, {Giomi}, {Mahabal}, {Mao}, {Miller}, {Monkewitz}, {Murphy}, {Ngeow}, {Nordin}, {Nugent}, {Ofek}, {Patterson}, {Penprase}, {Porter}, {Rauch}, {Rebbapragada}, {Reiley}, {Rigault}, {Rodriguez}, {van Roestel}, {Rusholme}, {van
  Santen}, {Schulze}, {Shupe}, {Singer}, {Soumagnac}, {Stein}, {Surace}, {Sollerman}, {Szkody}, {Taddia}, {Terek}, {Van Sistine}, {van Velzen}, {Vestrand}, {Walters}, {Ward}, {Ye}, {Yu}, {Yan}, \& {Zolkower}}]{ZTF19}
{Bellm}, E.~C., {Kulkarni}, S.~R., {Graham}, M.~J., {et~al.} 2019, PASP, 131, 018002, \dodoi{10.1088/1538-3873/aaecbe}

\bibitem[{{Ben-Ami} {et~al.}(2026){Ben-Ami}, {Ofek}, {Yaron}, {Polishook}, {Bichkovsky}, {Blumenzweig}, {Brody}, {Gal-Yam}, {Golod}, {Hallakoun}, {Hershko}, {Ironi}, {Rekhi}, {Sofer-Rimalt}, {Kuncarayakti}, {Mattila}, \& {Mazeh}}]{MAST26}
{Ben-Ami}, S., {Ofek}, E.~O., {Yaron}, O., {et~al.} 2026, in Society of Photo-Optical Instrumentation Engineers (SPIE) Conference Series, Vol. 14147, Ground-based and Airborne Telescopes XI, ed. H.~K. {Marshall}, T.~{Usuda}, B.~{Onodera}, \& J.~{Spyromilio} (SPIE), 1414728, \dodoi{10.1117/12.3105785}

\bibitem[{{Blagorodnova} {et~al.}(2025){Blagorodnova}, {Pejcha}, {Kami{\'n}ski}, {Cai}, {De}, {Elias-Rosa}, {Fuller}, {Ge}, {Jones}, {Justham}, {Karambelkar}, {Klencki}, {Mason}, {Metzger}, {Pastorello}, {Reguitti}, {R{\"o}pke}, {Shore}, \& {Valerin}}]{2025arXiv251214802B}
{Blagorodnova}, N., {Pejcha}, O., {Kami{\'n}ski}, T., {et~al.} 2025, arXiv e-prints, arXiv:2512.14802, \dodoi{10.48550/arXiv.2512.14802}

\bibitem[{{Boato} {et~al.}(2026){Boato}, {Greggio}, {Rebrysh}, {Bournane}, {Margin}, {Di Rosa}, {Ciroi}, {Ragazzoni}, {Arcidiacono}, {Dima}, \& et~al.}]{2026SPIE14147E..0DB}
{Boato}, G., {Greggio}, D., {Rebrysh}, O., {et~al.} 2026, in Society of Photo-Optical Instrumentation Engineers (SPIE) Conference Series, Vol. 14147, Ground-based and Airborne Telescopes XI, ed. H.~K. {Marshall}, T.~{Usuda}, B.~{Onodera}, \& J.~{Spyromilio} (SPIE), 141470D, \dodoi{10.1117/12.3105088}

\bibitem[{{Borucki} {et~al.}(2010){Borucki}, {Koch}, {Basri}, {Batalha}, {Brown}, {Caldwell}, {Caldwell}, {Christensen-Dalsgaard}, {Cochran}, {DeVore}, {Dunham}, {Dupree}, {Gautier}, {Geary}, {Gilliland}, {Gould}, {Howell}, {Jenkins}, {Kondo}, {Latham}, {Marcy}, {Meibom}, {Kjeldsen}, {Lissauer}, {Monet}, {Morrison}, {Sasselov}, {Tarter}, {Boss}, {Brownlee}, {Owen}, {Buzasi}, {Charbonneau}, {Doyle}, {Fortney}, {Ford}, {Holman}, {Seager}, {Steffen}, {Welsh}, {Rowe}, {Anderson}, {Buchhave}, {Ciardi}, {Walkowicz}, {Sherry}, {Horch}, {Isaacson}, {Everett}, {Fischer}, {Torres}, {Johnson}, {Endl}, {MacQueen}, {Bryson}, {Dotson}, {Haas}, {Kolodziejczak}, {Van Cleve}, {Chandrasekaran}, {Twicken}, {Quintana}, {Clarke}, {Allen}, {Li}, {Wu}, {Tenenbaum}, {Verner}, {Bruhweiler}, {Barnes}, \& {Prsa}}]{Kepler10}
{Borucki}, W.~J., {Koch}, D., {Basri}, G., {et~al.} 2010, Science, 327, 977, \dodoi{10.1126/science.1185402}

\bibitem[{{Braun} {et~al.}(2019){Braun}, {Bonaldi}, {Bourke}, {Keane}, \& {Wagg}}]{2019arXiv191212699B}
{Braun}, R., {Bonaldi}, A., {Bourke}, T., {Keane}, E., \& {Wagg}, J. 2019, arXiv e-prints, arXiv:1912.12699, \dodoi{10.48550/arXiv.1912.12699}

\bibitem[{{Castro Segura} {et~al.}(2026){Castro Segura}, {C{\'u}neo}, {Tombesi}, {Fijma}, {Corral-Santana}, {Veledina}, {Ambrifi}, {Buckley}, {Casella}, {Coppejans}, {de Martino}, \& {Scaringi}}]{2026arXiv260102028C}
{Castro Segura}, N., {C{\'u}neo}, V., {Tombesi}, F., {et~al.} 2026, arXiv e-prints, arXiv:2601.02028, \dodoi{10.48550/arXiv.2601.02028}

\bibitem[{{Chakraborty} {et~al.}(2024){Chakraborty}, {Burdge}, {Rappaport}, {Munday}, {Chen}, {Rodr{\'\i}guez-Gil}, {Dhillon}, {Hughes}, {Nelemans}, {Kara}, {Bellm}, {Brown}, {Castro Segura}, {Chen}, {Chickles}, {Dyer}, {Dekany}, {Drake}, {Garbutt}, {Graham}, {Green}, {Jarvis}, {Kennedy}, {Kerry}, {Kulkarni}, {Littlefair}, {Mahabal}, {Masci}, {McCormac}, {Parsons}, {Pelisoli}, {Pike}, {Prince}, {Riddle}, {van Roestel}, {Sahman}, {Wold}, \& {Sunny Wong}}]{Chakraborty24}
{Chakraborty}, J., {Burdge}, K.~B., {Rappaport}, S.~A., {et~al.} 2024, ApJ, 977, 262, \dodoi{10.3847/1538-4357/ad9563}

\bibitem[{{Chambers} {et~al.}(2016){Chambers}, {Magnier}, {Metcalfe}, {Flewelling}, {Huber}, {Waters}, {Denneau}, {Draper}, {Farrow}, {Finkbeiner}, {Holmberg}, {Koppenhoefer}, {Price}, {Rest}, {Saglia}, {Schlafly}, {Smartt}, {Sweeney}, {Wainscoat}, {Burgett}, {Chastel}, {Grav}, {Heasley}, {Hodapp}, {Jedicke}, {Kaiser}, {Kudritzki}, {Luppino}, {Lupton}, {Monet}, {Morgan}, {Onaka}, {Shiao}, {Stubbs}, {Tonry}, {White}, {Ba{\~n}ados}, {Bell}, {Bender}, {Bernard}, {Boegner}, {Boffi}, {Botticella}, {Calamida}, {Casertano}, {Chen}, {Chen}, {Cole}, {Deacon}, {Frenk}, {Fitzsimmons}, {Gezari}, {Gibbs}, {Goessl}, {Goggia}, {Gourgue}, {Goldman}, {Grant}, {Grebel}, {Hambly}, {Hasinger}, {Heavens}, {Heckman}, {Henderson}, {Henning}, {Holman}, {Hopp}, {Ip}, {Isani}, {Jackson}, {Keyes}, {Koekemoer}, {Kotak}, {Le}, {Liska}, {Long}, {Lucey}, {Liu}, {Martin}, {Masci}, {McLean}, {Mindel}, {Misra}, {Morganson}, {Murphy}, {Obaika}, {Narayan}, {Nieto-Santisteban}, {Norberg}, {Peacock}, {Pier}, {Postman}, {Primak}, {Rae}, {Rai},
  {Riess}, {Riffeser}, {Rix}, {R{\"o}ser}, {Russel}, {Rutz}, {Schilbach}, {Schultz}, {Scolnic}, {Strolger}, {Szalay}, {Seitz}, {Small}, {Smith}, {Soderblom}, {Taylor}, {Thomson}, {Taylor}, {Thakar}, {Thiel}, {Thilker}, {Unger}, {Urata}, {Valenti}, {Wagner}, {Walder}, {Walter}, {Watters}, {Werner}, {Wood-Vasey}, \& {Wyse}}]{PS1DR2}
{Chambers}, K.~C., {Magnier}, E.~A., {Metcalfe}, N., {et~al.} 2016, arXiv e-prints, arXiv:1612.05560, \dodoi{10.48550/arXiv.1612.05560}

\bibitem[{{Cherenkov Telescope Array Consortium} {et~al.}(2019{\natexlab{a}}){Cherenkov Telescope Array Consortium}, {Acharya}, {Agudo}, {Al Samarai}, {Alfaro}, {Alfaro}, {Alispach}, {Alves Batista}, {Amans}, {Amato}, \& et~al.}]{2019scta.book.....C}
{Cherenkov Telescope Array Consortium}, {Acharya}, B.~S., {Agudo}, I., {et~al.} 2019{\natexlab{a}}, {Science with the Cherenkov Telescope Array}, \dodoi{10.1142/10986}

\bibitem[{{Cherenkov Telescope Array Consortium} {et~al.}(2019{\natexlab{b}}){Cherenkov Telescope Array Consortium}, {Acharya}, {Agudo}, {Al Samarai}, {Alfaro}, {Alfaro}, {Alispach}, {Alves Batista}, {Amans}, {Amato}, {Ambrosi}, {Antolini}, {Antonelli}, {Aramo}, {Araya}, {Armstrong}, {Arqueros}, {Arrabito}, {Asano}, {Ashley}, {Backes}, {Balazs}, {Balbo}, {Ballester}, {Ballet}, {Bamba}, {Barkov}, {Barres de Almeida}, {Barrio}, {Bastieri}, {Becherini}, {Belfiore}, {Benbow}, {Berge}, {Bernardini}, {Bernardini}, {Bernardos}, {Bernl{\"o}hr}, {Bertucci}, {Biasuzzi}, {Bigongiari}, {Biland}, {Bissaldi}, {Biteau}, {Blanch}, {Blazek}, {Boisson}, {Bolmont}, {Bonanno}, {Bonardi}, {Bonavolont{\`a}}, {Bonnoli}, {Bosnjak}, {B{\"o}ttcher}, {Braiding}, {Bregeon}, {Brill}, {Brown}, {Brun}, {Brunetti}, {Buanes}, {Buckley}, {Bugaev}, {B{\"u}hler}, {Bulgarelli}, {Bulik}, {Burton}, {Burtovoi}, {Busetto}, {Canestrari}, {Capalbi}, {Capitanio}, {Caproni}, {Caraveo}, {C{\'a}rdenas}, {Carlile}, {Carosi}, {Carqu{\'\i}n}, {Carr},
  {Casanova}, {Cascone}, {Catalani}, {Catalano}, {Cauz}, {Cerruti}, {Chadwick}, {Chaty}, {Chaves}, {Chen}, {Chen}, {Chernyakova}, {Chikawa}, {Christov}, {Chudoba}, {Cie{\'s}lar}, {Coco}, {Colafrancesco}, {Colin}, {Conforti}, {Connaughton}, {Conrad}, {Contreras}, {Cortina}, {Costa}, {Costantini}, {Cotter}, {Covino}, {Crocker}, {Cuadra}, {Cuevas}, {Cumani}, {D'A{\`\i}}, {D'Ammando}, {D'Avanzo}, {D'Urso}, {Daniel}, {Davids}, {Dawson}, {Dazzi}, {De Angelis}, {de C{\'a}ssia dos Anjos}, {De Cesare}, {De Franco}, {de Gouveia Dal Pino}, {de la Calle}, {de los Reyes Lopez}, {De Lotto}, {De Luca}, {De Lucia}, {de Naurois}, {de O{\~n}a Wilhelmi}, {De Palma}, {De Persio}, {de Souza}, {Deil}, {Del Santo}, {Delgado}, {della Volpe}, {Di Girolamo}, {Di Pierro}, {Di Venere}, {D{\'\i}az}, {Dib}, {Diebold}, {Djannati-Ata{\"\i}}, {Dom{\'\i}nguez}, {Dominis Prester}, {Dorner}, {Doro}, {Drass}, {Dravins}, {Dubus}, {Dwarkadas}, {Ebr}, {Eckner}, {Egberts}, {Einecke}, {Ekoume}, {Els{\"a}sser}, {Ernenwein}, {Espinoza}, {Evoli},
  {Fairbairn}, {Falceta-Goncalves}, {Falcone}, {Farnier}, {Fasola}, {Fedorova}, {Fegan}, {Fernandez-Alonso}, {Fern{\'a}ndez-Barral}, {Ferrand}, {Fesquet}, {Filipovic}, {Fioretti}, {Fontaine}, {Fornasa}, {Fortson}, {Freixas Coromina}, {Fruck}, {Fujita}, {Fukazawa}, {Funk}, {F{\"u}{\ss}ling}, {Gabici}, {Gadola}, {Gallant}, {Garcia}, {Garcia L{\'o}pez}, {Garczarczyk}, {Gaskins}, {Gasparetto}, {Gaug}, {Gerard}, {Giavitto}, {Giglietto}, {Giommi}, {Giordano}, {Giro}, \& {Giroletti}}]{cta19}
---. 2019{\natexlab{b}}, {Science with the Cherenkov Telescope Array}, \dodoi{10.1142/10986}

\bibitem[{{CHIME/FRB Collaboration} {et~al.}(2021){CHIME/FRB Collaboration}, {Amiri}, {Andersen}, {Bandura}, {Berger}, {Bhardwaj}, {Boyce}, {Boyle}, {Brar}, {Breitman}, \& et~al.}]{2021ApJS..257...59C}
{CHIME/FRB Collaboration}, {Amiri}, M., {Andersen}, B.~C., {et~al.} 2021, ApJS, 257, 59, \dodoi{10.3847/1538-4365/ac33ab}

\bibitem[{{Chrimes} {et~al.}(2025){Chrimes}, {Sarin}, {Coppejans}, {Groot}, {Inkenhaag}, {Jonker}, {Killestein}, {Perley}, \& {Pursiainen}}]{2025arXiv251214832C}
{Chrimes}, A.~A., {Sarin}, N., {Coppejans}, D., {et~al.} 2025, arXiv e-prints, arXiv:2512.14832, \dodoi{10.48550/arXiv.2512.14832}

\bibitem[{{Claudi} {et~al.}(2019){Claudi}, {Campana}, {Schipani}, {Aliverti}, {Baruffolo}, {Ben-Ami}, {Biondi}, {Brucalassi}, {Capasso}, {Cosentino}, \& et~al.}]{2019frap.confE..78C}
{Claudi}, R., {Campana}, S., {Schipani}, P., {et~al.} 2019, in Frontier Research in Astrophysics - III. 28 May - 2 June 2018. Mondello (Palermo), 78, \dodoi{10.22323/1.331.0078}

\bibitem[{{Cordier} {et~al.}(2026){Cordier}, {Wei}, {Zhang}, {Basa}, {Atteia}, {Claret}, {Coleiro}, {Daigne}, {Dagoneau}, {Deng}, {Dong}, {Godet}, {G{\"o}tz}, {Han}, {Lachaud}, {Liang}, {Piron}, {Qiu}, {Schanne}, {Turpin}, {Vergani}, {Wang}, {Wu}, {Xin}, {Zhang}, {Bai}, {Crepaldi}, {Feng}, {Gonzalez}, {Huang}, {Li}, {Liu}, {Louvin}, {Mercier}, {Jaubert}, {Su}, {Wei}, {Zhang}, {Zhang}, \& {the SVOM collaboration}}]{SVOM26}
{Cordier}, B., {Wei}, J.~Y., {Zhang}, S.~N., {et~al.} 2026, Research in Astronomy and Astrophysics, 26, 104001, \dodoi{10.1088/1674-4527/ae75b8}

\bibitem[{{Corral-Santana} {et~al.}(2016){Corral-Santana}, {Casares}, {Mu{\~n}oz-Darias}, {Bauer}, {Mart{\'\i}nez-Pais}, \& {Russell}}]{BlackCat16}
{Corral-Santana}, J.~M., {Casares}, J., {Mu{\~n}oz-Darias}, T., {et~al.} 2016, A\&A, 587, A61, \dodoi{10.1051/0004-6361/201527130}

\bibitem[{Cusini {et~al.}(2022)Cusini, Berretta, Conca, Incoronato, Madonini, Maurina, Nonne, Riccardo, \& Villa}]{10.3389/fphy.2022.906671}
Cusini, I., Berretta, D., Conca, E., {et~al.} 2022, Frontiers in Physics, Volume 10 - 2022, \dodoi{10.3389/fphy.2022.906671}

\bibitem[{{Da{\v{g}}} {et~al.}(2026){Da{\v{g}}}, {O'Brien}, {Scaringi}, {Gill-Fleming}, \& {Crossley}}]{KIDspec}
{Da{\v{g}}}, M.~K., {O'Brien}, K., {Scaringi}, S., {Gill-Fleming}, J., \& {Crossley}, I. 2026, in Society of Photo-Optical Instrumentation Engineers (SPIE) Conference Series, Vol. 14149, Ground-based and Airborne Instrumentation for Astronomy XI, ed. J.~D.~R. {Vernet}, J.~J. {Bryant}, \& K.~{Motohara} (SPIE), 141492G, \dodoi{10.1117/12.3104736}

\bibitem[{{DeBoer} {et~al.}(2009){DeBoer}, {Gough}, {Bunton}, {Cornwell}, {Beresford}, {Johnston}, {Feain}, {Schinckel}, {Jackson}, {Kesteven}, \& et~al.}]{2009IEEEP..97.1507D}
{DeBoer}, D.~R., {Gough}, R.~G., {Bunton}, J.~D., {et~al.} 2009, IEEE Proceedings, 97, 1507, \dodoi{10.1109/JPROC.2009.2016516}

\bibitem[{{Ederoclite} {et~al.}(2025){Ederoclite}, {De Martino}, {Groot}, {Mason}, {Sala}, {Guerrero}, {Kupfer}, {Pala}, {Scaringi}, \& {Castro Segura}}]{2025arXiv251215821E}
{Ederoclite}, A., {De Martino}, D., {Groot}, P., {et~al.} 2025, arXiv e-prints, arXiv:2512.15821, \dodoi{10.48550/arXiv.2512.15821}

\bibitem[{{Euclid Collaboration} {et~al.}(2025){Euclid Collaboration}, {Mellier}, {Abdurro'uf}, {Acevedo Barroso}, {Ach{\'u}carro}, {Adamek}, {Adam}, {Addison}, {Aghanim}, {Aguena}, {Ajani}, {Akrami}, {Al-Bahlawan}, {Alavi}, {Albuquerque}, {Alestas}, {Alguero}, {Allaoui}, {Allen}, {Allevato}, {Alonso-Tetilla}, {Altieri}, {Alvarez-Candal}, {Alvi}, {Amara}, {Amendola}, {Amiaux}, {Andika}, {Andreon}, {Andrews}, {Angora}, {Angulo}, {Annibali}, {Anselmi}, {Anselmi}, {Arcari}, {Archidiacono}, {Aric{\`o}}, {Arnaud}, {Arnouts}, {Asgari}, {Asorey}, {Atayde}, {Atek}, {Atrio-Barandela}, {Aubert}, {Aubourg}, {Auphan}, {Auricchio}, {Aussel}, {Aussel}, {Avelino}, {Avgoustidis}, {Avila}, {Awan}, {Azzollini}, {Baccigalupi}, {Bachelet}, {Bacon}, {Baes}, {Bagley}, {Bahr-Kalus}, {Balaguera-Antolinez}, {Balbinot}, {Balcells}, {Baldi}, {Baldry}, {Balestra}, {Ballardini}, {Ballester}, {Balogh}, {Ba{\~n}ados}, {Barbier}, {Bardelli}, {Baron}, {Barreiro}, {Barrena}, {Barriere}, {Barros}, {Barthelemy}, {Bartolo}, {Basset},
  {Battaglia}, {Battisti}, {Baugh}, {Baumont}, {Bazzanini}, {Beaulieu}, {Beckmann}, {Belikov}, {Bel}, {Bellagamba}, {Bella}, {Bellini}, {Benabed}, {Bender}, {Benevento}, {Bennett}, {Benson}, {Bergamini}, {Bermejo-Climent}, {Bernardeau}, {Bertacca}, {Berthe}, {Berthier}, {Bethermin}, {Beutler}, {Bevillon}, {Bhargava}, {Bhatawdekar}, {Bianchi}, {Bisigello}, {Biviano}, {Blake}, {Blanchard}, {Blazek}, {Blot}, {Bosco}, {Bodendorf}, {Boenke}, {B{\"o}hringer}, {Boldrini}, {Bolzonella}, {Bonchi}, {Bonici}, {Bonino}, {Bonino}, {Bonvin}, {Bon}, {Booth}, {Borgani}, {Borlaff}, {Borsato}, {Bose}, {Botticella}, {Boucaud}, {Bouche}, {Boucher}, {Boutigny}, {Bouvard}, {Bouwens}, {Bouy}, {Bowler}, {Bozza}, {Bozzo}, {Branchini}, {Brando}, {Brau-Nogue}, {Brekke}, {Bremer}, {Brescia}, {Breton}, {Brinchmann}, {Brinckmann}, {Brockley-Blatt}, {Brodwin}, {Brouard}, {Brown}, {Bruton}, {Bucko}, {Buddelmeijer}, {Buenadicha}, {Buitrago}, {Burger}, {Burigana}, {Busillo}, {Busonero}, {Cabanac}, {Cabayol-Garcia}, {Cagliari}, {Caillat},
  {Caillat}, {Calabrese}, {Calabro}, {Calderone}, {Calura}, {Camacho Quevedo}, {Camera}, {Campos}, {Ca{\~n}as-Herrera}, {Candini}, {Cantiello}, {Capobianco}, {Cappellaro}, {Cappelluti}, {Cappi}, {Caputi}, {Cara}, {Carbone}, {Cardone}, {Carella}, {Carlberg}, {Carle}, {Carminati}, {Caro}, {Carrasco}, {Carretero}, {Carrilho}, {Carron Duque}, \& {Carry}}]{Euclid25}
{Euclid Collaboration}, {Mellier}, Y., {Abdurro'uf}, {et~al.} 2025, A\&A, 697, A1, \dodoi{10.1051/0004-6361/202450810}

\bibitem[{{Evans} {et~al.}(2021){Evans}, {Adhikari}, {Afle}, {Ballmer}, {Biscoveanu}, {Borhanian}, {Brown}, {Chen}, {Eisenstein}, {Gruson}, {Gupta}, {Hall}, {Huxford}, {Kamai}, {Kashyap}, {Kissel}, {Kuns}, {Landry}, {Lenon}, {Lovelace}, {McCuller}, {Ng}, {Nitz}, {Read}, {Sathyaprakash}, {Shoemaker}, {Slagmolen}, {Smith}, {Srivastava}, {Sun}, {Vitale}, \& {Weiss}}]{CE21}
{Evans}, M., {Adhikari}, R.~X., {Afle}, C., {et~al.} 2021, arXiv e-prints, arXiv:2109.09882, \dodoi{10.48550/arXiv.2109.09882}

\bibitem[{{Gaia Collaboration} {et~al.}(2016){Gaia Collaboration}, {Prusti}, {de Bruijne}, {Brown}, {Vallenari}, {Babusiaux}, {Bailer-Jones}, {Bastian}, {Biermann}, {Evans}, {Eyer}, {Jansen}, {Jordi}, {Klioner}, {Lammers}, {Lindegren}, {Luri}, {Mignard}, {Milligan}, {Panem}, {Poinsignon}, {Pourbaix}, {Randich}, {Sarri}, {Sartoretti}, {Siddiqui}, {Soubiran}, {Valette}, {van Leeuwen}, {Walton}, {Aerts}, {Arenou}, {Cropper}, {Drimmel}, {H{\o}g}, {Katz}, {Lattanzi}, {O'Mullane}, {Grebel}, {Holland}, {Huc}, {Passot}, {Bramante}, {Cacciari}, {Casta{\~n}eda}, {Chaoul}, {Cheek}, {De Angeli}, {Fabricius}, {Guerra}, {Hern{\'a}ndez}, {Jean-Antoine-Piccolo}, {Masana}, {Messineo}, {Mowlavi}, {Nienartowicz}, {Ord{\'o}{\~n}ez-Blanco}, {Panuzzo}, {Portell}, {Richards}, {Riello}, {Seabroke}, {Tanga}, {Th{\'e}venin}, {Torra}, {Els}, {Gracia-Abril}, {Comoretto}, {Garcia-Reinaldos}, {Lock}, {Mercier}, {Altmann}, {Andrae}, {Astraatmadja}, {Bellas-Velidis}, {Benson}, {Berthier}, {Blomme}, {Busso}, {Carry}, {Cellino}, {Clementini},
  {Cowell}, {Creevey}, {Cuypers}, {Davidson}, {De Ridder}, {de Torres}, {Delchambre}, {Dell'Oro}, {Ducourant}, {Fr{\'e}mat}, {Garc{\'\i}a-Torres}, {Gosset}, {Halbwachs}, {Hambly}, {Harrison}, {Hauser}, {Hestroffer}, {Hodgkin}, {Huckle}, {Hutton}, {Jasniewicz}, {Jordan}, {Kontizas}, {Korn}, {Lanzafame}, {Manteiga}, {Moitinho}, {Muinonen}, {Osinde}, {Pancino}, {Pauwels}, {Petit}, {Recio-Blanco}, {Robin}, {Sarro}, {Siopis}, {Smith}, {Smith}, {Sozzetti}, {Thuillot}, {van Reeven}, {Viala}, {Abbas}, {Abreu Aramburu}, {Accart}, {Aguado}, {Allan}, {Allasia}, {Altavilla}, {{\'A}lvarez}, {Alves}, {Anderson}, {Andrei}, {Anglada Varela}, {Antiche}, {Antoja}, {Ant{\'o}n}, {Arcay}, {Atzei}, {Ayache}, {Bach}, {Baker}, {Balaguer-N{\'u}{\~n}ez}, {Barache}, {Barata}, {Barbier}, {Barblan}, {Baroni}, {Barrado y Navascu{\'e}s}, {Barros}, {Barstow}, {Becciani}, {Bellazzini}, {Bellei}, {Bello Garc{\'\i}a}, {Belokurov}, {Bendjoya}, {Berihuete}, {Bianchi}, {Bienaym{\'e}}, {Billebaud}, {Blagorodnova}, {Blanco-Cuaresma}, {Boch},
  {Bombrun}, {Borrachero}, {Bouquillon}, {Bourda}, {Bouy}, {Bragaglia}, {Breddels}, {Brouillet}, {Br{\"u}semeister}, {Bucciarelli}, {Budnik}, {Burgess}, {Burgon}, {Burlacu}, {Busonero}, {Buzzi}, {Caffau}, {Cambras}, {Campbell}, {Cancelliere}, {Cantat-Gaudin}, {Carlucci}, {Carrasco}, {Castellani}, {Charlot}, {Charnas}, {Charvet}, {Chassat}, {Chiavassa}, {Clotet}, {Cocozza}, {Collins}, {Collins}, \& {Costigan}}]{Gaia16}
{Gaia Collaboration}, {Prusti}, T., {de Bruijne}, J.~H.~J., {et~al.} 2016, A\&A, 595, A1, \dodoi{10.1051/0004-6361/201629272}

\bibitem[{{Gezari}(2021)}]{TDEGezari}
{Gezari}, S. 2021, ARA\&A, 59, 21, \dodoi{10.1146/annurev-astro-111720-030029}

\bibitem[{{Groot} {et~al.}(2025){Groot}, {Scaringi}, \& {Elias-Rosa}}]{2025arXiv251217404G}
{Groot}, P.~J., {Scaringi}, S., \& {Elias-Rosa}, N. 2025, arXiv e-prints, arXiv:2512.17404, \dodoi{10.48550/arXiv.2512.17404}

\bibitem[{{Groot} {et~al.}(2024){Groot}, {Bloemen}, {Vreeswijk}, {van Roestel}, {Jonker}, {Nelemans}, {Klein-Wolt}, {Lepoole}, {Pieterse}, {Rodenhuis}, {Boland}, {Haverkorn}, {Aerts}, {Bakker}, {Balster}, {Bekema}, {Dijkstra}, {Dolron}, {Elswijk}, {van Elteren}, {Engels}, {Fokker}, {de Haan}, {Hahn}, {ter Horst}, {Lesman}, {Kragt}, {Morren}, {Nillissen}, {Pessemier}, {Raskin}, {de Rijke}, {Scheers}, {Schuil}, {Timmer}, {Antunes Amaral}, {Arancibia-Rojas}, {Arcavi}, {Blagorodnova}, {Biswas}, {Breton}, {Dawson}, {Dayal}, {De Wet}, {Duffy}, {Faris}, {Fausnaugh}, {Gal-Yam}, {Geier}, {Horesh}, {Johnston}, {Katusiime}, {Kelley}, {Kosakowski}, {Kupfer}, {Leloudas}, {Levan}, {Modiano}, {Mogawana}, {Munday}, {Paice}, {Patat}, {Pelisoli}, {Ramsay}, {Ranaivomanana}, {Ruiz-Carmona}, {Schaffenroth}, {Scaringi}, {Stoppa}, {Street}, {Tranin}, {Uzundag}, {Valenti}, {Veresvarska}, {Vuc̆kovi{\'c}}, {Wichern}, {Wijers}, {Wijnands}, \& {Zimmerman}}]{BG24}
{Groot}, P.~J., {Bloemen}, S., {Vreeswijk}, P.~M., {et~al.} 2024, PASP, 136, 115003, \dodoi{10.1088/1538-3873/ad8b6a}

\bibitem[{{Hernandez} {et~al.}(2024){Hernandez}, {Verteletskyi}, {Nasser}, \& {Aguayo-Ortiz}}]{2024MNRAS.528.4720H}
{Hernandez}, X., {Verteletskyi}, V., {Nasser}, L., \& {Aguayo-Ortiz}, A. 2024, MNRAS, 528, 4720, \dodoi{10.1093/mnras/stad3446}

\bibitem[{{Holman} {et~al.}(2026){Holman}, {Juri{\'c}}, {Napier}, {Kurlander}, {Heinze}, {Moeyens}, {Schwamb}, {Bernardinelli}, {Sullivan}, {Bellm}, {Protopapa}, {Chandler}, {Chow}, {Christensen}, {Cornwall}, {Eggl}, {Greenstreet}, {Li}, {Nourbakhsh}, {Robinson}, {Singh}, \& {Zilkova}}]{2026arXiv261038693H}
{Holman}, M.~J., {Juri{\'c}}, M., {Napier}, K.~J., {et~al.} 2026, arXiv e-prints, arXiv:2610.38693, \dodoi{10.48550/arXiv.2610.38693}

\bibitem[{{Ivezi{\'c}} {et~al.}(2019){Ivezi{\'c}}, {Kahn}, {Tyson}, {Abel}, {Acosta}, {Allsman}, {Alonso}, {AlSayyad}, {Anderson}, {Andrew}, {Angel}, {Angeli}, {Ansari}, {Antilogus}, {Araujo}, {Armstrong}, {Arndt}, {Astier}, {Aubourg}, {Auza}, {Axelrod}, {Bard}, {Barr}, {Barrau}, {Bartlett}, {Bauer}, {Bauman}, {Baumont}, {Bechtol}, {Bechtol}, {Becker}, {Becla}, {Beldica}, {Bellavia}, {Bianco}, {Biswas}, {Blanc}, {Blazek}, {Blandford}, {Bloom}, {Bogart}, {Bond}, {Booth}, {Borgland}, {Borne}, {Bosch}, {Boutigny}, {Brackett}, {Bradshaw}, {Brandt}, {Brown}, {Bullock}, {Burchat}, {Burke}, {Cagnoli}, {Calabrese}, {Callahan}, {Callen}, {Carlin}, {Carlson}, {Chandrasekharan}, {Charles-Emerson}, {Chesley}, {Cheu}, {Chiang}, {Chiang}, {Chirino}, {Chow}, {Ciardi}, {Claver}, {Cohen-Tanugi}, {Cockrum}, {Coles}, {Connolly}, {Cook}, {Cooray}, {Covey}, {Cribbs}, {Cui}, {Cutri}, {Daly}, {Daniel}, {Daruich}, {Daubard}, {Daues}, {Dawson}, {Delgado}, {Dellapenna}, {de Peyster}, {de Val-Borro}, {Digel}, {Doherty}, {Dubois},
  {Dubois-Felsmann}, {Durech}, {Economou}, {Eifler}, {Eracleous}, {Emmons}, {Fausti Neto}, {Ferguson}, {Figueroa}, {Fisher-Levine}, {Focke}, {Foss}, {Frank}, {Freemon}, {Gangler}, {Gawiser}, {Geary}, {Gee}, {Geha}, {Gessner}, {Gibson}, {Gilmore}, {Glanzman}, {Glick}, {Goldina}, {Goldstein}, {Goodenow}, {Graham}, {Gressler}, {Gris}, {Guy}, {Guyonnet}, {Haller}, {Harris}, {Hascall}, {Haupt}, {Hernandez}, {Herrmann}, {Hileman}, {Hoblitt}, {Hodgson}, {Hogan}, {Howard}, {Huang}, {Huffer}, {Ingraham}, {Innes}, {Jacoby}, {Jain}, {Jammes}, {Jee}, {Jenness}, {Jernigan}, {Jevremovi{\'c}}, {Johns}, {Johnson}, {Johnson}, {Jones}, {Juramy-Gilles}, {Juri{\'c}}, {Kalirai}, {Kallivayalil}, {Kalmbach}, {Kantor}, {Karst}, {Kasliwal}, {Kelly}, {Kessler}, {Kinnison}, {Kirkby}, {Knox}, {Kotov}, {Krabbendam}, {Krughoff}, {Kub{\'a}nek}, {Kuczewski}, {Kulkarni}, {Ku}, {Kurita}, {Lage}, {Lambert}, {Lange}, {Langton}, {Le Guillou}, {Levine}, {Liang}, {Lim}, {Lintott}, {Long}, {Lopez}, {Lotz}, {Lupton}, {Lust}, {MacArthur}, {Mahabal},
  {Mandelbaum}, {Markiewicz}, {Marsh}, {Marshall}, {Marshall}, {May}, {McKercher}, {McQueen}, {Meyers}, {Migliore}, {Miller}, \& {Mills}}]{LSST19}
{Ivezi{\'c}}, {\v{Z}}., {Kahn}, S.~M., {Tyson}, J.~A., {et~al.} 2019, ApJ, 873, 111, \dodoi{10.3847/1538-4357/ab042c}

\bibitem[{{Jonas} \& {MeerKAT Team}(2016)}]{2016mks..confE...1J}
{Jonas}, J., \& {MeerKAT Team}. 2016, in MeerKAT Science: On the Pathway to the SKA, 1, \dodoi{10.22323/1.277.0001}

\bibitem[{{Kagra Collaboration} {et~al.}(2019){Kagra Collaboration}, {Akutsu}, {Ando}, {Arai}, {Arai}, {Araki}, {Araya}, {Aritomi}, {Asada}, {Aso}, {Atsuta}, {Awai}, {Bae}, {Baiotti}, {Barton}, {Cannon}, {Capocasa}, {Chen}, {Chiu}, {Cho}, {Chu}, {Craig}, {Creus}, {Doi}, {Eda}, {Enomoto}, {Flaminio}, {Fujii}, {Fujimoto}, {Fukunaga}, {Fukushima}, {Furuhata}, {Haino}, {Hasegawa}, {Hashino}, {Hayama}, {Hirobayashi}, {Hirose}, {Hsieh}, {Huang}, {Ikenoue}, {Inoue}, {Ioka}, {Itoh}, {Izumi}, {Kaji}, {Kajita}, {Kakizaki}, {Kamiizumi}, {Kanbara}, {Kanda}, {Kanemura}, {Kaneyama}, {Kang}, {Kasuya}, {Kataoka}, {Kawai}, {Kawamura}, {Kawasaki}, {Kim}, {Kim}, {Kim}, {Kim}, {Kim}, {Kimura}, {Kinugawa}, {Kirii}, {Kitaoka}, {Kitazawa}, {Kojima}, {Kokeyama}, {Komori}, {Kong}, {Kotake}, {Kozu}, {Kumar}, {Kuo}, {Kuroyanagi}, {Lee}, {Lee}, {Lee}, {Leonardi}, {Lin}, {Lin}, {Liu}, {Liu}, {Majorana}, {Mano}, {Marchio}, {Matsui}, {Matsushima}, {Michimura}, {Mio}, {Miyakawa}, {Miyamoto}, {Miyamoto}, {Miyo}, {Miyoki}, {Morii},
  {Morisaki}, {Moriwaki}, {Morozumi}, {Musha}, {Nagano}, {Nagano}, {Nakamura}, {Nakamura}, {Nakano}, {Nakano}, {Nakao}, {Narikawa}, {Naticchioni}, {Nguyen Quynh}, {Ni}, {Nishizawa}, {Obuchi}, {Ochi}, {Oh}, {Oh}, {Ohashi}, {Ohishi}, {Ohkawa}, {Okutomi}, {Ono}, {Oohara}, {Ooi}, {Pan}, {Park}, {Pe{\~n}a Arellano}, {Pinto}, {Sago}, {Saijo}, {Saitou}, {Saito}, {Sakai}, {Sakai}, {Sakai}, {Sasai}, {Sasaki}, {Sasaki}, {Sato}, {Sato}, {Sato}, {Sekiguchi}, {Seto}, {Shibata}, {Shimoda}, {Shinkai}, {Shishido}, {Shoda}, {Somiya}, {Son}, {Suemasa}, {Suzuki}, {Suzuki}, {Tagoshi}, {Tahara}, {Takahashi}, {Takahashi}, {Takamori}, {Takeda}, {Tanaka}, {Tanaka}, {Tanaka}, {Tanioka}, {Tapia San Martin}, {Tatsumi}, {Tomaru}, {Tomura}, {Travasso}, {Tsubono}, {Tsuchida}, {Uchikata}, {Uchiyama}, {Uehara}, {Ueki}, {Ueno}, {Uraguchi}, {Ushiba}, {van Putten}, {Vocca}, {Wada}, {Wakamatsu}, {Watanabe}, {Xu}, {Yamada}, {Yamamoto}, {Yamamoto}, {Yamamoto}, {Yamamoto}, {Yamamoto}, {Yokogawa}, {Yokoyama}, {Yokozawa}, {Yoon}, {Yoshioka},
  {Yuzurihara}, {Zeidler}, \& {Zhu}}]{KAGRA19}
{Kagra Collaboration}, {Akutsu}, T., {Ando}, M., {et~al.} 2019, Nature Astronomy, 3, 35, \dodoi{10.1038/s41550-018-0658-y}

\bibitem[{{Kaiser} {et~al.}(2026){Kaiser}, {Guerin}, {Vakili}, {Berger}, {Nomerotski}, {Kulkov}, {Svihra}, {Santos}, {Carlile}, {Dravins}, {Funk}, {Saha}, {Walter}, {Borges Fernandes}, {Kim}, {Dunsky}, {Van Tilburg}, {Baryakhtar}, {Galanis}, {Wagoner}, {Dalal}, {Huang}, {Gammie}, \& {Murray}}]{2026arXiv260212717K}
{Kaiser}, R., {Guerin}, W., {Vakili}, F., {et~al.} 2026, arXiv e-prints, arXiv:2602.12717, \dodoi{10.48550/arXiv.2602.12717}

\bibitem[{{Koehn} {et~al.}(2026){Koehn}, {Wouters}, {Sadeh}, {Pang}, {Bulla}, {Van Den Broeck}, {Coughlin}, \& {Dietrich}}]{2026MNRAS.tmp.1798K}
{Koehn}, H., {Wouters}, T., {Sadeh}, G., {et~al.} 2026, MNRAS, \dodoi{10.1093/mnras/stag1948}

\bibitem[{{Krisciunas} {et~al.}(2007){Krisciunas}, {Semler}, {Richards}, {Schwarz}, {Suntzeff}, {Vera}, \& {Sanhueza}}]{2007PASP..119..687K}
{Krisciunas}, K., {Semler}, D.~R., {Richards}, J., {et~al.} 2007, PASP, 119, 687, \dodoi{10.1086/519564}

\bibitem[{{Kulkarni}(2020)}]{SRK20}
{Kulkarni}, S.~R. 2020, arXiv e-prints, arXiv:2004.03511, \dodoi{10.48550/arXiv.2004.03511}

\bibitem[{{Kulkarni} {et~al.}(2021){Kulkarni}, {Harrison}, {Grefenstette}, {Earnshaw}, {Andreoni}, {Berg}, {Bloom}, {Cenko}, {Chornock}, {Christiansen}, {Coughlin}, {Wuollet Criswell}, {Darvish}, {Das}, {De}, {Dessart}, {Dixon}, {Dorsman}, {El-Badry}, {Evans}, {Ford}, {Fremling}, {Gansicke}, {Gezari}, {Goetberg}, {Green}, {Graham}, {Heida}, {Ho}, {Jaodand}, {Johns-Krull}, {Kasliwal}, {Lazzarini}, {Lu}, {Margutti}, {Martin}, {Masters}, {McKernan}, {Naze}, {Nissanke}, {Parazin}, {Perley}, {Phinney}, {Piro}, {Raaijmakers}, {Rauw}, {Rodriguez}, {Sana}, {Senchyna}, {Singer}, {Spake}, {Stassun}, {Stern}, {Teplitz}, {Weisz}, \& {Yao}}]{UVEX21}
{Kulkarni}, S.~R., {Harrison}, F.~A., {Grefenstette}, B.~W., {et~al.} 2021, arXiv e-prints, arXiv:2111.15608, \dodoi{10.48550/arXiv.2111.15608}

\bibitem[{{Kupfer} {et~al.}(2025{\natexlab{a}}){Kupfer}, {Scaringi}, {Groot}, {G{\"a}nsicke}, {Pelisoli}, {Pala}, {van Roestel}, {Toonen}, {de Martino}, {Castro Segura}, {Buckley}, {Van Grootel}, {O'Brien}, \& {Nissanke}}]{2025arXiv251216988K}
{Kupfer}, T., {Scaringi}, S., {Groot}, P., {et~al.} 2025{\natexlab{a}}, arXiv e-prints, arXiv:2512.16988, \dodoi{10.48550/arXiv.2512.16988}

\bibitem[{{Kupfer} {et~al.}(2025{\natexlab{b}}){Kupfer}, {Scaringi}, {Pelisoli}, {Pala}, {Toonen}, {de Martino}, {Gall}, {Deshmukh}, {Van Grootel}, {Blondin}, \& {Nissanke}}]{2025arXiv251216991K}
{Kupfer}, T., {Scaringi}, S., {Pelisoli}, I., {et~al.} 2025{\natexlab{b}}, arXiv e-prints, arXiv:2512.16991, \dodoi{10.48550/arXiv.2512.16991}

\bibitem[{{Law} {et~al.}(2009){Law}, {Kulkarni}, {Dekany}, {Ofek}, {Quimby}, {Nugent}, {Surace}, {Grillmair}, {Bloom}, {Kasliwal}, {Bildsten}, {Brown}, {Cenko}, {Ciardi}, {Croner}, {Djorgovski}, {van Eyken}, {Filippenko}, {Fox}, {Gal-Yam}, {Hale}, {Hamam}, {Helou}, {Henning}, {Howell}, {Jacobsen}, {Laher}, {Mattingly}, {McKenna}, {Pickles}, {Poznanski}, {Rahmer}, {Rau}, {Rosing}, {Shara}, {Smith}, {Starr}, {Sullivan}, {Velur}, {Walters}, \& {Zolkower}}]{PTF09}
{Law}, N.~M., {Kulkarni}, S.~R., {Dekany}, R.~G., {et~al.} 2009, PASP, 121, 1395, \dodoi{10.1086/648598}

\bibitem[{{Law} {et~al.}(2022){Law}, {Corbett}, {Galliher}, {Gonzalez}, {Vasquez}, {Walters}, {Machia}, {Ratzloff}, {Ackley}, {Bizon}, {Clemens}, {Cox}, {Eikenberry}, {Howard}, {Glazier}, {Mann}, {Quimby}, {Reichart}, \& {Trilling}}]{2022PASP..134c5003L}
{Law}, N.~M., {Corbett}, H., {Galliher}, N.~W., {et~al.} 2022, PASP, 134, 035003, \dodoi{10.1088/1538-3873/ac4811}

\bibitem[{{Levan}(2018)}]{GRBLevan}
{Levan}, A.~J. 2018, {Gamma-Ray Bursts}, \dodoi{10.1088/2514-3433/aae164}

\bibitem[{{LIGO Scientific Collaboration} {et~al.}(2015){LIGO Scientific Collaboration}, {Aasi}, {Abbott}, {Abbott}, {Abbott}, {Abernathy}, {Ackley}, {Adams}, {Adams}, {Addesso}, {Adhikari}, {Adya}, {Affeldt}, {Aggarwal}, {Aguiar}, {Ain}, {Ajith}, {Alemic}, {Allen}, {Amariutei}, {Anderson}, {Anderson}, {Arai}, {Araya}, {Arceneaux}, {Areeda}, {Ashton}, {Ast}, {Aston}, {Aufmuth}, {Aulbert}, {Aylott}, {Babak}, {Baker}, {Ballmer}, {Barayoga}, {Barbet}, {Barclay}, {Barish}, {Barker}, {Barr}, {Barsotti}, {Bartlett}, {Barton}, {Bartos}, {Bassiri}, {Batch}, {Baune}, {Behnke}, {Bell}, {Bell}, {Benacquista}, {Bergman}, {Bergmann}, {Berry}, {Betzwieser}, {Bhagwat}, {Bhandare}, {Bilenko}, {Billingsley}, {Birch}, {Biscans}, {Biwer}, {Blackburn}, {Blackburn}, {Blair}, {Blair}, {Bock}, {Bodiya}, {Bojtos}, {Bond}, {Bork}, {Born}, {Bose}, {Brady}, {Braginsky}, {Brau}, {Bridges}, {Brinkmann}, {Brooks}, {Brown}, {Brown}, {Brown}, {Buchman}, {Buikema}, {Buonanno}, {Cadonati}, {Calder{\'o}n Bustillo}, {Camp}, {Cannon}, {Cao},
  {Capano}, {Caride}, {Caudill}, {Cavagli{\`a}}, {Cepeda}, {Chakraborty}, {Chalermsongsak}, {Chamberlin}, {Chao}, {Charlton}, {Chen}, {Cho}, {Cho}, {Chow}, {Christensen}, {Chu}, {Chung}, {Ciani}, {Clara}, {Clark}, {Collette}, {Cominsky}, {Constancio}, {Cook}, {Corbitt}, {Cornish}, {Corsi}, {Costa}, {Coughlin}, {Countryman}, {Couvares}, {Coward}, {Cowart}, {Coyne}, {Coyne}, {Craig}, {Creighton}, {Creighton}, {Cripe}, {Crowder}, {Cumming}, {Cunningham}, {Cutler}, {Dahl}, {Dal Canton}, {Damjanic}, {Danilishin}, {Danzmann}, {Dartez}, {Dave}, {Daveloza}, {Davies}, {Daw}, {DeBra}, {Del Pozzo}, {Denker}, {Dent}, {Dergachev}, {DeRosa}, {DeSalvo}, {Dhurandhar}, {D́{\i}az}, {Di Palma}, {Dojcinoski}, {Dominguez}, {Donovan}, {Dooley}, {Doravari}, {Douglas}, {Downes}, {Driggers}, {Du}, {Dwyer}, {Eberle}, {Edo}, {Edwards}, {Edwards}, {Effler}, {Eggenstein}, {Ehrens}, {Eichholz}, {Eikenberry}, {Essick}, {Etzel}, {Evans}, {Evans}, {Factourovich}, {Fairhurst}, {Fan}, {Fang}, {Farr}, {Farr}, {Favata}, {Fays}, {Fehrmann},
  {Fejer}, {Feldbaum}, {Ferreira}, {Fisher}, {Frei}, {Freise}, {Frey}, {Fricke}, {Fritschel}, {Frolov}, {Fuentes-Tapia}, {Fulda}, {Fyffe}, \& {Gair}}]{LIGO15}
{LIGO Scientific Collaboration}, {Aasi}, J., {Abbott}, B.~P., {et~al.} 2015, Classical and Quantum Gravity, 32, 074001, \dodoi{10.1088/0264-9381/32/7/074001}

\bibitem[{{Liu} {et~al.}(2026){Liu}, {Sun}, {Ge}, {Huang}, {Hu}, {Jia}, {Li}, {Li}, {Sun}, \& {Zou}}]{2026SPIE14147E..0NL}
{Liu}, J., {Sun}, M., {Ge}, L., {et~al.} 2026, in Society of Photo-Optical Instrumentation Engineers (SPIE) Conference Series, Vol. 14147, Ground-based and Airborne Telescopes XI, ed. H.~K. {Marshall}, T.~{Usuda}, B.~{Onodera}, \& J.~{Spyromilio} (SPIE), 141470N, \dodoi{10.1117/12.3106183}

\bibitem[{{Luo} {et~al.}(2016){Luo}, {Chen}, {Duan}, {Gong}, {Hu}, {Ji}, {Liu}, {Mei}, {Milyukov}, {Sazhin}, {Shao}, {Toth}, {Tu}, {Wang}, {Wang}, {Yeh}, {Zhan}, {Zhang}, {Zharov}, \& {Zhou}}]{TianQin16}
{Luo}, J., {Chen}, L.-S., {Duan}, H.-Z., {et~al.} 2016, Classical and Quantum Gravity, 33, 035010, \dodoi{10.1088/0264-9381/33/3/035010}

\bibitem[{{Maggiore} {et~al.}(2020){Maggiore}, {Van Den Broeck}, {Bartolo}, {Belgacem}, {Bertacca}, {Bizouard}, {Branchesi}, {Clesse}, {Foffa}, {Garc{\'\i}a-Bellido}, {Grimm}, {Harms}, {Hinderer}, {Matarrese}, {Palomba}, {Peloso}, {Ricciardone}, \& {Sakellariadou}}]{et20}
{Maggiore}, M., {Van Den Broeck}, C., {Bartolo}, N., {et~al.} 2020, JCAP, 2020, 050, \dodoi{10.1088/1475-7516/2020/03/050}

\bibitem[{{Maxted} {et~al.}(2026){Maxted}, {Bowman}, {Wilson}, {Rosu}, {Stassun}, {Murphy}, {Moharana}, {Triaud}, {Hahlin}, \& {Southworth}}]{2026arXiv260117179M}
{Maxted}, P. F.~L., {Bowman}, D.~M., {Wilson}, T.~G., {et~al.} 2026, arXiv e-prints, arXiv:2601.17179, \dodoi{10.48550/arXiv.2601.17179}

\bibitem[{{Mayor} \& {Queloz}(1995)}]{51Peg95}
{Mayor}, M., \& {Queloz}, D. 1995, Nature, 378, 355, \dodoi{10.1038/378355a0}

\bibitem[{{Mayor} {et~al.}(2003){Mayor}, {Pepe}, {Queloz}, {Bouchy}, {Rupprecht}, {Lo Curto}, {Avila}, {Benz}, {Bertaux}, {Bonfils}, {Dall}, {Dekker}, {Delabre}, {Eckert}, {Fleury}, {Gilliotte}, {Gojak}, {Guzman}, {Kohler}, {Lizon}, {Longinotti}, {Lovis}, {Megevand}, {Pasquini}, {Reyes}, {Sivan}, {Sosnowska}, {Soto}, {Udry}, {van Kesteren}, {Weber}, \& {Weilenmann}}]{HARPS03}
{Mayor}, M., {Pepe}, F., {Queloz}, D., {et~al.} 2003, The Messenger, 114, 20

\bibitem[{{Miraval Zanon} {et~al.}(2025){Miraval Zanon}, {Illiano}, {Ambrosino}, {de Martino}, {Baglio}, {Ballocco}, {Buckle}, {Coti Zelati}, {Del Santo}, {Groot}, {La Placa}, {Malacaria}, {Marino}, {Papitto}, {Rea}, {Sanna}, {Scaringi}, {Turner}, {Veledina}, \& {Zampieri}}]{2025arXiv251213960M}
{Miraval Zanon}, A., {Illiano}, G., {Ambrosino}, F., {et~al.} 2025, arXiv e-prints, arXiv:2512.13960, \dodoi{10.48550/arXiv.2512.13960}

\bibitem[{{Mogawana} {et~al.}(2026){Mogawana}, {Groot}, {Blagorodnova}, {Vreeswijk}, {Pieterse}, {Bloemen}, {Leloudas}, {Wijnands}, {Buckley}, {Woudt}, {Fender}, \& {Stappers}}]{Mogawana26}
{Mogawana}, O., {Groot}, P.~J., {Blagorodnova}, N., {et~al.} 2026, PASP, 138, 094207, \dodoi{10.1088/1538-3873/aea18d}

\bibitem[{{Nissanke} {et~al.}(2025){Nissanke}, {Sarin}, {Copperwheat}, {Antier}, {Berge}, {Bosch}, {Ghosh}, {Groot}, {Hallinan}, {Hinderer}, {Hotokezaka}, {Karydas}, {Kasliwal}, {Kini}, {Kotak}, {Kotera}, {Kowalski}, {Krauth}, {Krishna}, {Kupfer}, {Lampropoulos}, {Levan}, {Liodis}, {Marcotulli}, {Mooley}, {Piranomonte}, {Rea}, {Roth}, {Scaringi}, {Schulze}, {Suleiman}, {Tanvir}, {Zegarelli}, \& {Zhu}}]{2025arXiv251214546N}
{Nissanke}, S., {Sarin}, N., {Copperwheat}, C., {et~al.} 2025, arXiv e-prints, arXiv:2512.14546, \dodoi{10.48550/arXiv.2512.14546}

\bibitem[{{Onken} {et~al.}(2024){Onken}, {Wolf}, {Bessell}, {Chang}, {Luvaul}, {Tonry}, {White}, \& {Da Costa}}]{SMDR4}
{Onken}, C.~A., {Wolf}, C., {Bessell}, M.~S., {et~al.} 2024, arXiv e-prints, arXiv:2402.02015, \dodoi{10.48550/arXiv.2402.02015}

\bibitem[{{Pala} {et~al.}(2025){Pala}, {Raddi}, {Rebassa-Mansergas}, {G{\"a}nsicke}, {Anderson}, {Belloni}, {Binnenfeld}, {Breedt}, {Buckley}, {Cunningham}, {Ederoclite}, {Escorza}, {Korol}, {Kupfer}, {de Martino}, {Merc}, {Meza}, {Parsons}, {Pelisoli}, {Reindl}, {Rodr{\'\i}guez-Gil}, {Santos-Garc{\'\i}a}, {Scaringi}, {Szkody}, {Toloza}, {Torres}, {Uzundag}, \& {Zorotovic}}]{2025arXiv251214800P}
{Pala}, A.~F., {Raddi}, R., {Rebassa-Mansergas}, A., {et~al.} 2025, arXiv e-prints, arXiv:2512.14800, \dodoi{10.48550/arXiv.2512.14800}

\bibitem[{{Raskin} {et~al.}(2020){Raskin}, {Schwab}, {Vandenbussche}, {De Ridder}, {Lanthermann}, {P{\'e}rez Padilla}, {Tkachenko}, {Sana}, {Royer}, {Prins}, {Decin}, {Defr{\`e}re}, {Pember}, {Atkinson}, {Glasse}, {Pollacco}, {Tinetti}, {Guedel}, {Stuermer}, {Ribas}, {Brandeker}, {Buchhave}, {Halverson}, {Avila}, {Morren}, \& {Van Winckel}}]{MARVEL20}
{Raskin}, G., {Schwab}, C., {Vandenbussche}, B., {et~al.} 2020, in Society of Photo-Optical Instrumentation Engineers (SPIE) Conference Series, Vol. 11447, Ground-based and Airborne Instrumentation for Astronomy VIII, ed. C.~J. {Evans}, J.~J. {Bryant}, \& K.~{Motohara}, 114473K, \dodoi{10.1117/12.2560895}

\bibitem[{{Rauer} {et~al.}(2025){Rauer}, {Aerts}, {Cabrera}, {Deleuil}, {Erikson}, {Gizon}, {Goupil}, {Heras}, {Walloschek}, {Lorenzo-Alvarez}, {Marliani}, {Martin-Garcia}, {Mas-Hesse}, {O'Rourke}, {Osborn}, {Pagano}, {Piotto}, {Pollacco}, {Ragazzoni}, {Ramsay}, {Udry}, {Appourchaux}, {Benz}, {Brandeker}, {G{\"u}del}, {Janot-Pacheco}, {Kabath}, {Kjeldsen}, {Min}, {Santos}, {Smith}, {Suarez}, {Werner}, {Aboudan}, {Abreu}, {Acu{\~n}a}, {Adams}, {Adibekyan}, {Affer}, {Agneray}, {Agnor}, {Aguirre B{\o}rsen-Koch}, {Ahmed}, {Aigrain}, {Al-Bahlawan}, {Alcacera Gil}, {Alei}, {Alencar}, {Alexander}, {Alfonso-Garz{\'o}n}, {Alibert}, {Allende Prieto}, {Almeida}, {Alonso Sobrino}, {Altavilla}, {Althaus}, {Alvarez Trujillo}, {Amarsi}, {Ammler-von Eiff}, {Am{\^o}res}, {Andrade}, {Antoniadis-Karnavas}, {Ant{\'o}nio}, {Aparicio del Moral}, {Appolloni}, {Arena}, {Armstrong}, {Aroca Aliaga}, {Asplund}, {Audenaert}, {Auricchio}, {Avelino}, {Baeke}, {Bailli{\'e}}, {Balado}, {Ballber Balaguer{\'o}}, {Balestra}, {Ball}, {Ballans},
  {Ballot}, {Barban}, {Barbary}, {Barbieri}, {Barcel{\'o} Forteza}, {Barker}, {Barklem}, {Barnes}, {Barrado Navascues}, {Barragan}, {Baruteau}, {Basu}, {Baudin}, {Baumeister}, {Bayliss}, {Bazot}, {Beck}, {Belkacem}, {Bellinger}, {Benatti}, {Benomar}, {B{\'e}rard}, {Bergemann}, {Bergomi}, {Bernardo}, {Biazzo}, {Bignamini}, {Bigot}, {Billot}, {Binet}, {Biondi}, {Biondi}, {Birch}, {Bitsch}, {Bluhm Ceballos}, {B{\'o}di}, {Bogn{\'a}r}, {Boisse}, {Bolmont}, {Bonanno}, {Bonavita}, {Bonfanti}, {Bonfils}, {Bonito}, {Bonomo}, {B{\"o}rner}, {Boro Saikia}, {Borreguero Mart{\'\i}n}, {Borsa}, {Borsato}, {Bossini}, {Bouchy}, {Bou{\'e}}, {Boufleur}, {Boumier}, {Bourrier}, {Bowman}, {Bozzo}, {Bradley}, {Bray}, {Bressan}, {Breton}, {Brienza}, {Brito}, {Brogi}, {Brown}, {Brown}, {Brun}, {Bruno}, {Bruns}, {Buchhave}, {Bugnet}, {Buldgen}, {Burgess}, {Busatta}, {Busso}, {Buzasi}, {Caballero}, {Cabral}, {Cabrero Gomez}, {Calderone}, {Cameron}, {Cameron}, {Campante}, {Campos Gestal}, {Canto Martins}, {Cara}, {Carone}, {Carrasco},
  {Casagrande}, {Casewell}, {Cassisi}, {Castellani}, {Castro}, {Catala}, {Catal{\'a}n Fern{\'a}ndez}, {Catelan}, {Cegla}, {Cerruti}, {Cessa}, {Chadid}, {Chaplin}, {Charpinet}, {Chiappini}, {Chiarucci}, {Chiavassa}, {Chinellato}, {Chirulli}, {Christensen-Dalsgaard}, {Church}, {Claret}, {Clarke}, {Claudi}, {Clermont}, {Coelho}, {Coelho}, {Cogato}, {Colom{\'e}}, {Condamin}, {Conde Garc{\'\i}a}, \& {Conseil}}]{PLATO25}
{Rauer}, H., {Aerts}, C., {Cabrera}, J., {et~al.} 2025, Experimental Astronomy, 59, 26, \dodoi{10.1007/s10686-025-09985-9}

\bibitem[{{Ravi} {et~al.}(2023){Ravi}, {Catha}, {Chen}, {Connor}, {Faber}, {Lamb}, {Hallinan}, {Harnach}, {Hellbourg}, {Hobbs}, \& et~al.}]{2023ApJ...949L...3R}
{Ravi}, V., {Catha}, M., {Chen}, G., {et~al.} 2023, ApJL, 949, L3, \dodoi{10.3847/2041-8213/acc4b6}

\bibitem[{{Ricker} {et~al.}(2015){Ricker}, {Winn}, {Vanderspek}, {Latham}, {Bakos}, {Bean}, {Berta-Thompson}, {Brown}, {Buchhave}, {Butler}, {Butler}, {Chaplin}, {Charbonneau}, {Christensen-Dalsgaard}, {Clampin}, {Deming}, {Doty}, {De Lee}, {Dressing}, {Dunham}, {Endl}, {Fressin}, {Ge}, {Henning}, {Holman}, {Howard}, {Ida}, {Jenkins}, {Jernigan}, {Johnson}, {Kaltenegger}, {Kawai}, {Kjeldsen}, {Laughlin}, {Levine}, {Lin}, {Lissauer}, {MacQueen}, {Marcy}, {McCullough}, {Morton}, {Narita}, {Paegert}, {Palle}, {Pepe}, {Pepper}, {Quirrenbach}, {Rinehart}, {Sasselov}, {Sato}, {Seager}, {Sozzetti}, {Stassun}, {Sullivan}, {Szentgyorgyi}, {Torres}, {Udry}, \& {Villasenor}}]{TESS15}
{Ricker}, G.~R., {Winn}, J.~N., {Vanderspek}, R., {et~al.} 2015, Journal of Astronomical Telescopes, Instruments, and Systems, 1, 014003, \dodoi{10.1117/1.JATIS.1.1.014003}

\bibitem[{{Roelofs} {et~al.}(2010){Roelofs}, {Rau}, {Marsh}, {Steeghs}, {Groot}, \& {Nelemans}}]{roelofs2010}
{Roelofs}, G. H.~A., {Rau}, A., {Marsh}, T.~R., {et~al.} 2010, ApJL, 711, L138, \dodoi{10.1088/2041-8205/711/2/L138}

\bibitem[{{Roth} {et~al.}(2026){Roth}, {Roje}, {Vje{\v{s}}nica}, {Cikota}, {Brown}, {Kretlow}, {Azzaro}, {Reinhart}, {Aceituno}, {Dawson}, {Denker}, {En{\ss}lin}, {Geier}, {Granzer}, {Kupfer}, {Li Causi}, {Nissanke}, {Ortiz}, {Pedichini}, {Poppenh{\"a}ger}, \& {Schwope}}]{ORCA26}
{Roth}, M.~M., {Roje}, P., {Vje{\v{s}}nica}, S., {et~al.} 2026, Astronomische Nachrichten, 347, e70110, \dodoi{10.1002/asna.70110}

\bibitem[{{Ruan} {et~al.}(2020){Ruan}, {Guo}, {Cai}, \& {Zhang}}]{Taiji20}
{Ruan}, W.-H., {Guo}, Z.-K., {Cai}, R.-G., \& {Zhang}, Y.-Z. 2020, International Journal of Modern Physics A, 35, 2050075, \dodoi{10.1142/S0217751X2050075X}

\bibitem[{{Sanders}(1997)}]{1997ApJ...480..492S}
{Sanders}, R.~H. 1997, ApJ, 480, 492, \dodoi{10.1086/303980}

\bibitem[{{Sarin} {et~al.}(2025){Sarin}, {Levan}, {Tanvir}, {Scaringi}, {Ragosta}, {Melandri}, {Groot}, {O'Brien}, {Lasky}, {Nissanke}, {Heger}, \& {Schulze}}]{2025arXiv251215010S}
{Sarin}, N., {Levan}, A., {Tanvir}, N., {et~al.} 2025, arXiv e-prints, arXiv:2512.15010, \dodoi{10.48550/arXiv.2512.15010}

\bibitem[{{Saunders} {et~al.}(2026){Saunders}, {Cudennec}, {Bacon}, {Dierickx}, {de Jong}, {Lee}, \& {Gausachs}}]{2026arXiv260808297S}
{Saunders}, W., {Cudennec}, C., {Bacon}, R., {et~al.} 2026, arXiv e-prints, arXiv:2608.08297, \dodoi{10.48550/arXiv.2608.08297}

\bibitem[{{Scaringi} {et~al.}(2022){Scaringi}, {Groot}, {Knigge}, {Bird}, {Breedt}, {Buckley}, {Cavecchi}, {Degenaar}, {de Martino}, {Done}, {Fratta}, {I{\l}kiewicz}, {Koerding}, {Lasota}, {Littlefield}, {Manara}, {O'Brien}, {Szkody}, \& {Timmes}}]{Scaringi22}
{Scaringi}, S., {Groot}, P.~J., {Knigge}, C., {et~al.} 2022, Nature, 604, 447, \dodoi{10.1038/s41586-022-04495-6}

\bibitem[{{Scaringi} {et~al.}(2025){Scaringi}, {de Martino}, {Pala}, {Sanna}, {Groot}, {O'Brien}, {Ederoclite}, {Castro Segura}, {Coppejans}, {Ilkiewicz}, {Casella}, {Buckley}, {Kupfer}, {Rea}, {Kubra DAG}, {Tampo}, {Zhang}, {Ford}, {Veresvarska}, \& {Wynn}}]{2025arXiv251213802S}
{Scaringi}, S., {de Martino}, D., {Pala}, A.~F., {et~al.} 2025, arXiv e-prints, arXiv:2512.13802, \dodoi{10.48550/arXiv.2512.13802}

\bibitem[{{Schutz}(1986)}]{Schutz86}
{Schutz}, B.~F. 1986, Nature, 323, 310, \dodoi{10.1038/323310a0}

\bibitem[{{Shvartzvald} {et~al.}(2024){Shvartzvald}, {Waxman}, {Gal-Yam}, {Ofek}, {Ben-Ami}, {Berge}, {Kowalski}, {B{\"u}hler}, {Worm}, {Rhoads}, {Arcavi}, {Maoz}, {Polishook}, {Stone}, {Trakhtenbrot}, {Ackermann}, {Aharonson}, {Birnholtz}, {Chelouche}, {Guetta}, {Hallakoun}, {Horesh}, {Kushnir}, {Mazeh}, {Nordin}, {Ofir}, {Ohm}, {Parsons}, {Pe'er}, {Perets}, {Perdelwitz}, {Poznanski}, {Sadeh}, {Sagiv}, {Shahaf}, {Soumagnac}, {Tal-Or}, {Santen}, {Zackay}, {Guttman}, {Rekhi}, {Townsend}, {Weinstein}, \& {Wold}}]{Ultrasat24}
{Shvartzvald}, Y., {Waxman}, E., {Gal-Yam}, A., {et~al.} 2024, ApJ, 964, 74, \dodoi{10.3847/1538-4357/ad2704}

\bibitem[{{Snellen} {et~al.}(2010){Snellen}, {de Kok}, {de Mooij}, \& {Albrecht}}]{2010Natur.465.1049S}
{Snellen}, I. A.~G., {de Kok}, R.~J., {de Mooij}, E. J.~W., \& {Albrecht}, S. 2010, Nature, 465, 1049, \dodoi{10.1038/nature09111}

\bibitem[{{Steeghs} {et~al.}(2022){Steeghs}, {Galloway}, {Ackley}, {Dyer}, {Lyman}, {Ulaczyk}, {Cutter}, {Mong}, {Dhillon}, {O'Brien}, {Ramsay}, {Poshyachinda}, {Kotak}, {Nuttall}, {Pall{\'e}}, {Breton}, {Pollacco}, {Thrane}, {Aukkaravittayapun}, {Awiphan}, {Burhanudin}, {Chote}, {Chrimes}, {Daw}, {Duffy}, {Eyles-Ferris}, {Gompertz}, {Heikkil{\"a}}, {Irawati}, {Kennedy}, {Killestein}, {Kuncarayakti}, {Levan}, {Littlefair}, {Makrygianni}, {Marsh}, {Mata-Sanchez}, {Mattila}, {Maund}, {McCormac}, {Mkrtichian}, {Mullaney}, {Noysena}, {Patel}, {Rol}, {Sawangwit}, {Stanway}, {Starling}, {Str{\o}m}, {Tooke}, {West}, {White}, \& {Wiersema}}]{GOTO22}
{Steeghs}, D., {Galloway}, D.~K., {Ackley}, K., {et~al.} 2022, MNRAS, 511, 2405, \dodoi{10.1093/mnras/stac013}

\bibitem[{{Stein} {et~al.}(2021){Stein}, {van Velzen}, {Kowalski}, {Franckowiak}, {Gezari}, {Miller-Jones}, {Frederick}, {Sfaradi}, {Bietenholz}, {Horesh}, {Fender}, {Garrappa}, {Ahumada}, {Andreoni}, {Belicki}, {Bellm}, {B{\"o}ttcher}, {Brinnel}, {Burruss}, {Cenko}, {Coughlin}, {Cunningham}, {Drake}, {Farrar}, {Feeney}, {Foley}, {Gal-Yam}, {Golkhou}, {Goobar}, {Graham}, {Hammerstein}, {Helou}, {Hung}, {Kasliwal}, {Kilpatrick}, {Kong}, {Kupfer}, {Laher}, {Mahabal}, {Masci}, {Necker}, {Nordin}, {Perley}, {Rigault}, {Reusch}, {Rodriguez}, {Rojas-Bravo}, {Rusholme}, {Shupe}, {Singer}, {Sollerman}, {Soumagnac}, {Stern}, {Taggart}, {van Santen}, {Ward}, {Woudt}, \& {Yao}}]{stein21}
{Stein}, R., {van Velzen}, S., {Kowalski}, M., {et~al.} 2021, Nature Astronomy, 5, 510, \dodoi{10.1038/s41550-020-01295-8}

\bibitem[{Su {et~al.}(2026)Su, Zhao, Zhao, Cui, Zuo, Bai, Zhang, Zhao, Luo, Jiao, Hu, Xu, Wang, Liang, Li, Chu, Zai, Hu, Zhang, Yuan, \& Gong}]{10.1117/12.3104704}
Su, D., Zhao, G., Zhao, G., {et~al.} 2026, in Ground-based and Airborne Telescopes XI, ed. H.~K. Marshall, T.~Usuda, B.~Onodera, \& J.~Spyromilio, Vol. 14147, International Society for Optics and Photonics (SPIE), 141470K, \dodoi{10.1117/12.3104704}

\bibitem[{{Tingay} {et~al.}(2013){Tingay}, {Goeke}, {Bowman}, {Emrich}, {Ord}, {Mitchell}, {Morales}, {Booler}, {Crosse}, {Wayth}, \& et~al.}]{2013PASA...30....7T}
{Tingay}, S.~J., {Goeke}, R., {Bowman}, J.~D., {et~al.} 2013, PASA, 30, e007, \dodoi{10.1017/pasa.2012.007}

\bibitem[{{Tonry} {et~al.}(2018){Tonry}, {Denneau}, {Heinze}, {Stalder}, {Smith}, {Smartt}, {Stubbs}, {Weiland}, \& {Rest}}]{ATLAS18}
{Tonry}, J.~L., {Denneau}, L., {Heinze}, A.~N., {et~al.} 2018, PASP, 130, 064505, \dodoi{10.1088/1538-3873/aabadf}

\bibitem[{{Uzundag} {et~al.}(2025){Uzundag}, {Pelisoli}, {Charpinet}, {Corsico}, {Althaus}, {Van Grootel}, {Randall}, {Kupfer}, \& {Raddi}}]{2025arXiv251215814U}
{Uzundag}, M., {Pelisoli}, I., {Charpinet}, S., {et~al.} 2025, arXiv e-prints, arXiv:2512.15814, \dodoi{10.48550/arXiv.2512.15814}

\bibitem[{{van Haarlem} {et~al.}(2013){van Haarlem}, {Wise}, {Gunst}, {Heald}, {McKean}, {Hessels}, {de Bruyn}, {Nijboer}, {Swinbank}, {Fallows}, \& et~al.}]{2013A&A...556A...2V}
{van Haarlem}, M.~P., {Wise}, M.~W., {Gunst}, A.~W., {et~al.} 2013, A\&A, 556, A2, \dodoi{10.1051/0004-6361/201220873}

\bibitem[{{Vatedka} {et~al.}(2025){Vatedka}, {Tyagi}, {Vadodariya}, {Agrawal}, {Chatterjee}, {Mahendran Chellammal}, {Jayasurya}, {Kumar}, {Sharan}, {Tadepalli}, {Kallubavi}, {Garg}, {Sharma}, \& {Justin}}]{2025JATIS..11c5001V}
{Vatedka}, R., {Tyagi}, A., {Vadodariya}, K., {et~al.} 2025, Journal of Astronomical Telescopes, Instruments, and Systems, 11, 035001, \dodoi{10.1117/1.JATIS.11.3.035001}

\bibitem[{{Vernet} {et~al.}(2011){Vernet}, {Dekker}, {D'Odorico}, {Kaper}, {Kjaergaard}, {Hammer}, {Randich}, {Zerbi}, {Groot}, {Hjorth}, {Guinouard}, {Navarro}, {Adolfse}, {Albers}, {Amans}, {Andersen}, {Andersen}, {Binetruy}, {Bristow}, {Castillo}, {Chemla}, {Christensen}, {Conconi}, {Conzelmann}, {Dam}, {de Caprio}, {de Ugarte Postigo}, {Delabre}, {di Marcantonio}, {Downing}, {Elswijk}, {Finger}, {Fischer}, {Flores}, {Fran{\c{c}}ois}, {Goldoni}, {Guglielmi}, {Haigron}, {Hanenburg}, {Hendriks}, {Horrobin}, {Horville}, {Jessen}, {Kerber}, {Kern}, {Kiekebusch}, {Kleszcz}, {Klougart}, {Kragt}, {Larsen}, {Lizon}, {Lucuix}, {Mainieri}, {Manuputy}, {Martayan}, {Mason}, {Mazzoleni}, {Michaelsen}, {Modigliani}, {Moehler}, {M{\o}ller}, {Norup S{\o}rensen}, {N{\o}rregaard}, {P{\'e}roux}, {Patat}, {Pena}, {Pragt}, {Reinero}, {Rigal}, {Riva}, {Roelfsema}, {Royer}, {Sacco}, {Santin}, {Schoenmaker}, {Spano}, {Sweers}, {Ter Horst}, {Tintori}, {Tromp}, {van Dael}, {van der Vliet}, {Venema}, {Vidali}, {Vinther}, {Vola},
  {Winters}, {Wistisen}, {Wulterkens}, \& {Zacchei}}]{XSH11}
{Vernet}, J., {Dekker}, H., {D'Odorico}, S., {et~al.} 2011, A\&A, 536, A105, \dodoi{10.1051/0004-6361/201117752}

\bibitem[{{Vincentelli} {et~al.}(2025){Vincentelli}, {Casella}, {Veledina}, {Ambrifi}, {Baglio}, {Buckley}, {Castro Segura}, {Cavecchi}, {de Martino}, {del Santo}, {Gandhi}, {Iliano}, {La Paca}, {Malacaria}, {Marino}, {O'Brien}, {Rea}, {Sanna}, {Scaringi}, {Shahbaz}, \& {Zampieri}}]{2025arXiv251215832V}
{Vincentelli}, F.~M., {Casella}, P., {Veledina}, A., {et~al.} 2025, arXiv e-prints, arXiv:2512.15832, \dodoi{10.48550/arXiv.2512.15832}

\bibitem[{{Wang} {et~al.}(2023){Wang}, {Liu}, {Cai}, {Geng}, {Fang}, {He}, {Jiang}, {Jiang}, {Kong}, {Li}, {Li}, {Luo}, {Pan}, {Wu}, {Yang}, {Yu}, {Zheng}, {Zhu}, {Cai}, {Chen}, {Chen}, {Dai}, {Fan}, {Fan}, {Fang}, {He}, {Hu}, {Hu}, {Jin}, {Jiang}, {Li}, {Li}, {Li}, {Liang}, {Lin}, {Liu}, {Liu}, {Liu}, {Liu}, {Liu}, {Lou}, {Qu}, {Sheng}, {Shi}, {Shu}, {Su}, {Sun}, {Wang}, {Wang}, {Wang}, {Wang}, {Wei}, {Wei}, {Xue}, {Yan}, {Yang}, {Yuan}, {Yuan}, {Zhang}, {Zhang}, {Zhao}, \& {Zhao}}]{WFST23}
{Wang}, T., {Liu}, G., {Cai}, Z., {et~al.} 2023, Science China Physics, Mechanics, and Astronomy, 66, 109512, \dodoi{10.1007/s11433-023-2197-5}

\bibitem[{{Waxman} \& {Katz}(2017)}]{WaxmanKatz17}
{Waxman}, E., \& {Katz}, B. 2017, in Handbook of Supernovae, ed. A.~W. {Alsabti} \& P.~{Murdin}, 967, \dodoi{10.1007/978-3-319-21846-5_33}

\bibitem[{{Weisskopf} {et~al.}(2022){Weisskopf}, {Soffitta}, {Baldini}, {Ramsey}, {O'Dell}, {Romani}, {Matt}, {Deininger}, {Baumgartner}, {Bellazzini}, {Costa}, {Kolodziejczak}, {Latronico}, {Marshall}, {Muleri}, {Bongiorno}, {Tennant}, {Bucciantini}, {Dovciak}, {Marin}, {Marscher}, {Poutanen}, {Slane}, {Turolla}, {Kalinowski}, {Di Marco}, {Fabiani}, {Minuti}, {La Monaca}, {Pinchera}, {Rankin}, {Sgro'}, {Trois}, {Xie}, {Alexander}, {Allen}, {Amici}, {Andersen}, {Antonelli}, {Antoniak}, {Attin{\`a}}, {Barbanera}, {Bachetti}, {Baggett}, {Bladt}, {Brez}, {Bonino}, {Boree}, {Borotto}, {Breeding}, {Brienza}, {Bygott}, {Caporale}, {Cardelli}, {Carpentiero}, {Castellano}, {Castronuovo}, {Cavalli}, {Cavazzuti}, {Ceccanti}, {Centrone}, {Citraro}, {D'Amico}, {D'Alba}, {Di Gesu}, {Del Monte}, {Dietz}, {Di Lalla}, {Persio}, {Dolan}, {Donnarumma}, {Evangelista}, {Ferrant}, {Ferrazzoli}, {Ferrie}, {Footdale}, {Forsyth}, {Foster}, {Garelick}, {Gunji}, {Gurnee}, {Head}, {Hibbard}, {Johnson}, {Kelly}, {Kilaru}, {Lefevre},
  {Roy}, {Loffredo}, {Lorenzi}, {Lucchesi}, {Maddox}, {Magazzu}, {Maldera}, {Manfreda}, {Mangraviti}, {Marengo}, {Marrocchesi}, {Massaro}, {Mauger}, {McCracken}, {McEachen}, {Mize}, {Mereu}, {Mitchell}, {Mitsuishi}, {Morbidini}, {Mosti}, {Nasimi}, {Negri}, {Negro}, {Nguyen}, {Nitschke}, {Nuti}, {Onizuka}, {Oppedisano}, {Orsini}, {Osborne}, {Pacheco}, {Paggi}, {Painter}, {Pavelitz}, {Pentz}, {Piazzolla}, {Perri}, {Pesce-Rollins}, {Peterson}, {Pilia}, {Profeti}, {Puccetti}, {Ranganathan}, {Ratheesh}, {Reedy}, {Root}, {Rubini}, {Ruswick}, {Sanchez}, {Sarra}, {Santoli}, {Scalise}, {Sciortino}, {Schroeder}, {Seek}, {Sosdian}, {Spandre}, {Speegle}, {Tamagawa}, {Tardiola}, {Tobia}, {Thomas}, {Valerie}, {Vimercati}, {Walden}, {Weddendorf}, {Wedmore}, {Welch}, {Zanetti}, \& {Zanetti}}]{2022JATIS...8b6002W}
{Weisskopf}, M.~C., {Soffitta}, P., {Baldini}, L., {et~al.} 2022, Journal of Astronomical Telescopes, Instruments, and Systems, 8, 026002, \dodoi{10.1117/1.JATIS.8.2.026002}

\bibitem[{{Williams} {et~al.}(2026){Williams}, {G{\"a}nsicke}, {Stone}, {Koester}, {Davies}, {Tong}, {Wilson}, {Sahu}, {Swan}, {Beatty}, {Ram{\'\i}rez}, {Cunningham}, \& {Long}}]{2026NatAs.tmp..221W}
{Williams}, J.~T., {G{\"a}nsicke}, B.~T., {Stone}, N.~C., {et~al.} 2026, Nature Astronomy, \dodoi{10.1038/s41550-026-02983-7}

\bibitem[{{Yoon} {et~al.}(2025){Yoon}, {Tian}, \& {Chae}}]{2025ApJ...992..102Y}
{Yoon}, Y., {Tian}, Y., \& {Chae}, K.-H. 2025, ApJ, 992, 102, \dodoi{10.3847/1538-4357/ae0190}

\bibitem[{{Yuan} {et~al.}(2022){Yuan}, {Zhang}, {Chen}, \& {Ling}}]{EP22}
{Yuan}, W., {Zhang}, C., {Chen}, Y., \& {Ling}, Z. 2022, in Handbook of X-ray and Gamma-ray Astrophysics, ed. C.~{Bambi} \& A.~{Sangangelo}, 86, \dodoi{10.1007/978-981-16-4544-0_151-1}

\bibitem[{{Zhang} {et~al.}(2019){Zhang}, {Santangelo}, {Feroci}, {Xu}, {Lu}, {Chen}, {Feng}, {Zhang}, {Brandt}, {Hernanz}, {Baldini}, {Bozzo}, {Campana}, {De Rosa}, {Dong}, {Evangelista}, {Karas}, {Meidinger}, {Meuris}, {Nandra}, {Pan}, {Pareschi}, {Orleanski}, {Huang}, {Schanne}, {Sironi}, {Spiga}, {Svoboda}, {Tagliaferri}, {Tenzer}, {Vacchi}, {Zane}, {Walton}, {Wang}, {Winter}, {Wu}, {in't Zand}, {Ahangarianabhari}, {Ambrosi}, {Ambrosino}, {Barbera}, {Basso}, {Bayer}, {Bellazzini}, {Bellutti}, {Bertucci}, {Bertuccio}, {Borghi}, {Cao}, {Cadoux}, {Campana}, {Ceraudo}, {Chen}, {Chen}, {Chevenez}, {Civitani}, {Cui}, {Cui}, {Dauser}, {Del Monte}, {Di Cosimo}, {Diebold}, {Doroshenko}, {Dovciak}, {Du}, {Ducci}, {Fan}, {Favre}, {Fuschino}, {G{\'a}lvez}, {Gao}, {Ge}, {Gevin}, {Grassi}, {Gu}, {Gu}, {Han}, {Hong}, {Hu}, {Ji}, {Jia}, {Jiang}, {Kennedy}, {Kreykenbohm}, {Kuvvetli}, {Labanti}, {Latronico}, {Li}, {Li}, {Li}, {Li}, {Li}, {Limousin}, {Liu}, {Liu}, {Lu}, {Luo}, {Macera}, {Malcovati}, {Martindale},
  {Michalska}, {Meng}, {Minuti}, {Morbidini}, {Muleri}, {Paltani}, {Perinati}, {Picciotto}, {Piemonte}, {Qu}, {Rachevski}, {Rashevskaya}, {Rodriguez}, {Schanz}, {Shen}, {Sheng}, {Song}, {Song}, {Sgro}, {Sun}, {Tan}, {Uttley}, {Wang}, {Wang}, {Wang}, {Wang}, {Wang}, {Wang}, {Watts}, {Wen}, {Wilms}, {Xiong}, {Yang}, {Yang}, {Yang}, {Yu}, {Zhang}, {Zampa}, {Zampa}, {Zdziarski}, {Zhang}, {Zhang}, {Zhang}, {Zhang}, {Zhang}, {Zhang}, {Zhang}, {Zhang}, {Zhao}, {Zheng}, {Zhou}, {Zorzi}, \& {Zwart}}]{2019SCPMA..6229502Z}
{Zhang}, S., {Santangelo}, A., {Feroci}, M., {et~al.} 2019, Science China Physics, Mechanics, and Astronomy, 62, 29502, \dodoi{10.1007/s11433-018-9309-2}

\bibitem[{{Zimmerman} {et~al.}(2024){Zimmerman}, {Irani}, {Chen}, {Gal-Yam}, {Schulze}, {Perley}, {Sollerman}, {Filippenko}, {Shenar}, {Yaron}, \& et~al.}]{2024Natur.627..759Z}
{Zimmerman}, E.~A., {Irani}, I., {Chen}, P., {et~al.} 2024, Nature, 627, 759, \dodoi{10.1038/s41586-024-07116-6}

\end{thebibliography}
\end{document}